\documentclass[12pt]{article}
\usepackage{graphicx}
\usepackage{amsfonts}
\usepackage{amssymb,amsmath}
\usepackage{latexsym}
\usepackage{color}
\usepackage{cite}
\usepackage{bm}
\usepackage{ulem}
\input{colordvi.tex}

\renewcommand{\thefootnote}{\#\arabic{footnote}}

\begin{document}

\renewcommand{\thepage}{\arabic{page}}
\setcounter{page}{1}
\renewcommand{\thefootnote}{\#\arabic{footnote}}

%%%%%%%%%%%%%%%%%%%%%%%%%%%%%%%%%%%%%%%%%%%%%%%%%%%%%%%%%%%%%%%
\begin{titlepage}
%%%%%%%%%%%%%%%%%%%%%%%%%%%%%%%%%%%%%%%%%%%%%%%%%%%%%%%%%%%%%%%
\begin{center}

\vskip .5in

{\Large \bf 
Constraints on the neutrino parameters by future cosmological 21cm line 
and precise CMB polarization observations
}

\vskip .45in

{\large
Yoshihiko~Oyama$\,^1$, 
Kazunori~Kohri$\,^{2,3}$ and
Masashi~Hazumi$\,^{2,3,4}$
}

\vskip .45in

{\it
$^1$
Institute for Cosmic Ray Research, The University of Tokyo, 5-1-5 Kashiwanoha, Kashiwa, Chiba 277-8582, Japan\\
$^2$
Institute of Particle and Nuclear Studies, KEK, 1-1 Oho, Tsukuba, Ibaraki 305-0801, Japan \\
$^3$
The Graduate University for Advanced Studies (SOKENDAI), 1-1 Oho, Tsukuba, Ibaraki 305-0801, Japan \\
$^4$
Kavli IPMU (WPI), UTIAS, The University of Tokyo, 5-1-5 Kashiwanoha, Kashiwa, Chiba 277-8583, Japan
}

\end{center}

\vskip .4in

\begin{abstract}
Observations of the 21 cm line radiation coming from the epoch of reionization have 
a great capacity to study the cosmological growth of the Universe. 
Besides, CMB polarization produced by gravitational lensing has 
a large amount of information about the growth of matter fluctuations at late time. 
In this paper, we investigate their sensitivities to
the impact of neutrino property on the growth of density fluctuations, 
such as the total neutrino mass, the effective number of neutrino species (extra radiation),
and the neutrino mass hierarchy.
%
%We will show that by combining a precise CMB polarization observations 
%%%% 2015/10/28
We show that by combining a precise CMB polarization observation
%%%%
%
such as Simons Array with a 21 cm line observation 
such as Square kilometer Array (SKA) phase~1 
and a baryon acoustic oscillation observation
(Dark Energy Spectroscopic Instrument:DESI) 
%we can measure the impact of non-zero neutrino mass on the growth of density fluctuation
we can measure effects of non-zero neutrino mass on the growth of density fluctuation
if the total neutrino mass is larger than $0.1$~eV.
Additionally, the combinations can strongly improve errors of 
the bounds on the effective number of neutrino species
$\sigma (N_{\nu}) \sim  0.06-0.09$ at 95 \% C.L..
Finally, by using SKA phase~2,
we can determine the neutrino mass hierarchy at 95 \% C.L. 
%if the total neutrino mass is smaller than 0.1 eV.
if the total neutrino mass is similar to or smaller than 0.1 eV.
%
 %95 \% C.L
% to be  $ \Delta\Sigma m_{\nu}
%\sim  0.12$~eV and $ \Delta N_{\nu}\sim  0.38$ at  2$\sigma$.%95 \% C.L, respectively.
\end{abstract}
\end{titlepage}

\setcounter{footnote}{0}
%%%%%%%%%%%%%%%%%%%%%%%%%%%%%%%%%%%%%%%%%%%%%%%%%%%%%%%%%%%%%%%
%%%%%%%%%%%%%%%%%%%%%%%%%%%%%%%%%%%%%%%%%%%%%%%%%%%%%%%%%%%%%%%

%%%%%%%%%%%%%%%%%%%%%%%%%%%%%%%%%%%%%%%%%%%%%%%%%%%%%%%%%%%%%%%
\section{Introduction}
%%%%%%%%%%%%%%%%%%%%%%%%%%%%%%%%%%%%%%%%%%%%%%%%%%%%%%%%%%%%%%%

Due to the discovery of non-zero neutrino masses by Super-Kamiokande 
through neutrino oscillation experiments in 1998, 
the standard model of particle physics was forced to 
be modified so as to theoretically include the neutrino masses.

So far only the mass-squared differences of neutrino species
have been measured by neutrino oscillation experiments, 
which are reported  to be $\Delta m^2_{21}\equiv m_{2}^2 -m_{1}^2
=7.59^{+0.19}_{-0.21}\times 10^{-5} {\rm eV}^2$~\cite{Aharmim:2008kc}
and $\Delta m^2_{32}\equiv m_{3}^2 -m_{2}^2
=2.43^{+0.13}_{-0.13}\times 10^{-3} {\rm eV}^2$~\cite{Adamson:2008zt}.
However, absolute values and their hierarchical structure 
(normal or inverted) have not been obtained yet
although information on them is indispensable to build new particle physics models.

In particle physics, some new ideas and new future experiments
have been proposed to measure the absolute values
and/or determine the hierarchy of neutrino masses, 
e.g., through tritium beta decay in KATRIN experiment~\cite{KATRIN}, 
neutrinoless double-beta decay~\cite{GomezCadenas:2010gs}, 
atmospheric neutrinos in the proposed iron calorimeter at INO~\cite{INO,Blennow:2012gj}
and the upgrade of the IceCube detector (PINGU) \cite{Akhmedov:2012ah},
and long-baseline oscillation experiments, 
e.g., NO$\nu$A~\cite{Ayres:2004js},
J-PARC to Korea (T2KK)~\cite{Ishitsuka:2005qi,Hagiwara:2005pe} 
or Oki island (T2KO)~\cite{Badertscher:2008bp},
and CERN to Super-Kamiokande
with high energy (5~GeV) neutrino beam~\cite{Agarwalla:2012zu}.

On the other hand, 
such nonzero neutrino masses affect cosmology significantly
through suppression of growth of density fluctuation 
%due to the following reason.  
%
%In general relativity, 
%the density perturbation of relativistic particles
%hardly evolve at all on sub-horizon scales
%before they become non-relativistic. 
%
%Relativistic neutrinos erase 
because relativistic neutrinos have large large thermal velocity
and erase their own density fluctuations up to horizon scales 
due to their free streaming behavior.
%because they have large thermal velocity.
%
%Therefore, the density perturbations of neutrino
%hardly evolve at all on sub-horizon scales
%before they become non-relativistic,
%and their energy density does not contribute to 
%the gravitational growth of density perturbation of non-relativistic matter
%on small scales.
%
By measuring power spectra of density fluctuations,
we can constrain the total neutrino mass
$\Sigma~m_{\nu}$ \cite{Komatsu:2008hk,Komatsu:2010fb,Reid:2009xm,
Hannestad:2010yi,Elgaroy:2003yh,Reid:2009nq,Crotty:2004gm,
Goobar:2006xz,Seljak:2004xh,Ichikawa:2004zi,Seljak:2006bg,Fukugita:2006rm,
Ichiki:2008ye,Thomas:2009ae,RiemerSorensen:2011fe,
Hamann:2010pw,Saito:2010pw,Ade:2013zuv,
%%%%%%%2015/10/27 added%%%%%%%%
DiValentino:2013mt,Giusarma:2014zza,DiValentino:2015wba,DiValentino:2015ola,Ade:2015xua}
%%%%%%%%%%%%%%%%%%%%%%%%%%%%%%
%
and the effective number of neutrino species $N_{\nu}$
\cite{Komatsu:2008hk,Komatsu:2010fb,Reid:2009xm,Reid:2009nq,Crotty:2004gm,
Pierpaoli:2003kw,Crotty:2003th,Seljak:2006bg,Hamann:2010pw,Ade:2013zuv,
%%%%%%%2015/10/27 added%%%%%%%%
DiValentino:2013mt,Said:2013hta,Giusarma:2014zza,DiValentino:2015wba,
DiValentino:2015ola,Ade:2015xua} 
%%%%%%%%%%%%%%%%%%%%%%%%%%%%%%%
%
through observations of cosmic microwave background (CMB) 
anisotropies and large-scale structure (LSS).
%
%by using observations of cosmic microwave background (CMB) 
%anisotropies and large-scale structure (LSS),   
%
%The robust upper bound on $\Sigma~m_{\nu}$ has been obtained to be
%
%%%%%%%%%%%2015/10/27 added ref
%
We can also obtain additional information about $N_{\nu}$
from the CMB bispectrum \cite{DiValentino:2013uia}.
Furthermore, we can constraint on the neutrino isocurvature perturbations 
from CMB observations \cite{DiValentino:2011sv,DiValentino:2014eea}.
%
%%%%%%%%%%%%%%%%%%%%%%%%%%%%%%%
%
The robust upper bound on $\Sigma~m_{\nu}$ obtained so far is
$\Sigma m_{\nu} < 0.23$~eV (95 $\%$ C.L.) 
by the CMB observation by Planck (see Ref.~\cite{Ade:2013zuv,Ade:2015xua}).
%
%For forecasts for future CMB observations, see also
%Refs.~\cite{Lesgourgues:2005yv,dePutter:2009kn,Abazajian:2013oma,Wu:2014hta}.
%
%%%%%%%%%%%2015/10/27 added ref
%
For forecasts for future CMB observations, see also
Refs.~\cite{Lesgourgues:2005yv,dePutter:2009kn,Abazajian:2013oma,Wu:2014hta}.
For current constraints and forecasts 
for combinations of cosmological and laboratory experiments 
(oscillation and neutrinoless double-beta decay),
see also Ref.\cite{Gerbino:2015ixa}.
%
%%%%%%%%%%%%%%%%%%%%%%%

Moreover, by observing the power spectrum of 
cosmological 21 cm line radiation fluctuation, 
we will be able to obtain useful information on the neutrino
%masses~\cite{McQuinn:2005hk,Loeb:2008hg,Pritchard:2008wy,Pritchard:2009zz,Abazajian:2011dt,Oyama:2012tq}. 
masses~\cite{McQuinn:2005hk,Loeb:2008hg,Pritchard:2008wy,Pritchard:2009zz,Abazajian:2011dt,Oyama:2012tq,Oyama:2015vva},
and its  asymmetry (lepton number asymmetry) \cite{Kohri:2014hea}.
%
%%%%%%%%%%%2015/10/27 added ref
%
Besides, we can obtain further information about the neutrino masses
from observations of the HI distribution in the post-reionization era
\cite{Villaescusa-Navarro:2015cca}.
%
%%%%%%%%%%%%%%%%
%
That is because the 21 cm line radiation is emitted 
(1) long after the recombination (at a redshift $z \ll 10^3$ ) and 
(2) before an onset of the LSS formation. 
The former condition (1) gives us information on
smaller neutrino mass ($\lesssim 0.1 $~eV),
and the latter condition (2) means we
can treat only a linear regime of the matter perturbation, 
which can be analytically calculated unlike the LSS case.

In actual analyses, it is essential that we combine 
data of the 21 cm line with those of CMB
because the constrained cosmological parameter space 
is complementary to each other. 
For example, the former is quite sensitive to the dark energy density, 
but the latter is relatively insensitive to it. 
On the other hand, the former has only a mild sensitivity to
the normalization of matter perturbation, 
but the latter has an obvious sensitivity to it by definition. 
In pioneering work by~\cite{Pritchard:2008wy}, 
the authors tried to make a forecast for constraint on 
the neutrino mass hierarchy by combining Planck with 
future 21 cm line observations 
in case of relatively degenerate neutrino masses 
$\Sigma m_{\nu}\sim 0.3 $~eV.
Additionally, in our previous work~\cite{Oyama:2012tq},
we investigate the detectability of the mass hierarchy 
by combination of a futuristic 21 cm line observation 
(Omniscope~\cite{Tegmark:2008au,Tegmark:2009kv})
and CMB polarization observation (\textsc{Polarbear} or CMBPol)
when the total neutrino mass is relatively small value 
$\Sigma m_{\nu} \leq 0.3 $~eV.
The CMB B-mode polarization produced by CMB lensing
gives us more detailed information on the matter power spectrum 
\cite{Kohri:2013mxa} at later epochs.
In that work, we found that the combinations have 
enough sensitivity to distinguishing the neutrino mass hierarchy
when total neutrino mass is $\Sigma m_{\nu} \sim 0.1$~eV or less.

In this paper, as for 21 cm line observation,
we particularly focus on Square Kilometer Array (SKA)~\cite{SKA},
which is planned in the 2020s and 
%more realistic observation than Omniscope,
more realistic observation than Omniscope in near future,
and forecast allowed parameter regions for 
the neutrino masses, the effective number of neutrino species 
and the mass hierarchy.
In our analysis, we consider combinations of SKA
with ground-based CMB polarization observations,
such as \textsc{Polarbear}-2 or Simons Array,
which can measure the gravitational lensing of CMB
with a high degree of accuracy.
Besides, we take into account 
%including the information of baryon acoustic oscillation (BAO) observation,
including the information of baryon acoustic oscillation (BAO) observations,
such as Dark Energy Spectroscopic Instrument (DESI) \cite{DESI:web}.
%
%and how the detectability is improved.
%

This paper is organized as follows. 
In Section~\ref{sec:21cm}, we briefly review 21 cm line observation,
analytical methods used in this paper,
and how neutrino properties are constrained by the 21 cm line observation. 
In Section~\ref{sec:CMB} and \ref{sec:BAOFisher},
we explain analytical methods of CMB and BAO observation, respectively.
We show our results in Section~\ref{sec:results}, 
and Section~\ref{sec:conclusion} is devoted to our conclusion.

%%%%%%%%%%%%%%%%%%%%%%%%%%%%%%%%%%%%%%%%%%%%%%%%%%%%%%%%%%%%%%%
\section{21 cm line observation}
\label{sec:21cm}
%%%%%%%%%%%%%%%%%%%%%%%%%%%%%%%%%%%%%%%%%%%%%%%%%%%%%%%%%%%%%%%

In this section, we briefly review basic methods to use the 21 cm line
observation as a cosmological probe.
%For further details, we refer readers to 
For further details, we refer to 
Refs.~\cite{Furlanetto:2006jb,Pritchard:2011xb}.

%%%%%%%%%%%%%%%%%%%%%%%%%%%%%%%%%%%%%%%%%%%%%%%%%%%%%%%%%%%%%%%
\subsection{Power spectrum of 21 cm  radiation}
%%%%%%%%%%%%%%%%%%%%%%%%%%%%%%%%%%%%%%%%%%%%%%%%%%%%%%%%%%%%%%%

%The 21 cm line of the neutral hydrogen atom is emitted by  hyperfine
%splitting of the 1S ground state  due to an interaction of  magnetic
%moments of proton and  electron. 
The 21 cm line of neutral hydrogen atom is emitted by transition
between the hyperfine levels of the 1S ground state,
and the hyperfine structure is induced by an interaction of magnetic moments 
between proton and electron.
%
%The energy difference of the hyperfine structure is $\Delta E \sim 5.8\times 10^{-6}$eV,
%and this energy corresponds to the frequency $\nu_{21} \simeq 1.4$GHz (the wave length
%is $\lambda \simeq 21$cm). Therefore this spectral line is called the 21 cm line
%(see Fig.\ref{fig:21cmline}).
%
It can be observed as the differential brightness temperature relative to the
CMB temperature $T_{\rm CMB}$:

\begin{eqnarray}
\Delta T_{b} \left(\mbox{\boldmath $r$},z \right) 
&=&
\frac{3c^{3}hA_{21}}{32\pi k_{B}\nu_{21}^{2}}
\frac
{x_{HI}(\mbox{\boldmath $r$},z)n_{H}(\mbox{\boldmath $r$},z) } {(1+z)H(z)}
\left(
1-\frac{T_{{\rm CMB}}(\mbox{\boldmath $r$},z) }{T_{S}(\mbox{\boldmath $r$},z)}
\right)
%\nonumber\\
%&&
%\times
\left(
1-\frac{1+z}{H(z)}
\frac{dv_{p\|}(\mbox{\boldmath $r$},z)}{dr_{\|}}
\right), \label{eq:obsbrightness2}
\end{eqnarray}
%
%
%where $A_{21} \simeq 2.869 \times 10^{-15}{\rm s^{-1}}$ 
where 
$\mbox{\boldmath $r$}$ is the comoving coordinate of the 21 cm radiation,
$z$ represents the redshift,
$A_{21} \simeq 2.869 \times 10^{-15}{\rm s^{-1}}$ 
is the spontaneous decay rate of 21 cm line transition,
$\nu_{21} \simeq 1.42$ GHz is the 21 cm line frequency,
$n_{H}$ is the number density of hydrogen
and $x_{HI}$ is the fraction of neutral hydrogen.
$T_S$ is the spin temperature, which is
defined by $n_1/ n_0 = 3 \exp ( - T_{21} / T_S)$,
where $n_0$ and $n_1$ are the number densities of singlet
and triplet states of neutral hydrogen atom, respectively.
Here $T_{21}= hc / k_B
\lambda_{21}$ is the temperature corresponding to 21 cm line,
and $\lambda_{21}$ is its wavelength.
$dv_{p\parallel} / dr_{\parallel}$ is the gradient of peculiar velocity 
along the line of sight.

From now on, we assume that $T_S \gg T_{\rm CMB}$ 
because we focus on the epoch of reionization (EOR)
during which this condition is well satisfied.
In general, the brightness temperature is sensitive 
to details of inter-galactic medium (IGM).
However, with a few reasonable assumptions, 
we can eliminate this dependence from the 21 cm line 
brightness temperature~\cite{Madau:1996cs,Furlanetto:2006tf,Pritchard:2008da}.
At the epoch of reionization long after star formation begins,
%X-ray background produced by early stellar remnants has heated the IGM.
X-ray background produced by early stellar remnants heats the IGM.
Therefore, gas kinetic temperature $T_{K}$ becomes much higher than
the CMB temperature $T_{{\rm CMB}}$.  
Furthermore, the star formation produces 
a large amount of Ly$\alpha$ photons sufficient to couple
$T_{S}$ to $T_{K}$ through the Wouthuysen-Field effect~\cite{Wouthuysen:1952,Field:1958}.
In this scenario, 
%we are justified in taking $T_{{\rm CMB}} \ll T_{K} \sim T_{S} $ 
%at $z \lesssim 10$, and $\Delta T_{b}$ does not depend on $T_{S}$.
$T_{{\rm CMB}} \ll T_{K} \sim T_{S} $ are justified
at $z \lesssim 10$, and $\Delta T_{b}$ does not depend on $T_{S}$.

Next, we are going to consider fluctuations of $\Delta T_b ({\bm r})$. 
By expanding the hydrogen number density $n_H$ 
and the ionization fraction $x_i$ ($x_i=1-x_{HI}$) as 
$n_H({\bm r}) = \bar{n}_H (1 + \delta ({\bm r}))$ 
and $x_i({\bm r}) = \bar{x}_i (1 + \delta_x ({\bm r}) )$, 
we can rewrite Eq.~\eqref{eq:obsbrightness2} to be
\begin{equation}
%\label{ }
\Delta T_b ({\bm r})  
=  
\Delta \bar{T}_b \left( 1 - \bar{x}_i ( 1+ \delta_x ({\bm r}) ) \right) 
(1+ \delta ({\bm r}) )
\left(
1-\frac{1+z}{H(z)}
\frac{dv_{p\|}(\mbox{\boldmath $r$},z)}{dr_{\|}}
\right),
\end{equation}
where we assume that $T_S \gg T_{\rm cmb}$ 
and neglect the term %of spin temperature dependence.
including spin temperature.
$\Delta\bar{T}_b$ is the spatially averaged 
differential brightness temperature at redshift $z$ and given by
\begin{equation}
%\label{ }
\Delta \bar{T}_b%(z)
\simeq 26.8 
\left(\frac{1-Y_p}{1-0.25} \right)
\left( \frac{\Omega_b h^2}{0.023} \right)
\left( \frac{0.15}{\Omega_m h^2} \frac{1+z}{10} \right)^{1/2}~{\rm mK},
\end{equation}
where $Y_p$ is the primordial $^4$He mass fraction. %Helium fraction.

By denoting the fluctuation of $\Delta T_b$ as 
$\delta (\Delta T_b ({\bm x})) \equiv \Delta T_b({\bm x}) - \bar{x}_{H}\Delta \bar{T}_b$,
the 21 cm line power
spectrum $P_{21} ({\bm k})$ in the $k$-space is defined by
\begin{equation}
\label{eq:power}
\left\langle
\delta (\Delta T^\ast_b ({\bm k})) \delta (\Delta T_b ({\bm k}'))  \right\rangle
= (2\pi)^3 \delta^3 ( \bm{k-k'}) P_{21} ({\bm k}).
\end{equation}
By treating the peculiar velocity $ \delta_v \equiv (1+z)(dv_{p\|} / dr_{\|})/H(z)$
as a perturbation and using that its Fourier transform is given by 
$ \delta_v  ({\bm k})= -\mu^2 \delta  ({\bm k})$ %with $\mu = \hat{\bm k}\cdot \hat{\bm n}$
%being the cosine of the angle between the wave vector and the line of
%sight, the power spectrum can be written as
($\mu = \hat{\bm k}\cdot \hat{\bm n}$
is the cosine of the angle between the wave vector and the line of sight), 
the power spectrum can be written as
\begin{equation}
%\label{ }
P_{21} ({\bm k}) 
=
P_{\mu^0} (k) + \mu^2 P_{\mu^2} (k)  + \mu^4 P_{\mu^4} (k), 
\end{equation}
where  $k = |{\bm k}|$ and
\begin{eqnarray}
P_{\mu^0}  & = & \mathcal{P}_{\delta \delta} - 2 \mathcal{P}_{x\delta} + \mathcal{P}_{xx},
 \\
P_{\mu^2}   & = & 2 \left( \mathcal{P}_{\delta \delta} - \mathcal{P}_{x\delta} \right),  
\\
P_{\mu^4}  & = &   \mathcal{P}_{\delta \delta}. 
\end{eqnarray}
Here, $ \mathcal{P}_{\delta \delta} \equiv (\Delta \bar{T}_b)^2 \bar x_{HI}^2P_{\delta \delta},
\mathcal{P}_{x \delta} \equiv (\Delta \bar{T}_b)^2 \bar{x}_i \bar{x}_{HI} 
P_{x \delta} $ and $ \mathcal{P}_{x x} \equiv (\Delta \bar{T}_b)^2 \bar{x}_i^2 P_{ x x} $,
where $P_{\delta\delta}, P_{x\delta}$ and
$P_{xx}$ are the power spectra defined in the same manner as
Eq.~\eqref{eq:power} for 
the fluctuation of hydrogen number density $\delta$ 
and that of ionization fraction $\delta_x$.
%
%Because $\delta$ represents the fluctuation of hydrogen number density,
%$P_{\delta\delta}$ traces that of matter, which includes the
Therefore,
$P_{\delta\delta}$ traces the fluctuation of matter, which includes
information on cosmological parameters such as neutrino mass.
%primordial power spectrum.  

$P_{x\delta}$ and $P_{xx}$ can be neglected 
as long as we consider the era when the IGM is completely neutral. 
However, after the reionization starts,
these two spectra significantly contribute 
to the 21 cm line power spectrum.  
Although a rigorous evaluation of these power spectra
may need some numerical simulations, 
we adopt the treatment given in Ref.~\cite{Mao:2008ug}, 
where it is assumed that 
$\mathcal{P}_{x\delta}$ and $\mathcal{P}_{xx}$ have specific forms
which match simulations of radiative transfer 
in Refs.~\cite{McQuinn:2006et,McQuinn:2007dy}.  
The explicit forms of the power spectra are parametrized to be %given by 
\begin{eqnarray}
\label{eq:Pxx}
\mathcal{P}_{x x}  (k) 
& = & 
b_{xx}^2 \left[ 1 + \alpha_{xx} (k R_{xx}) + (k R_{xx})^2 \right]^{-\gamma_{xx} / 2} \mathcal{P}_{\delta\delta} (k), \\
\label{eq:Pxdelta}
\mathcal{P}_{x \delta} (k) 
& = &  
b_{x\delta}^2 ~e^{ - \alpha_{x\delta} (k R_{x\delta}) - (k R_{x\delta})^2} \mathcal{P}_{\delta\delta} (k),
\end{eqnarray}
where $b_{xx}$, $b_{x\delta}$, $\alpha_{xx}$, $\gamma_{xx}$ and $\alpha_{x\delta}$
are parameters which characterize the amplitudes and the shapes of the spectra.
$R_{xx}$ and $R_{x\delta}$ represent the effective size of ionized bubbles.  
%
%The values of these parameters which we use in our analysis 
%are listed in Table~\ref{tab:Pxx_xdelta}.
%
In our analysis,
we adopt the values listed in Table~\ref{tab:Pxx_xdelta}
as the fiducial values of these parameters.

%%%%%%%%%%%%%%%%%%%%%%%%%%%%%%%%%%%%%%%%%%%%%%%%%%%%%%%%%%%%%%%
\begin{table}[t]
  \centering 
  \begin{tabular}{ccccccccc}
\hline \hline
~~$z$~~ & ~~$\bar{x}_H$~~
& ~~$b_{xx}^2$~~ & ~~$R_{xx}$~~  & ~~$\alpha_{xx}$~~ & ~~$\gamma_{xx}$~~ 
& ~~$b_{x\delta}^2$~~ & ~~$R_{x\delta}$~~  & ~~$\alpha_{x\delta}$~~ \\ 
&  &  & $[{\rm Mpc}]$& & & &$[{\rm Mpc}]$ \\
\hline
$9.2$  &  $0.9$ & $0.208$ & $1.24$ & $-1.63$ & $0.38$ & $0.45$ & $0.56$   & $-0.4$ \\ 
$8.0$  &  $0.7$ & $2.12$   & $1.63$ & $-0.1$   & $1.35$ & $1.47$ & $0.62$   & $0.46$ \\ 
$7.5$  &  $0.5$ & $9.9$     & $1.3$   & $1.6$    & $2.3$   & $3.1$   & $0.58$   & $2.0$ \\ 
$7.0$  &  $0.3$ & $77.0$   & $3.0$   & $4.5$    & $2.05$ & $8.2$   & $0.143$ & $28.0$ \\
\hline \hline
\end{tabular}
\caption{Fiducial values of the parameters in $\mathcal{P}_{xx}(k)$
and $\mathcal{P}_{x\delta}(k)$ 
(See Eqs.~\eqref{eq:Pxx} and \eqref{eq:Pxdelta}) \cite{Mao:2008ug}.
}\label{tab:Pxx_xdelta}
\end{table}
%%%%%%%%%%%%%%%%%%%%%%%%%%%%%%%%%%%%%%%%%%%%%%%%%%%%%%%%%%%%%%%

%We note that 
%observations of 21 cm line radiation 
%do not directly measure the wave number ${\bm k}$ 
%nor the power spectrum in the $k$-space $P_{21} ({\bm k})$.
%
%Instead, an experiment measures 
%the angular location on the sky and the frequency,
%which can be specified by the following vector
%
We note that
%the wave number ${\bm k}$ 
the power spectrum in the $k$-space $P_{21} ({\bm k})$
are not directrly measured by 21 cm line observations.
Instead, 
the angular location on the sky and the frequency 
are measured by an experiment,
and they 
%the angular location on the sky and the frequency,
can be specified by the following vector
\begin{equation}
%\label{ }
{\bm \Theta} = \theta_x \hat{e}_x + \theta_y \hat{e}_y + \Delta f \hat{e}_z 
\equiv  {\bm \Theta}_\perp + \Delta f \hat{e}_z,
\end{equation}
where $\Delta f$ represents the frequency difference 
from the central redshift $z$ of a given redshift bin.
Then, we can define the Fourier dual of ${\bm \Theta}$ as
\begin{equation}
%\label{ }
{\bm u} \equiv u_x \hat{e}_x + u_y \hat{e}_y + u_\parallel \hat{e}_z 
\equiv {\bm u}_\perp + u_\parallel \hat{e}_z.
\end{equation}
Notice that $u_\parallel$ has the unit of time
since it is the Fourier dual of $\Delta f$.
%
%Assuming that the sky is flat\footnote{
By using the flat-sky approximation~\footnote{
%%$$$$$$$$$$$$
  Even if we consider all-sky experiments, the flat-sky approximation
  can be valid as long as we analyze 
  the data in a lot of small patches of the sky \cite{Mao:2008ug}.
%%%%%%%%%%%%%%
}, 
we can linearize the relation between ${\bm r}$ and ${\bm \Theta}$,
and it is written as
%
%Denoting the vector perpendicular to the line of sight as 
%${\bm  r}_\perp$, we have the relations
%
\begin{equation}
%\label{ }
{\bm \Theta}_\perp = {\bm r}_\perp / d_A (z), 
\qquad
\Delta f = \Delta r_\parallel / y(z),
\end{equation}
where ${\bm  r}_\perp$
is the vector perpendicular to the line of sight,
$\Delta r_\parallel$ is the comoving distance interval 
corresponding to the frequency intervals $\Delta f$,
$d_A (z)$ is the comoving angular diameter distance, 
and $y(z) \equiv \lambda_{21} (1+z)^2 / H(z)$. 
Then, the relation between ${\bm k}$ and ${\bm u}$ %are
can be written as
\begin{equation}
%\label{ }
{\bm u}_\perp = d_A {\bm k}_\perp, 
\qquad
u_\parallel = y k_\parallel.
\end{equation}
The power spectrum of $\Delta T_b$ in the $u$-space can be defined 
in the same manner as the treatment in the $k$-space.
By replacing ${\bm k}$ with ${\bm u}$ in Eq.~\eqref{eq:power}
and using the relation between ${\bm k}$ and ${\bm u}$, 
%the power spectra in each space are connected as
those two spectra  are related each other by
\begin{equation}
%\label{ }
P_{21} ({\bm u})  = \frac{1}{d_A(z)^2  y (z)} P_{21} ({\bm k}).
\end{equation}
We use the $u$-space power spectrum in the following analysis
because this quantity is directly measurable 
without any cosmological assumptions.

%We perform our analyses in terms of $P_{21}(\bold{u},z)$ since this
%quantity is directly measurable without any cosmological assumptions.

%%%%%%%%%%%%%%%%%%%%%%%%%%%%%%%%%%%%%%%%%%%%%%%%%%%%%%%%%%%%%%%
\subsection{Forecasting methods}\label{subsec:Forecast}
%%%%%%%%%%%%%%%%%%%%%%%%%%%%%%%%%%%%%%%%%%%%%%%%%%%%%%%%%%%%%%%

%%%%%%%%%%%%%%%%%%%%%%%%%%%%%%%%%%%%%%%%%%%%%%%%%%%%%%%%%%%%%%%
\subsubsection{Fisher matrix of 21 cm line observation}\label{subsubsec:Forecast}
%%%%%%%%%%%%%%%%%%%%%%%%%%%%%%%%%%%%%%%%%%%%%%%%%%%%%%%%%%%%%%%

Here, we provide a brief review of the Fisher matrix analysis 
for the 21 cm observations. 
In order to forecast errors of cosmological parameters, 
we use the Fisher matrix analysis \cite{Tegmark:1996bz}.
The Fisher matrix of 21 cm line observations is given by \cite{McQuinn:2005hk}
\begin{equation}
\label{eq:Fisher_21}
F^{({\rm 21cm})}_{\alpha\beta} = 
\sum_{\rm pixels}  
\frac{1}{[ \delta P_{21}({\bm u}) ]^2} 
\left( \frac{\partial P_{21} ({\bm u})}{\partial \theta_{\alpha}} \right)
\left( \frac{\partial P_{21} ({\bm u})}{\partial \theta_{\beta}} \right),
\end{equation}
where $\delta P_{21}({\bm u})$ is the error 
in the power spectrum measurements for a Fourier pixel ${\bm u}$, 
%and $\theta_{i}$ represents cosmological parameters.  
and $\theta_{i}$ represents a cosmological parameter with its index "$i$".  
The 1 $\sigma$ error of the parameter $\theta_{i}$ 
is evaluated by the Fisher matrix,
and it is given by
%
%The Fisher matrix determines the 1 $\sigma$  errors of 
%the parameter $\theta_{i}$ to be
%%
\begin{align}
    \Delta \theta_{\alpha} \geq \sqrt{(F^{-1})_{\alpha\alpha}}.
\end{align}
When we differentiate $P_{21}({\bm u})$ with respect to 
the cosmological parameters, 
we fix ${\cal P}_{\delta \delta}(k)$ 
in Eqs.~(\ref{eq:Pxx})~and~(\ref{eq:Pxdelta})
so that we get conservative evaluations for %evaluation of 
errors of cosmological parameters.
In this situation,
the information of the matter distribution only 
comes from the ${\cal P}_{\delta \delta}(k)$ 
terms in $P_{\mu^0},P_{\mu^2}$ and $P_{\mu^4}$.
The error of the power spectrum
$\delta P_{21}({\bm u})$ consists of sample variances and experimental
noises, and is written by 
\begin{equation}
%\label{ }
\delta P_{21}({\bm u}) = \frac{ P_{21}({\bm u})  + P_N (u_{\perp}) }{N_c^{1/2} },
\label{eq:variance_21cm}
\end{equation}
%
%The first term on the right hand side gives the one from the sample variance.
where the first term on the right hand side 
represents the contribution from sample variance,
and the second term gives  that of experimental noise, respectively.
Here, $N_c = 2 \pi k_\perp \Delta k_\perp \Delta k_\parallel V(z) /(2\pi)^3$
is the number of independent cells in an annulus summing over the azimuthal angle,
$V(z) = d_A(z)^2 y(z) B \times {\rm FoV} $ is the survey volume,
where $B$ is the bandwidth,
and 
% FoV$= \lambda^2 /A_e$ 
FoV $\propto \lambda^{2}$ is the field of view of an interferometer.
%

%%%%%%%%%%%%%%%%%%%%%%%%%%%%%%%%%%%%%%%%%%%%%%%%%%%%%%%%%%%%%%%
%\section{Specifications of the experiments}\label{sec:21cm_spec}
\subsubsection{Specifications of the experiment}\label{sec:21cm_spec}
%%%%%%%%%%%%%%%%%%%%%%%%%%%%%%%%%%%%%%%%%%%%%%%%%%%%%%%%%%%%%%%

%%%%%%%%%%%%%%%%%%%%%%%%%%%%%%%%%%%%%%%%%%%%%%%%%%%%%%%%%%%%%%%
\begin{table}[t]
\centering 
\begin{tabular}{cccccccc}
\hline \hline
   & $N_{\rm ant}$& $A_e (z=8)$& $L_{\rm min}$
   & $L_{\rm max}$& ${\rm FoV}(z=8)$ & $t_{0}$&  $z$ \\ 
   & &$[{\rm m}^2]$&$[{\rm m}]$
   & $[{\rm km}]$ &$[{\rm deg}^2]$&[hour per field]&\\
\hline
SKA1  & $911 \times 1/2 $ & $443$ & $35$ & $6$ & $13.12 $  & 1000 & $6.8-10$ \\
\hline
SKA2  & $911 \times 4   $ & $443$ & $35$ & $6$ & $13.12 $  & 1000 & $6.8-10$ \\
\hline \hline
\end{tabular}
\caption{
Specifications for 21 cm line experiments adopted in the current analysis.
We assume that for SKA phase~1 (SKA1) (re-baseline design),
the number of antennae is half as many as
%the originally planned SKA1, which has 911 antennae,
that of the originally planned SKA1, which has 911 antennae,
and for SKA phase~2 (SKA2), 
the number of antennae is 4 times as many as that of originally planned SKA1.
%
%Hence, we take its noise power spectrum  to be 1/16 of 
%the original SKA1, and the other specifications to be the same values.
Additionally, for SKA 1 and 2, we assume that 
%these experiments observe multiple fields,
multiple fields are observed by using
these experiments,
and the number of fields is $N_{{\rm filed}}=4$ or $8$
in our analyses.
%
%\cite{Mellema:2012ht}.
%
Then, the effective field of view is
${\rm FoV_{SKA}} = 13.21 \times N_{{\rm filed}}$ $[{\rm deg}^2]$.
}
\label{tab:21obs}
\end{table}
%%%%%%%%%%%%%%%%%%%%%%%%%%%%%%%%%%%%%%%%%%%%%%%%%%%%%%%%%%%%%%%

%Here, we show the specifications 
%of the 21 cm line observations which are considered in this thesis.
%of the 21 cm line observation which is considered in this paper.

We show the specifications 
%of the 21 cm line observations which are considered in this thesis.
of the 21 cm line observation which is considered in this paper.
%
%%%%%%%%%%%%%%%%%%%%%%%%%%%%%%%%%%%%%%%%%%%%%%%%%%%%%%%%%%%%%%%
%\subsection{Specifications of experiments}
%%%%%%%%%%%%%%%%%%%%%%%%%%%%%%%%%%%%%%%%%%%%%%%%%%%%%%%%%%%%%%%
%
%%%%%%%%%%%%%%%%%
\subsubsection*{Survey range}
%%%%%%%%%%%%%%%%%
%
In our analyses, we consider the redshift range $ z = 6.75 - 10.05$,
which we divide into 4 bins: $z = 6.75 - 7.25, 7.25 - 7.75, 7.75-8.25$
and $8.25 -  10.05$.  For the wave number, 
we set its minimum cut off $k_{{\rm min} \parallel} = 2 \pi / (y B) $ 
to avoid foreground contaminations \cite{McQuinn:2005hk},
and take its maximum value $ k_{\rm max} = 2~{\rm Mpc}^{-1}$ 
in order not to be affected by a nonlinear effect which
becomes important on $k \ge k_{\rm max}$. 
For methods of foreground removals, see also recent discussions about
the independent component analysis (ICA) algorithm
(FastICA)~\cite{Chapman:2012yj} which will be developed in terms of the
ongoing LOFAR observation~\cite{LOFAR}.

%%%%%%%%%%%%%%%%%%
\subsubsection*{Noise power spectrum}
%%%%%%%%%%%%%%%%%%

The noise power spectrum, $P_N (u_{\perp})$ appeared  %denoted as $P_N (u_{\perp})$ 
in Eq.~(\ref{eq:variance_21cm}) is given by
%
%From Eq.(\ref{eq:noisevariance2}), the noise power spectrum of a interferometer 
%is given by
%
\begin{equation}
P_N (u_{\perp}) 
= \left( \frac{\lambda^{2} (z) T_{\rm sys} (z)  }{A_e (z)} \right)^2 
\frac{1}{t_0 n(u_\perp)},
\end{equation}
where, the system temperature $T_{\rm sys}$ is estimated to be
$T_{\rm sys} = T_{{\rm sky}} + T_{{\rm rcvr}}$,
and is dominated by the sky temperature due to synchrotron radiation.
Here, $T_{{\rm sky}} = 60 (\lambda/[m])^{2.55} $ [K] 
is the sky temperature, and
$T_{{\rm rcvr}} = 0.1T_{{\rm sky}} + 40 $[K] 
is the receiver noise \cite{SKA}.
In addition,
the  effective collecting area is proportional to the square of the 
observed wave length $A_e \propto \lambda^{2} $.
The number density of the baseline $n(u_\perp)$ 
depends on an actual %realization of 
antenna distribution.

To obtain the future cosmological constraints from 21 cm experiments,
we consider SKA (phase~1 and phase~2) \cite{SKA,Mellema:2012ht}
%and Omniscope \cite{Tegmark:2009kv,Tegmark:2008au},
,
whose specifications are shown in Table~\ref{tab:21obs}.
%
%In the analysis of  the total neutrino mass,
%the neutrino number of species and the neutrino mass
%hierarchy (the Chapter \ref{Chap:result_mass}), 
%we only estimate the sensitivity of SKA.
%In that of the lepton asymmetry of the Universe
%(the Chapter \ref{chap:result_lepton}),
%we take account of both the experiments.
%
In order to calculate the number density of the baseline $n(u_\perp)$,
%we assume a realization of antenna distributions for these arrays as follows.
we have to determine a realization of antenna distributions.
Recently, SKA phase~1 re-baseline design was determined,
and its total collecting area is one-half as large as 
that of the originally planned SKA phase~1.
Therefore, for SKA1 (re-baseline design),
we assume that the number of antennae is half as many as 
that of the originally planned SKA1, which has 911 antennae, 
and for SKA phase~2 (SKA2),  the number of antennae is 
4 times as many as that of the originally planned SKA1.
%
%in the same manner as Ref. \cite{Mao:2008ug}. 

The number density of the baseline of the originally planned SKA1
is determined as follows. 
We take 95\% (866) of the total antennae (stations)
distributed with a core region of radius 3000 m,
and the distribution has an antenna density profile 
of the originally planned SKA1 $\rho_{{\rm origSKA1}}(r)$ 
($r$: a radius from center of the array) as follows \cite{Kohri:2013mxa},
\begin{align}
\rho_{{\rm origSKA1}}(r) = 
\left\{
\begin{array}{lll}
 \rho_{0}r^{-1},     &\rho_{0} \equiv \frac{13}{16\pi\left(\sqrt{10}-1\right) }  \ {\rm m}^{-2}
& \hspace{60pt} r \leq 400 \ {\rm m},\\
 \rho_{1}r^{-3/2},  &\rho_{1} \equiv \rho_{0} \times 400^{1/2}, & \ \ \ 400 \ {\rm m} < r \leq 1000 \ {\rm m}, \\
 \rho_{2}r^{-7/2},  &\rho_{2} \equiv \rho_{1} \times 1000^{2}, & \ \ 1000 \ {\rm m} < r \leq 1500 \ {\rm m}, \\
 \rho_{3}r^{-9/2},  &\rho_{3} \equiv \rho_{2} \times 1500 ,          & \ \ 1500 \ {\rm m} < r \leq 2000 \ {\rm m}, \\
 \rho_{4}r^{-17/2},&\rho_{4} \equiv \rho_{3} \times 2000^{4}, & \ \ 2000  \ {\rm m} < r \leq 3000 \ {\rm m}. \\
\end{array}
\right.
\end{align}
Here, we assume an azimuthally symmetric distribution of antennae in SKA.
In this analysis, we ignore measurements from the sparse distribution of 
the remaining 5\% of the total antennae that are outside this core region.
This distribution agrees with the specification of 
the originally planned SKA1 baseline design.

By using this distribution, we can calculate the number density of baseline 
of the originally planned SKA1 $n_{{\rm origSKA1}}(u_\perp)$.
For SKA1 (re-baseline design) and SKA2,
we assume that these number densities of baseline are
\begin{eqnarray}
n_{{\rm SKA1}}(u_\perp) 
&=& n_{{\rm origSKA1}}(u_\perp) \times \left( \frac{1}{2} \right)^{2}, \\
n_{{\rm SKA2}}(u_\perp)
&=& n_{{\rm origSKA1}}(u_\perp) \times 4^{2},
\end{eqnarray}
where $n_{{\rm SKA1}}(u_\perp)$ 
is the number density of baseline of SKA1,
and $n_{{\rm SKA2}}(u_\perp)$ is that of SKA2, respectively.

%For SKA phase2, we assume that it has the 10 times larger
%total collecting area than the phase1.
%Hence, we take its noise power spectrum  to be 1/100 of the phase1.
%
%We assume that the other specifications of SKA phase2 are 
%the same as values of the phase1.
%
%For Omniscope , which is a future square-kilometer collecting area array
%optimized for 21 cm tomography, we take all of antennae distributed 
%with a filled nucleus in the same manner as Ref. \cite{Mao:2008ug}.

%%%%%%%%%%%%%%%%%%%%%%%%%%%%%%%%%%%%%%
\subsection{Contribution of residual foregrounds}
%%%%%%%%%%%%%%%%%%%%%%%%%%%%%%%%%%%%%%

Here, we consider the some existing residual foregrounds.
For the 21 cm line observation, 
we take account of the most dominant galactic foreground, 
namely the synchrotron radiation.
We assume that
the foreground subtraction can be done down to a given level,
and treat the contribution of residual foreground 
as the Gaussian random field.
Then, we introduce the following 
effective noise including the contribution of residual foreground \cite{Oyama:2015vva},
%
%$\Delta V^{{\rm RFg}}$,
%
\begin{eqnarray}
P_{N,{\rm eff}} (\mbox{\boldmath $u$}_{\perp})
      &=&     
      \left(
      \frac{\lambda^{2}T_{sys}}{A_{e}}
      \right)^{2}
      \frac{1}{n_{b}(\mbox{\boldmath $u$}_{\perp})t_{0}}
      + (\sigma^{{\rm RFg}}_{{\rm 21cm}}\times 1 \ {\rm MHz})
      C^{{\rm Fg}}(\mbox{\boldmath $u$}_{\perp},\nu_{*}),
      \label{eq:effective_noise_power}
\end{eqnarray}
where % we regards $\nu_{m} \sim \nu_{*}$,
$\nu_{*}$ is the central frequency value in the frequency band,
and $C^{{\rm Fg}}(\mbox{\boldmath $u$}_{\perp},\nu)$ 
is the power of foreground.
In Eq.(\ref{eq:effective_noise_power}),
we introduce a foreground removal parameter 
$\sigma^{{\rm RFg}}_{{\rm 21cm}}$, %which is defined as
%$B C^{{\rm RFg}}(\mbox{\boldmath $u$}_{\perp i},\nu) 
%=
%(\sigma^{{\rm RFg}}\times 1{\rm MHz})C^{{\rm Fg}}(\mbox{\boldmath $u$}_{\perp i},\nu)$
%(B/{\rm MHz}) C^{{\rm RFg}}(\mbox{\boldmath $u$}_{\perp i},\nu) $
%$B C^{{\rm RFg}}(\mbox{\boldmath $u$}_{\perp i},\nu) 
%=
%(\sigma^{{\rm RFg}}\times 1{\rm MHz})C^{{\rm Fg}}(\mbox{\boldmath $u$}_{\perp i},\nu)$
%
and assume 
$\sigma^{{\rm RFg}}_{{\rm 21cm}}=10^{-7}$ 
%(this value corresponds to $0.03$\% at the signal)
%in our analysis.
in forecasts of constraints on 
$\Sigma m_{\nu}$ and $N_{\nu}$ (Sec.~\ref{subsec:const_N_nu}),
and $\sigma^{{\rm RFg}}_{{\rm 21cm}}=10^{-8}$
%(this value corresponds to $0.01$\% at the signal)
in those of constrains on $\Sigma m_{\nu}$ and the mass hierarchy 
(Sec.~\ref{subsec:const_hie}).
When we include the contribution of residual foregrounds, % in our analysis,
we use this effective noise as the 21cm noise power spectrum.
%
%From now on, we introduce the foreground removal parameter 

As long as we use the flat sky approximation,
the $u$ space variable $\mbox{\boldmath $u$}_{\perp}$
is related to the multipole $\ell$ of angular power spectrum,
$|\mbox{\boldmath $u$}_{\perp}|=\ell/2\pi$.
In our analysis, 
we use the scale dependence of synchrotron radiation 
%$C_{\ell}^{S,X}(\nu)$ %(Eq.(\ref{eq:synchrotron_rad}))
as
the foreground power
$C^{{\rm Fg}}(\mbox{\boldmath $u$}_{\perp},\nu)=C_{\ell}^{S}(\nu) $,
and we model the synchrotron foreground $C_{\ell}^{S}(\nu)$  as
\begin{eqnarray}
C_{\ell}^{S}(\nu) 
&=&
A_{S}
\left(
\frac{\nu}{\nu_{S,0}}
\right)^{2\alpha_{S}}
\left(
\frac{\ell}{\ell_{S,0}}
\right)^{\beta_{S}}.
\label{eq:synchrotron_rad_21cm}
\end{eqnarray}
where %$X={\rm EE, TE, BB}$,
$\alpha_{S}=-3$, $\beta_{S}=-2.6$,
$\nu_{S,0}=30$ GHz, $\ell_{S,0}=350$,
$A_{S}=4.7\times10^{-5}\mu{\rm K}^2$.
%$\alpha_{D}=2.2$, $\nu_{D,0}=94$ GHz,
%$\ell_{D,0}=10$, $A_{D}=1.0 \mu {\rm K}^2$,
%$\beta_{D}^{X}=-2.5$.
%and $p$ is the dust polarization fraction $p=15\%$.
%
These choices are the values used in the Refs.~\cite{Baumann:2008aq,Verde:2005ff},
and match the parameters of foregrounds observed
by WMAP \cite{Page:2006hz}, DASI \cite{Leitch:2004gd} .
%and IRAS \cite{Finkbeiner:1999aq}.

%%%%%%%%%%%%%%%%%%%%%%%%%%%%%%%%%%%%%%%%%%%%%%%%%%%%%%%%%%%%%%%
\subsection{Effects of neutrino mass on matter power spectrum}
%%%%%%%%%%%%%%%%%%%%%%%%%%%%%%%%%%%%%%%%%%%%%%%%%%%%%%%%%%%%%%%

The massive neutrinos affect the growth of matter density
fluctuation mainly due to the following two physical
mechanisms.~\cite{Lesgourgues:2006nd,Wong:2011ip}.  
%
%First of all, a massive neutrino $\nu_{i}$ 
%(even with its light mass $m_{\nu_{i}} \lesssim0.3$ eV ) 
%becomes non-relativistic at $T \sim m_{\nu_{i}}$ 
%and has contributed to the energy density of cold dark matter (CDM), 
%
First of all, %a massive neutrino $\nu_{i}$ 
%(even with its light mass $m_{\nu_{i}} \lesssim0.3$ eV ) 
a massive neutrino becomes non-relativistic 
at $3T_{\nu_{i}} \sim m_{\nu_{i}}$,
and contributes to the energy density of cold dark matter (CDM).
Therefore, the matter-radiation equality time
and the expansion rate of the universe
are affected by the transition between relativistic and
non-relativistic neutrino.
When we consider the total mass of neutrinos 
$\Sigma m_{\nu}~(\lesssim 0.2-0.3~{\rm eV})$, 
only the latter impact is effective
because such light neutrinos do not become non-relativistic
before the matter-radiation equality time.
Secondly, the matter density fluctuation on small scales
are suppressed due to the free-streaming effect of neutrino.
As long as neutrinos are relativistic, 
they travel at speed of light, 
and their free-streaming scales are 
approximately equal to the Hubble horizon. 
Then, their own fluctuations 
within the free-streaming scales are erased,
and their energy densities do not 
contribute to the growth of matter density fluctuation.

In comparison with the standard $\Lambda$CDM models 
in which three massless active neutrinos are assumed, 
we can introduce two more freedoms.
A first additional freedom is 
the effective number of neutrino species $N_{\nu}$, 
which represents generations of relativistic neutrinos
before the matter-radiation equality epoch.
$N_{\nu}$ can include other relativistic components,
and may not be equal to three. 
A second additional freedom is the hierarchy of neutrino masses %neutrino mass hierarchy. 
%It is clear that 
%a change of $N_{\nu}$ affects the epoch of matter-radiation equality.
The difference of the  hierarchy of neutrino masses  affects 
both the free-streaming scales and the expansion rate 
of the Universe~\cite{Lesgourgues:2004ps}.  
%as was mentioned above~\cite{Lesgourgues:2004ps}.  
%
In terms of 21 cm line observations,
the minimum cutoff of the wave number is given by
$k_{{\rm min}}=2\pi/(yB)\sim 6\times10^{-2} h{\rm Mpc}^{-1}$
(see \ref{sec:21cm_spec}). 
%while the wave number corresponding to the
%
However, the wave number corresponding to the
neutrino free-streaming scale is 
$k_{{\rm free}}\lesssim 10^{-2} h{\rm Mpc}^{-1}$.
Therefore, the main feature due to the difference of the mass hierarchy  
comes from the modification on the cosmic expansion rate
when we focus on the 21 cm line observation. 
%when we fix the total matter density at the present time.
%
%Additionally, we separately study  following two cases:
In this paper, we separately study the following two cases:

%
%%%%%%%%%%%%%%%%%%%%%%%%%%%%
\subsubsection*{(A) Effective number of neutrino species}
%%%%%%%%%%%%%%%%%%%%%%%%%%%%

In this analysis, 
we add the effective number of neutrino species $N_{\nu}$ 
to the fiducial parameter set,
and the fiducial value of this parameter is set to be $N_{\nu}=3.046$.
%
%%%%revised version
Generally $N_{\nu}$ represents three species of massive neutrinos 
plus an extra relativistic component.
%In this analysis, we assumed three species of massive neutrinos + 
%an extra relativistic component.
%%%%%%%%%%%%%%%%%%%%%%%%%%%%

%%%%%%%%%%%%%%%%%%%%%%%%%%%%
\subsubsection*{(B) Neutrino mass hierarchy}
%%%%%%%%%%%%%%%%%%%%%%%%%%%%

The normal and inverted mass hierarchies mean  
$m_{1}<m_{2} \ll m_{3}$ 
and $m_{3} \ll m_{1}<m_{2}$, respectively. 
In a cosmological context, many different parameterizations of the
mass hierarchy have been proposed
\cite{Takada:2005si,Slosar:2006xb,DeBernardis:2009di,Jimenez:2010ev}.
In our analysis, we adopt $r_{\nu} \equiv 
(m_{3} - m_{1})/\Sigma m_{\nu}$~\cite{Jimenez:2010ev} 
as an additional parameter to 
discriminate the true neutrino mass hierarchy from the other. 
%between the normal and inverted hierarchies. 
%We add $r_{\nu}$ to the fiducial set of the parameters.  
%
$r_{\nu}$ becomes positive for the normal hierarchy, 
and negative for the inverted hierarchy.
Besides, the difference between $r_{\nu}$ of  these two hierarchies
becomes larger as the total mass $\Sigma m_{\nu}$ becomes smaller.
Therefore, $r_{\nu}$ is particularly useful 
for distinguishing the mass hierarchies.  
%
%In Fig \ref{fig:hie_ellipse}, we plot behaviors of $r_{\nu}$ as 
In Fig \ref{fig:hie_ellipse_fsky0016}-\ref{fig:hie_ellipse_m8}, 
we plot behaviors of $r_{\nu}$ as a function of $\Sigma m_{\nu}$. 

Note that there is a lowest value of
$\Sigma m_{\nu}$ which depends on the mass hierarchy
by the neutrino oscillation experiments. 
The lowest value is
$\Sigma m_{\nu}\sim$0.1 eV for the inverted hierarchy
or $\Sigma m_{\nu}\sim$0.06 eV for the normal hierarchy. 
Therefore, if we obtain a clear constraint like 
$\Sigma m_{\nu} \ll $~0.10~eV, 
we can determine
that the mass hierarchy is obviously 
normal without any ambiguities. 
However, we can discriminate the
mass hierarchy even when 
%the mass hierarchy is 
%inverted and 0.10~eV~$ \lesssim \Sigma m_{\nu}$ 
the neutrino mass $\Sigma m_{\nu}$  is larger than 0.10~eV
if we use $r_{\nu}$, as will be shown later.

%As will be shown later, however, we can discriminate the
%hierarchy even when 0.10~eV~$ \lesssim \Sigma m_{\nu}$.

%%%%%%%%%%%%%%%%%%%%%%%%%%%%%%%%%%%%%%%%%%%%%%%%%%%%%%%%%%%%%%%
\section{CMB }
\label{sec:CMB}
%%%%%%%%%%%%%%%%%%%%%%%%%%%%%%%%%%%%%%%%%%%%%%%%%%%%%%%%%%%%%%%

%%%%%%%%%%%%%%%%%%%%%%%%%%%%%%%%%%%%%%%%%%%%%%%%%%%%%%%%%%%%%%%
\subsection{CMB and neutrino}
%%%%%%%%%%%%%%%%%%%%%%%%%%%%%%%%%%%%%%%%%%%%%%%%%%%%%%%%%%%%%%%

%CMB power spectra are sensitive to neutrino masses.   There are three
%effects that provide detectable signals for the neutrino masses: 
%(1) the transition from relativistic neutrino to nonrelativistic one, 
%(2)smoothing of the matter perturbation by its free-streaming in small
%scales, 
%and (3) variation of lensed CMB power spectra.  

In this paper, 
we focus on not only the observations of the 21 cm line 
but also the CMB observations, especially CMB B-mode polarization 
produced by gravitational lensing of the matter fluctuation.
%
%Although 21 cm line experiments have 
%strong power to probe the matter power spectrum, 
Although the 21 cm line observation is 
a power probe of the matter power spectrum, 
particularly, on small scales, 
observations of CMB greatly help to determine 
other cosmological parameters 
%such as energy densities of dark matter, 
%baryon and dark energy, and so on.
%
such as energy densities of the dark matter, 
baryons and dark energy.

Besides, CMB power spectra are sensitive to neutrino masses
through the CMB lensing.
Future precise CMB experiments are expected to 
set stringent constraints on
the sum of the neutrino masses and
the effective number of neutrino species~\cite{Wu:2014hta,Abazajian:2013oma}. 
%In addition, it is notable that we are able to determine the neutrino mass hierarchy.
%
%the sum of neutrino masses~\cite{Lesgourgues:2006nd,Wong:2011ip,Wu:2014hta,Abazajian:2013oma}. 
%the sum of the neutrino masses~\cite{Wu:2014hta,Abazajian:2013oma}. 
%In addition, it is notable that we are able to detect the effective
%  %number of neutrino species~\cite{Wu:2014hta,Chacko:2003dt,Abazajian:2013oma} and determine the
%number of neutrino species~\cite{Wu:2014hta,Abazajian:2013oma} and determine the
%neutrino mass hierarchy.
%
Therefore, we propose to combine the CMB experiments with 
the 21 cm line observations.  

%%%%%%%%%%%%%%%%%%%%%%%%%%%%%%%%%%%%%%%%%%%%%%%%%%%%%%%%%%%%%%%
\subsection{Fisher matrix of CMB}
%%%%%%%%%%%%%%%%%%%%%%%%%%%%%%%%%%%%%%%%%%%%%%%%%%%%%%%%%%%%%%%

We evaluate errors of cosmological parameters
by using the Fisher matrix of CMB,
which is given by~\cite{Tegmark:1996bz}.
\begin{equation}
F_{\alpha \beta}^{\rm (CMB)}
= \sum_{\ell}
\frac{\left( 2\ell+1\right)}{2}
\mathrm{Tr}
\left[
  C_{\ell}^{-1}
  \frac{\partial C_{\ell}}{\partial \theta_{\alpha}}
  C_{\ell}^{-1}
  \frac{\partial C_{\ell}}{\partial \theta_{\beta}}
\right],
\label{eq:Fisher_CMB}
\end{equation}
\begin{align}
    C_{\ell} =  \left(
\begin{array}{ccc}
\hspace{5pt} C_{\ell}^{\mathrm{TT}} 
+ N_{\ell}^{\mathrm{TT}} \hspace{5pt} &
\hspace{5pt} C_{\ell}^{\mathrm{TE}} \hspace{5pt} &
\hspace{5pt} C_{\ell}^{\mathrm{Td}} \hspace{5pt} \\
\hspace{5pt} C_{\ell}^{\mathrm{TE}} \hspace{5pt} &
\hspace{5pt} C_{\ell}^{\mathrm{EE}} 
+N_{\ell}^{\mathrm{EE}} \hspace{5pt} &
0 \hspace{5pt} \\
\hspace{5pt} C_{\ell}^{\mathrm{Td}} \hspace{5pt} &
0 \hspace{5pt} &
\hspace{5pt} C_{\ell}^{\mathrm{dd}} 
+ N_{\ell}^{\mathrm{dd}} \hspace{5pt}
\end{array}
\right).
\end{align}
Here $\ell$ is th multipole of angular power spectra,
$C_{\ell}^{\mathrm{X}} 
\left(\mathrm{X}=\mathrm{TT, EE, TE} \right)$
are the CMB power spectra,
$C_{\ell}^{\mathrm{dd}}$ is the deflection angle spectrum,
$C_{\ell}^{\mathrm{Td}}$ is the
cross correlation between the deflection angle and the temperature,
$N_{\ell}^{\mathrm{X'}}$ 
$\left(\mathrm{X'}=\mathrm{TT, EE} \right)$
and $N_{\ell}^{\mathrm{dd}}$
are the noise power spectra, 
where $C_{\ell}^{\mathrm{dd}}$ is calculated by a lensing
potential~\cite{Okamoto:2003zw} and is related with
the lensed CMB power spectra.
The noise power spectra of CMB 
%$N_{\ell}^{\mathrm{X'}}$ are expressed by using both a beam size
$N_{\ell}^{\mathrm{X'}}$ are expressed with a beam size
$\sigma_{\mathrm{beam}}(\nu)=$ $\theta
_{\mathrm{FWHM}}(\nu)/\sqrt{8\ln 2}$,
where 
$\sqrt{8\ln 2}\sigma_{{\rm beam}}$ means 
the full width at half maximum of the Gaussian distribution,
and instrumental sensitivity
$\Delta _{\mathrm{X'}}(\nu)$ by
\begin{eqnarray}
  N_{\ell}^{\mathrm{X'}}
  = \left[ 
    \sum_{i} \frac1{n_{\ell}^{\mathrm{X'}}(\nu_{i})}
    \right]^{-1},
    \label{eq:noise_CMB}
    \end{eqnarray}
where $\nu_{i}$ is an observing frequency and
\begin{align}
n_{\ell}^{\mathrm{X'}}(\nu)=
\Delta ^2_{\mathrm{X'}}(\nu)\exp\left[\ell(\ell+1) \sigma ^2_{\mathrm{beam}}(\nu)\right].
\end{align}
%%
%The noise of the deflection angle $N_{\ell}^{\mathrm{dd}}$
%s estimated by the noise of CMB $N_{\ell}^{\mathrm{X}}$
%and the power spectra $C_{\ell}^{\mathrm{X}}$.
%
%Besides, 
%we estimate the noise power spectrum of deflection angle $N^{dd}_l$
The noise power spectrum of deflection angle 
%$N^{dd}_l$
%by assuming that we use lensing reconstruction with the
$N^{dd}_l$ is obtained assuming lensing  reconstruction with the
quadratic estimator \cite{Okamoto:2003zw},
%and compute it by using a public code FUTURCMB \cite{paper:FUTURCMB},
%which adopts the estimator.
which is computed with FUTURCMB \cite{paper:FUTURCMB}.
In this algorithm,
%$N_{\ell}^{dd}$ is estimated by 
%the noise $N_{\ell}^{X'}$ and 
%
$N_{\ell}^{dd}$ is estimated from
the noise $N_{\ell}^{X'}$, 
and 
lensed and unlensed power spectra of CMB temperature, 
E-mode and B-mode polarizations.

%Since there are a minimum $\ell_{min}$ 
%and a maximum multipole $\ell_{min}$
%and an observed sky region is limited to part of the all sky,
%we introduce a sky coverage 
%(a ratio of the observed region to the all sky),
%and the Fisher matrix is rewritten as
%
Finally, the Fisher matrix in Eq.(\ref{eq:Fisher_CMB}) is modified as follows 
by taking the multipole range 
[$\ell_{min}$, $\ell_{max}$] 
and the fraction of the observed sky $f_{{\rm sky}}$ into account,
\begin{equation}
F_{\alpha \beta}^{\rm (CMB)}
= \sum_{\ell = \ell_{min}}^{\ell_{max}}
\frac{\left( 2\ell+1\right)}{2}f_{\mathrm{sky}}
\mathrm{Tr}
\left[
  C_{\ell}^{-1}
  \frac{\partial C_{\ell}}{\partial \theta_{\alpha}}
  C_{\ell}^{-1}
  \frac{\partial C_{\ell}}{\partial \theta_{\beta}}
\right].
\end{equation}
%

%%%%%%%%%%%%%%%%%%%%%%%%%%%%%%%
%\section[Treatment of residual foregrounds]
%{Treatment of residual foregrounds 
%\normalsize{\cite{Baumann:2008aq,Verde:2005ff}}}
%
%\subsection[Treatment of residual foregrounds]
%{Treatment of residual foregrounds 
\subsection[Residual foregrounds]
{Residual foregrounds 
\normalsize{\cite{Baumann:2008aq,Verde:2005ff}}}
%%%%%%%%%%%%%%%%%%%%%%%%%%%%%%%

%In this section, we show treatments of 
%Here, we show treatments of 
%residual foregrounds of CMB in our analysis.
%
%For observation of CMB, 
%we consider two dominant galactic foregrounds,
%the one is the synchrotron radiation,
%the other is the dust emission.
%
We consider synchrotron radiation 
and dust emission in our galaxy 
as the dominant sources of foregrounds.
%
%These foregrounds are subtracted in pixel space of CMB maps.
%
These foregrounds are subtracted from each sky pixel of the CMB map.
%
%Here we assume that
%foreground subtraction can be done down to 
%a given level ($1\%$ level in the power spectra of CMB).
%
Here, we assume that 
the foreground subtraction can be performed at 
a certain level 
demonstrated in previous foreground separation studies 
\cite{Katayama:2011eh}; 
%we assume 1\% level in the power spectra of CMB.
we assume 2\% level in CMB maps.
%
%
%Below we model the residual foregrounds 
%which can not be subtracted in the CMB maps.
%Note that 
%we only consider the residual foreground of 
%CMB polarization maps, not that of temperature,
%because it has already been precisely measured 
%by WMAP or Planck.
%
We then model the residual foregrounds in the CMB maps. 
Note that 
we only consider the residual foreground of 
CMB polarization maps
as distinct from temperature.
%not that of temperature, 
That is because it
%because it 
has already been precisely measured 
by WMAP and Planck.

%We assume that the scale dependence of the
We model the synchrotron 
$C_{\ell}^{S, X}$ and 
dust $C_{\ell}^{D, X}$ foregrounds as
\begin{eqnarray}
C_{\ell}^{S, X}(\nu) 
&=&
A_{S}
\left(
\frac{\nu}{\nu_{S,0}}
\right)^{2\alpha_{S}}
\left(
\frac{\ell}{\ell_{S,0}}
\right)^{\beta_{S}}, \label{eq:synchrotron_rad} \\
C_{\ell}^{D, X}(\nu) 
&=& 
p^2 A_{D}
\left(
\frac{\nu}{\nu_{D,0}}
\right)^{2\alpha_{D}}
\left(
\frac{\ell}{\ell_{D,0}}
\right)^{\beta_{D}^{X}}
\left[
\frac{e^{h\nu_{D,0}/k_{B}T_{{\rm dust}}}-1}{e^{h\nu/k_{B}T_{\rm dust}}-1}
\right]^2,
\end{eqnarray}
where $X={\rm EE, TE, BB}$,
$\alpha_{S}=-3$, $\beta_{S}=-2.6$,
$\nu_{S,0}=30$ GHz, $\ell_{S,0}=350$,
$A_{S}=4.7\times10^{-5}\mu{\rm K}^2$,
$\alpha_{D}=2.2$, $\nu_{D,0}=94$ GHz,
$\ell_{D,0}=10$, $A_{D}=1.0 \mu {\rm K}^2$,
$\beta_{D}^{X}=-2.5$, and the temperature of the dust grains $T_{\rm dust}=18 {\rm K}$.
%and $p$ is the dust polarization fraction $p=15\%$.
%
These choices are the values used in the Refs.~\cite{Baumann:2008aq,Verde:2005ff},
and match the parameters of foregrounds observed
by WMAP \cite{Page:2006hz}, DASI \cite{Leitch:2004gd} 
and IRAS \cite{Finkbeiner:1999aq}.
 The dust polarization fraction, $p$,  is assumed to be $15\%$,
which matches the  latest measurement by Planck.
%
%Furthermore we assume these residual foregrounds 
%are the Gaussian and modeled as follows,
We then assume that residual foregrounds are modeled as follows,
\begin{eqnarray}
C^{X,{\rm RFg}}_{\ell}(\nu)=
\left[
C_{\ell}^{S,X}(\nu) + C_{\ell}^{D,X}(\nu)
\right]
\sigma^{{\rm RFg}}_{{\rm CMB}} + n_{\ell}^{{\rm RFg},X}(\nu),
\end{eqnarray}
where 
$\sigma^{{\rm RFg}}_{{\rm CMB}}$ is the foreground residual parameter
of CMB observations.
We assume $\sigma^{{\rm RFg}}_{{\rm CMB}}=4\times 10^{-4}$ 
(this value corresponds to 2\% in the CMB maps),
and $n_{\ell}^{{\rm RFg},X}$ is the noise power spectrum
of the foreground template maps,
which is created by taking map differences and 
thus are somewhat affected by the instrumental noise.
We assume that
this noise power spectrum of the template maps is given by
\begin{eqnarray}
n_{\ell}^{{\rm RFg},X}(\nu) = 
\frac{N_{\ell}^{{\rm RFg}, \mathrm{X}}}{N_{{\rm chan}}(N_{{\rm chan}}-1)/4}
\left\{
\left(
\frac{\nu}{\nu_{S,{\rm ref}}}
\right)^{2\alpha_{S}}
+
\left(
\frac{\nu}{\nu_{D,{\rm ref}}}
\right)^{2\alpha_{D}}
\right\},
\end{eqnarray}
where  
$N_{{\rm chan}}$ is the total frequency channels
which are used for the foreground removal,
$\nu_{D,{\rm ref}}$ and $\nu_{S,{\rm ref}}$ are
the highest and lowest frequency channel included 
in the removal, respectively,
and $N_{\ell}^{{\rm RFg}, \mathrm{X}}$ is defined by
%
%%%%2016/01/12
\begin{eqnarray}
N_{\ell}^{{\rm RFg}, \mathrm{X}}
\equiv \left[ 
    \sum_{i} ^{N_{\rm chan}}\frac1{n_{\ell}^{\mathrm{X}}(\nu_{i})}
    \right]^{-1},
    \label{eq:noise_CMB_forground}
\end{eqnarray}
%%%%%%
%
where $i$ represents a frequency channel which is used for the removal.
Furthermore, we introduce the following 
effective noise power spectrum 
including the residual foregrounds,
\begin{eqnarray}
N^{{\rm eff},X}_{\ell}
=
\left[
\sum_{i}
\frac{1}{n^{X}_{\ell}(\nu_{i}) + C^{X,{\rm RFg}}_{\ell}(\nu_{i})}
\right]^{-1},
\end{eqnarray}
where $i$ represents a frequency channel which is used for the observation of CMB.
When we include the effects due to 
the residual foregrounds in our analysis,
we use this effective noise as the CMB noise power spectrum Eq.(\ref{eq:noise_CMB}).

By making the modifications given above
to FUTURCMB \cite{paper:FUTURCMB},
we 
%calculate the 
%estimated errors 
estimate the errors
of the deflection
angle of CMB and use them in our Fisher matrix analysis.

%%%%%%%%%%%%%%%%%%%%%%%%%%%%%%%
%\section{Specifications of experiments}
\subsection{Specifications of the experiments}
%%%%%%%%%%%%%%%%%%%%%%%%%%%%%%%

%%%%%%%%%%%%%%%%%%%%%%%%%%%%%%%%%%%%%%%%%%%%%%%%%%%%%%%%%%%%%%%%%%%%%%
\begin{table}[tp]
\begin{center}
\begin{tabular}{c|ccccccc}
\hline
\hline
\shortstack{Experiment\\ \,}&
\shortstack{$\nu$ \\ $[\mathrm{GHz}]$}&
\shortstack{$\Delta _{\mathrm{TT}}$\\ $[\mathrm{\mu K-'}]$}&
\shortstack{$\Delta _{\mathrm{PP}}$\\ $[\mathrm{\mu K-'}]$}& 
\shortstack{$\theta_{\mathrm{FWHM}}$\\ $[\mathrm{-'}]$}&
\shortstack{$f_{\mathrm{sky}}$\\ $ $} &
\shortstack{$\ell_{{\rm min}}$\\ $ $} &
\shortstack{$\ell_{{\rm max}}$\\ $ $} 
\\
\hline
\hline
Planck & 30 & 145 & 205  & 33  &      &   &      \\ 
       & 44 & 150 & 212  & 23  &      &   &      \\ 
       & 70 & 137 & 195  & 14  &      &   &      \\
       & 100& 64.6& 104  & 9.5 & 0.65 & 2 & 3000 \\
       & 143& 42.6& 80.9 & 7.1 &      &   &      \\
       & 217& 65.5& 134  & 5   &      &   &      \\ 
       & 353& 406 & 406  & 5   &      &   &      \\ 
\hline
\hline
\textsc{Polarbear}-2 (PB-2)  & 95  & - & 3.09 & 5.2 & 0.016 & 25 & 3000 \\ 
$f_{{\rm sky}}=0.016$ & 150 & - & 3.09 & 3.5 &       &    &      \\
\hline
PB-2 & 95  & - & 10.9 & 5.2 & 0.2   & 25 & 3000 \\ 
$f_{{\rm sky}}=0.2$   & 150 & - & 10.9 & 3.5 &       &    &      \\
\hline
PB-2 & 95  & - & 19.7 & 5.2 & 0.65  & 25 & 3000 \\ 
$f_{{\rm sky}}=0.65$  & 150 & - & 19.7 & 3.5 &       &    &      \\
\hline
\hline
Simons Array (SA)     & 95  & - & 2.18 & 5.2 &       &    &      \\ 
$f_{{\rm sky}}=0.016$ & 150 & - & 1.78 & 3.5 & 0.016 & 25 & 3000 \\
                      & 220 & - & 6.34 & 2.7 &       &    &      \\
\hline
SA                    & 95  & - & 7.72 & 5.2 &       &    &      \\ 
$f_{{\rm sky}}=0.2$   & 150 & - & 6.30 & 3.5 & 0.2   & 25 & 3000 \\
                      & 220 & - & 21.5 & 2.7 &       &    &      \\
\hline
SA                    & 95  & - & 13.9 & 5.2 &       &    &      \\ 
$f_{{\rm sky}}=0.65$  & 150 & - & 11.4 & 3.5 & 0.65  & 25 & 3000 \\
                      & 220 & - & 38.8 & 2.7 &       &    &      \\                 
\hline
\hline
\end{tabular}
\caption{
Experimental specifications of Planck, \textsc{Polarbear}-2
and Simons Array assumed in our analysis. 
Here $\nu$ is the observation frequency, 
$\Delta_{\rm TT}$ is the temperature sensitivity per $1'\times1'$
pixel, $\Delta_{\rm PP}=\Delta_{\rm EE}=\Delta_{\rm BB}$ 
is the polarization (E-mode and B-mode) sensitivity 
per $1'\times1'$ pixel,
$\theta_{\rm FWHM}$ is the angular resolution defined to be the full width at
half-maximum, and $f_{\rm sky}$ is the observed fraction of the sky.
For the Planck experiment, we assume that 
the three frequency bands ($70, 100, 143$ GHz)
are only used for the observation of CMB,
and the other bands ($30,44,217,353$ GHz)  are used for foregrounds removal.
For Simons Array, we consider two cases:
One case is that 220 GHz band is used for the observation of CMB,
and the other is that the band is used for the foreground removal.
%220 GHz band is used for the observation of CMB;
%the band is used for the foreground removal.
}
\label{tab:cmb_obs_mass}
%\label{tab:sensitivity}
\end{center}
\end{table}
%%%%%%%%%%%%%%%%%%%%%%%%%%%%%%%%%%%%%%%%%%%%%%%%%%%%%%%%%%%%%%%%%%%%%

%Now in this section, we show the specifications 
Now, we show the specifications 
of 
%the observations of CMB.
the CMB observations.
%which are considered in this thesis.
%
%%%%%%%%%%%%%%%%%%%%%%%%%%%
%\subsection{Analysis of the neutrino mass 
%and the mass hierarchy}
%%%%%%%%%%%%%%%%%%%%%%%%%%%
%
%In the analysis of the total neutrino mass,
%the number of the neutrino species,
%and the mass hierarchy (the Chapter \ref{Chap:result_mass}),
%to obtain the future constraints, 
In order to obtain the future constraints, 
we consider Planck~\cite{Planck:2006aa},
\textsc{Polarbear}-2 and Simons Array,
whose experimental specifications 
are summarized in Table~\ref{tab:cmb_obs_mass}.
The latter two experiments are 
ground-based precise CMB polarization observations.

%In the case of Planck and \textsc{Polarbear}-2 or Simons Array, 
For the analysis about Planck and \textsc{Polarbear}-2 (or Simons Array) ,
we combine both the experiments, 
%and assume that the 1.6\% region of the whole sky is observed 
%by both the experiments, 
%and the remaining  $63.4\%(=65\%-1.6\%)$ region is observed 
%by Planck only. 
%
and assume that a part of the whole sky $(f_{{\rm sky}}\times 100\%)$ 
is observed by both the experiments, 
and the remaining region $(65\%-f_{{\rm sky}}\times 100\%)$ 
is observed only by Planck. 
Therefore, we evaluate a total Fisher matrix of CMB
$F^{({\rm CMB})}$ by summing the two Fisher matrices,
\begin{align}
F^{({\rm CMB})} = F^{({\rm Planck})}(65\%-f_{{\rm sky}}\times 100\%)
+F^{({\rm Planck + PB-2 \ or \ SA})}(f_{{\rm sky}}\times 100\%),
\end{align}
%where $F^{({\rm Planck})}$ is the Fisher matrix of the 
%region observed by Planck only,
%and $F^{({\rm Planck + PB-2 \ or \ SA})}$
%is that by both Planck and \textsc{Polarbear}-2 (PB-2)
%or Simons Array (SA). 
%
where 
$F^{({\rm Planck + PB-2 \ or \ SA})}$
is the Fisher matrix of the 
region observed by both Planck and \textsc{Polarbear}-2 (PB-2)
[or Simons Array (SA)], 
and $F^{({\rm Planck})}$ is that by Planck only. 

In addition, we calculate noise power spectra 
$N_{\ell}^{\mathrm{X,Planck+PB-2 \ or \ SA}}$ of 
the CMB polarization ($X = {\rm EE}$ or ${\rm BB}$)
in $F^{\mathrm{Planck+PB-2 \ or \ SA}}$ with 
the following operation. 
\begin{description}
\item[(1)] $2\leq \ell < 25$%, $2000<\ell \leq 3000$
\begin{align}
N_{\ell}^{\mathrm{X,Planck + PB-2 \ or \ SA}} 
= N_{\ell}^{\mathrm{X,Planck}}
\end{align}
\item[(2)] $25\leq \ell \leq 3000$
\begin{align}
N_{\ell}^{\mathrm{X,Planck + PB-2 \ or \ SA}} 
= [1/N_{\ell}^{\mathrm{X,Planck}} 
+ 1/N_{\ell}^{\mathrm{X,PB-2 \ or \ SA}}]^{-1}
\end{align}
\end{description}
Since we assume that 
the CMB temperature fluctuation observed by
\textsc{Polarbear}-2 or Simons Array 
is not used for constraints on the cosmological parameters,
the temperature noise power spectrum 
$N_{\ell}^{\mathrm{TT,\,Planck+PB-2 \ or \ SA}}$ is equal to 
% $N_{\ell}^{\mathrm{TT,\,Planck}}$ . 
{\bf $N_{\ell}^{\mathrm{TT,\,Planck}}$}.
This reason is that the CMB temperature fluctuation 
observed by Planck reaches almost cosmic variance limit.
(Planck observation greatly helps to determine other cosmological parameters 
such as the normalization of power spectra,
energy densities of the dark matter, baryons and dark energy).
Therefore, the constraints are not strongly improved
if we include the CMB temperature fluctuation
observed by \textsc{Polarbear}-2 or Simons Array.

%%%%%%%%%%%%%%%%%%%%%%%%%%%%%%%%%%%%%%%%%%%%%%%%%%%%%%%%%%%%%%%%%%%%%
\section{BAO}
\label{sec:BAOFisher}
%%%%%%%%%%%%%%%%%%%%%%%%%%%%%%%%%%%%%%%%%%%%%%%%%%%%%%%%%%%%%%%%%%%%%

In this section, we briefly explain 
analysis methods about 
the baryon acoustic oscillation (BAO).
In the early Universe,
baryons and photons are strongly coupled
and their fluctuations (Fourier components) of 
the mixed fluid oscillate by the pressure of radiation.
At the time of the decouple between them,
a characteristic peak feature remains at
the sound horizon.
The scale can be used for a standard ruler of distance. 
Therefore, we can get the information of 
the distance and the Hubble expansion rate
by measurements of the BAO scale for matter fluctuations.
In this paper, we especially consider galaxy surveys
for the BAO observation.

%%%%%%%%%%%%%%%%%%%%%%%%%%%%%%
%\section[Fisher matrix of BAO]
%{Fisher matrix of BAO \normalsize{\cite{Albrecht:2006um}}}
\subsection[Fisher matrix of BAO]
{Fisher matrix of BAO \normalsize{\cite{Albrecht:2006um}}}
%%%%%%%%%%%%%%%%%%%%%%%%%%%%%%

Here we introduce the Fisher matrix of BAO experiments.
The observables of BAO are the comoving angular diameter
distance $d_{A}(z)$ and the Hubble parameter $H(z)$
(and more specifically, $\ln(d_{A}(z))$ and $\ln(H(z))$
are the observables).
The Fisher matrix is given by 
\begin{eqnarray}
F^{({\rm BAO}) \ d,H}_{\alpha \beta} &=&
\sum_{i} \frac{1}{\sigma_{d,H}^2(z_{i})+(\sigma_{s}^i)^{2}} 
\frac{\partial f_{i}^{d,H}}{\partial \theta_{\alpha}}
\frac{\partial f_{i}^{d,H}}{\partial \theta_{\beta}}, \\
f_{i}^{d} &=& \ln(d_{A}(z_{i})), \\
f_{i}^{H} &=& \ln(H(z_{i})),
\end{eqnarray}
where $\sigma_{d}(z_{i})$ and $\sigma_{H}(z_{i})$
are the variances of $\ln(d_{A}(z_{i}))$
and $\ln(H(z_{i}))$ in the BAO observation respectively,
$\sigma_{s}^{i}$ is the error of the systematics,
%and we consider that bins in redshift space of width $\Delta z_{i}$
%centered on $z_{i}$. 
%
and we assume that 
the observed redshift range is divided into bins,
whose width and central redshift values
are represented by $\Delta z_{i}$ and $z_{i}$, respectively.
Here, $i$ is the index of the redshift bins.

Their variances  $\ln(d_{A}(z_{i}))$
and $\ln(H(z_{i}))$ are determined by the fitting formulae
of BAO presented by \cite{Blake:2005jd}, and they are given by
\begin{eqnarray}
\sigma_{d}(z_{i}) 
&=&
x^{d}_{0} \frac{4}{3}
\sqrt{\frac{V_{0}}{V_{i}}}
f_{nl}(z_{i}), \\
\sigma_{d}(z_{i})
&=&
x^{H}_{0} \frac{4}{3}
\sqrt{\frac{V_{0}}{V_{i}}}
f_{nl}(z_{i}).
\end{eqnarray}
Here, $V_{i}$ is the comoving survey volume
and expressed as
\begin{eqnarray}
V_{i} =
\frac{(d_{A}(z_{i}))^{2}}{H(z_{i})}\Omega_{{\rm sky}}\Delta z_{i},
\end{eqnarray}
where $\Omega_{{\rm sky}}$ is the survey solid angle.
$f_{nl}(z_{i})$ is the non-linear evolution factor,
which represents the erasure of baryon oscillation features by the non-linear evolution of density fluctuations.
In our analysis,
we use the following function for $f_{nl}(z_{i})$,
\begin{eqnarray}
f_{nl}(z_{i}) = \left\{
\begin{array}{c}
1 \hspace{40pt} z>z_{m},\\
\left(
\frac{z_{m}}{z_{i}}
\right)^{\gamma} \hspace{10pt} z<z_{m}.
\end{array}
\right.
\end{eqnarray}
where $z_{m}$ is the redshift at which the improvement in the baryon oscillation 
accuracy saturates (for fixed survey volume and number density).
Additionally, in the analysis of the BAO observation,
we use the following parameters,
\begin{subequations}
\begin{eqnarray}
x_{0}^{d}&=&0.0085, \\
x_{0}^{H}&=&0.0148, \\
V_{0}    &=& \frac{2.16}{h^3} {\rm Gpc}^3, \\
\gamma &=& \frac{1}{2}, \\
z_{m}  &=& 1.4,
\end{eqnarray}
\end{subequations}
where $h\equiv H_{0}/(100{\rm km}/s/{\rm Mpc})$ 
is the dimensionless Hubble parameter.
According to \cite{Albrecht:2006um},
we assume the following systematic error,
\begin{eqnarray}
\sigma_{s}^{i} = 0.01 \times \sqrt{\frac{0.5}{\Delta z_{i}}}.
\end{eqnarray}

The set of  cosmological parameters 
related to the BAO observation
are only $(\Omega_{m}h^2, \Omega_{\Lambda})$
or $(h, \Omega_{\Lambda})$
when we assume that the Universe is flat and
the dark energy is the cosmological constant.

%%%%%%%%%%%%%%%%%%%%%%%%%%%%%%
%\section{Specifications of the observation of BAO}
\subsection{Specification of the BAO observation }
%%%%%%%%%%%%%%%%%%%%%%%%%%%%%%

We estimate the sensitivity of the BAO observation
only in the analysis of the neutrino mass, 
the number of neutrino species and the mass hierarchy.
In the analysis, we focus on the 
Dark Energy Spectroscopic Instrument (DESI) \cite{DESI:web,Font-Ribera:2013rwa},
which is a future large volume galaxy survey. 
The survey redshift range is 
$0.1<z<1.9$ 
(we do not include 
the Ly-$\alpha$ forest at $1.9<z$ for simplicity)
and the sold angle is
$\Omega_{sky}=14000 [{\rm deg}^2]$.
In our analysis, we divide the redshift range
into 18 bins, in other words $\Delta z_{i}=0.1$ \cite{Wu:2014hta}.

Additionally,
in the same manner as \cite{Wu:2014hta},
when we combine BAO with the other observations,
%we put 1\% prior on the present Hubble parameter $H_{0}$,
we add a 1\% $H_{0}$ prior to the Fisher matrix.
The prior of the Hubble parameter is achievable in the next decade.
The Fisher matrix of the Hubble prior is given by 
\begin{eqnarray}
F^{(H_{0} \ {\rm prior})}_{\theta_{\alpha}\theta_{\beta}} 
= \left\{
\begin{array}{c}
 \hspace{-40pt} \frac{1}{(1\%\times H_{0,{\rm fid}})^2}, 
 \hspace{30pt} \theta_{\alpha}=\theta_{\beta}=H_{0},\\
 \hspace{20pt}0,
 \hspace{60pt} {\rm the \ other \ components},
\end{array}
\right.
\end{eqnarray}
where $H_{0,{\rm fid}}$ is the fiducial value of $H_{0}$.
If we choose the Hubble parameter as 
a dependent parameter,
it is necessary to translate the Fisher matrix
into that of the chosen parameter space.
Under the transformation of a parameter space
$\theta \longrightarrow \tilde{\theta}$,
the translated Fisher matrix is give by \cite{Albrecht:2006um}
\begin{eqnarray}
\tilde{F}_{l,m} = 
\frac{\partial \theta_{j}}{\partial \tilde{\theta}_{l}}
\frac{\partial \theta_{k}}{\partial \tilde{\theta}_{m}}
F_{jk}.
\end{eqnarray}
By using this formula,
under the translation of 
$(h, \Omega_{\Lambda})$ $\longrightarrow$
$(\Omega_{m}h^2, \Omega_{\Lambda})$,
the Fisher matrix in the new parameter space is written as
\begin{align}
\tilde{F}^{H_{0} \ {\rm prior}} = 
\left(
\begin{array}{cc}
\tilde{F}_{\Omega_{m}h^2 \Omega_{m}h^2} &
\tilde{F}_{\Omega_{m}h^2 \Omega_{\Lambda}} \\
\tilde{F}_{\Omega_{m}h^2 \Omega_{\Lambda}} & 
\tilde{F}_{\Omega_{\Lambda} \Omega_{\Lambda}}
\end{array}
\right)
=
\frac{1}{(1\%\times H_{0,{\rm fid}})^2}
\left(\frac{1}{2\Omega_{m}h^2}\right)^2
\left(
\begin{array}{cc}
h^2 & h^4 \\
h^4 & h^6
\end{array}
\right).
\end{align}

%%%%%%%%%%%%%%%%%%%%%%%%%%%%%%%%%%%%%%%%%%%%%%%%%%%%%%%%%%%%%%%%%%%%%
\section{Results}
\label{sec:results}
%%%%%%%%%%%%%%%%%%%%%%%%%%%%%%%%%%%%%%%%%%%%%%%%%%%%%%%%%%%%%%%%%%%%%

%%%%%%%%%%%%%%%%%%%%%%%%%%%%%%%%%%%%%%%%%%%%%%%%%%%%%%%%%%%%%%%%%%%%%
\subsection{Future constraints}
%\section{Future constraints}
%%%%%%%%%%%%%%%%%%%%%%%%%%%%%%%%%%%%%%%%%%%%%%%%%%%%%%%%%%%%%%%%%%%%%

In this section, we present our results for projected constraints 
by the 21cm line, CMB and BAO observations on cosmological parameters, 
paying particular attention to parameters related to neutrino
(the total neutrino mass, the effective number of neutrino species 
and the neutrino mass hierarchy). 
When we calculate the Fisher matrices, we choose the following basic set
of cosmological parameters:
the energy density of matter $\Omega_{m}h^{2}$, baryons $\Omega_{b}h^{2}$
and dark energy $\Omega_{\Lambda}$, 
the scalar spectral index $n_{s}$, the scalar fluctuation amplitude $A_{s}$ 
(the pivot scale is taken to be $k_{{\rm pivot}}=$ $0.05 \ {\rm Mpc}^{-1}$), 
the reionization optical depth $\tau$, 
%the helium fraction $Y_{{\rm p}}$ 
the primordial value of the $^4$He mass fraction $Y_{{\rm p}}$ 
and the total neutrino mass $\Sigma m_{\nu} = m_{1}+m_{2}+m_{3}$.
Fiducial values of these parameters (except for $\Sigma m_{\nu}$) are adopted 
to be $(\Omega_{m}h^{2},\Omega_{b}h^{2},\Omega_{\Lambda},n_{s},A_{s},\tau,Y_{\rm p})$
$=( 0.1417, 0.02216, 0.6914, 0.9611, 2.214\times 10^{-9}, 0.0952, 0.25)$,
which are the best fit values of the Planck result \cite{Ade:2013zuv}.

Here, we numerically  evaluate how we can determine 
the effective number of neutrino species (in section \ref{subsec:const_hie}), 
and the neutrino mass hierarchy (in section \ref{subsec:const_hie}),  
by combining the  21 cm line observations (SKA phase~1 or phase~2)
with the CMB experiments (Planck + \textsc{Polarbear}-2 (PB-2) or Simons Array (SA)) 
and the BAO observation (DESI).
In the former analysis, we fix the neutrino mass hierarchy to be the normal one,
and set the fiducial value of the total neutrino mass $\Sigma m_{\nu}$
and the effective number of neutrino species $N_{\nu}$ 
to be $\Sigma m_{\nu} = 0.1$ or $0.06 \ {\rm eV}$ and $N_{\nu} = 3.046$.
%
%On the other hand, in the latter analysis, we fix $N_{\nu}$ to be $3.046$,
Next, in the latter analysis, we fix $N_{\nu}$ to be $3.046$,
and set the fiducial values of the $\Sigma m_{\nu}$ and the mass hierarchy parameter $r_{\nu}$ to be 
$(\Sigma m_{\nu},r_{\nu})=(0.06 \ {\rm eV}, 0.82)$ (normal hierarchy) or
$(\Sigma m_{\nu},r_{\nu})=(0.1 \ {\rm eV}, -0.46)$ (inverted hierarchy).

To obtain Fisher matrices we use CAMB \cite{Lewis:1999bs,CAMB} 
%%%%%%%%%%%%%%%%
\footnote{In this analysis, we use non-linear power spectra
for the calculations by performing a public code HALOFIT~\cite{Lewis:1999bs,CAMB}.
}
%%%%%%%%%%%
for calculations of CMB anisotropies $C_{l}$ and matter power spectra $P_{\delta \delta}(k)$.
In order to combine the CMB experiments the 21 cm line and the BAO observations,
 we calculate the combined Fisher matrix to be
\begin{equation}
%\label{ }
F_{\alpha\beta} 
= F^{\rm (21cm)}_{\alpha\beta} + F^{\rm (CMB)}_{\alpha\beta} 
+ F^{\rm (BAO)}_{\alpha\beta},
\end{equation}
In this paper, we do not use information for a possible correlation 
between fluctuations of the 21 cm and the CMB.

%%%%%%%%%%%%%%%%%%%%%%%%%%%%%%%%%%%%%%%%%%%%%%%%%%%%%%%%%%%%%%%%%%%%%
\subsection{Constraints on $\Sigma m_{\nu}$ and $N_{\nu}$}
%\section{Constraints on $\Sigma m_{\nu}$ and $N_{\nu}$}
\label{subsec:const_N_nu}
%%%%%%%%%%%%%%%%%%%%%%%%%%%%%%%%%%%%%%%%%%%%%%%%%%%%%%%%%%%%%%%%%%%%%

%%%%%%%%%%%%%%%%%FIGURE%%%%%%%%%%%%%%%%%%%%
\begin{figure*}[htbp]
 \begin{center}
   \includegraphics[bb=34 127 579 665 ,width=1\linewidth]{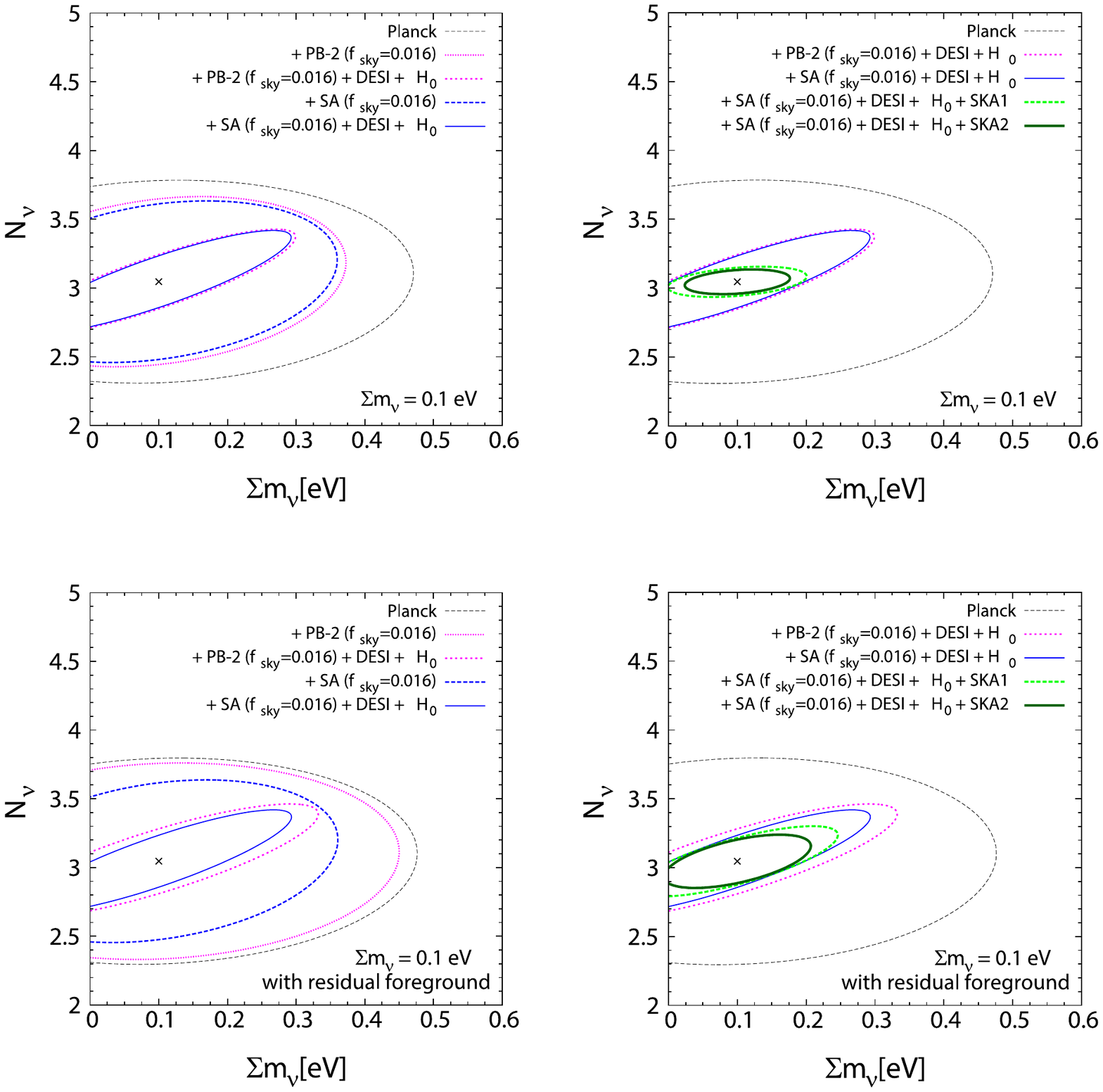}
   %\pdffile{Nnu.pdf ,width=1.0\hsize}
   \caption{ 
   Contours of 95\% C.L. forecasts in $\Sigma m_{\nu}$-$N_{\nu}$ plane.
   Fiducial values of neutrino parameters, $N_{\nu}$ and $\Sigma m_{\nu}$, 
   are taken to be $N_{\nu} = 3.046$ and $\Sigma m_{\nu} = 0.1$~eV.
   In the left two panels, the contours are the constraints 
   by adopting Planck (outer dashed line), 
   Planck combined with \textsc{Polarbear}-2 (PB-2) ($f_{{\rm sky}}=0.016$) (outer dotted line) 
   or Simons Array (SA) (inner thick dashed line),
   Planck + BAO(DESI) + Hubble prior + \textsc{Polarbear}-2 ($f_{{\rm sky}}=0.016$) 
   (inner thick dotted line) 
   or Simons Array (thin solid line), respectively.
   In the right two panels, they are the constraints 
   by adopting Planck (outer dashed line), 
   Planck + BAO(DESI) + Hubble prior combined with \textsc{Polarbear}-2 
   ($f_{{\rm sky}}=0.016$) (dotted line) or Simons Array (outer thin solid line),
   Planck + BAO(DESI) + Hubble prior + Simons Array 
   combined with SKA phase~1 ($N_{{\rm filed}}=4$) (inner thick dashed line) 
   or phase~2 ($N_{{\rm filed}}=4$) (inner thick line), respectively.
%   Same as Fig.\ref{fig:Nnu01_fsky02}, 
%   but sky coverages of \textsc{Polarbear}-2 and Simons Array
%   are $f_{sky}=0.016$
}\label{fig:Nnu01_fsky0016}
 \end{center}
\end{figure*}
%%%%%%%%%%%%%%%%%%%%%%%%%%%%%%%%%%%%%%%%%

%%%%%%%%%%%%%%%%%FIGURE%%%%%%%%%%%%%%%%%%%%
\begin{figure*}[htbp]
 \begin{center}
   \includegraphics[bb= 34 127 579 665,width=1\linewidth]{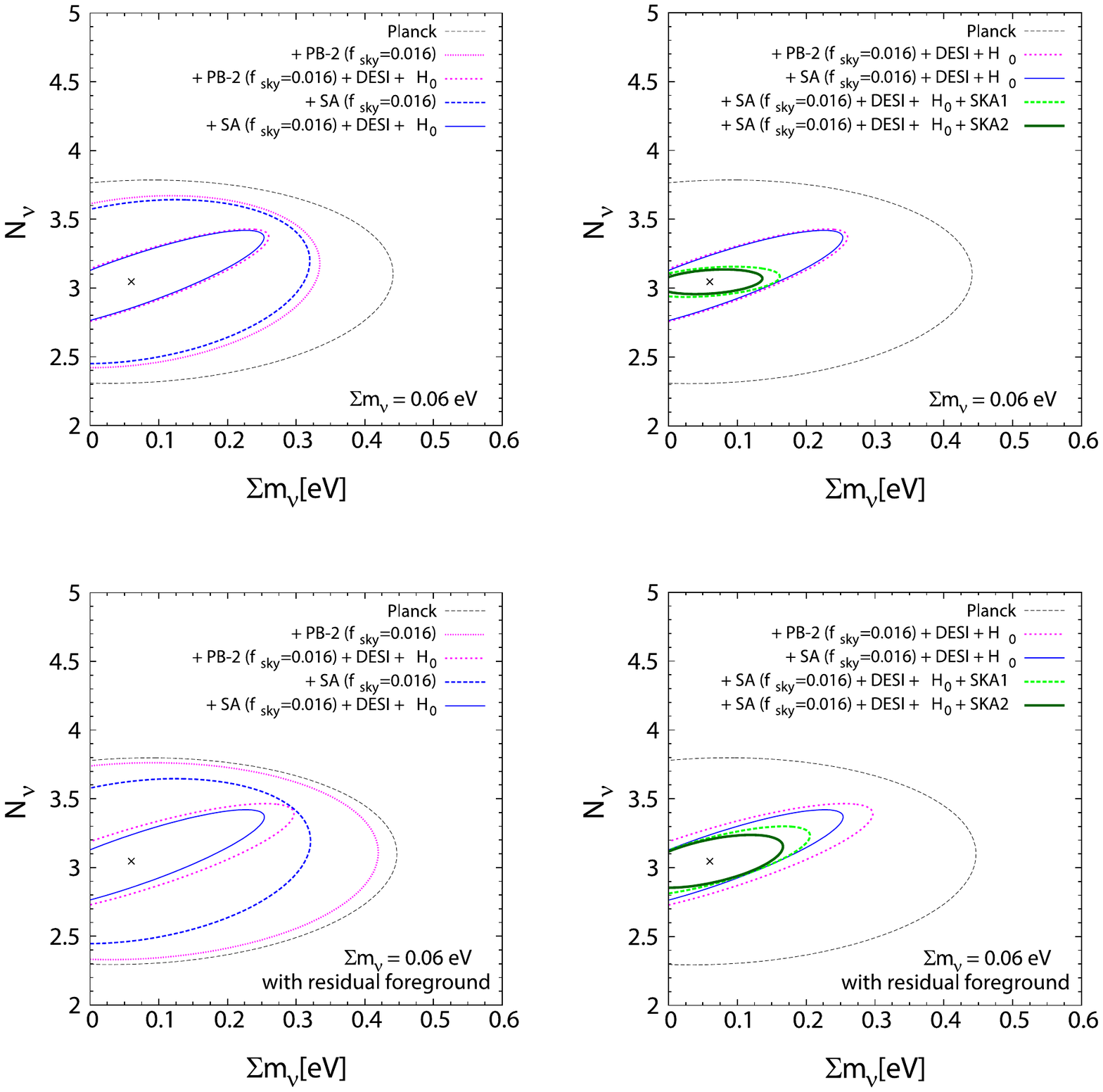}
   %\pdffile{Nnu.pdf ,width=1.0\hsize}
   \caption{ 
   Same as Fig.\ref{fig:Nnu01_fsky0016}, but 
   the fiducial values of neutrino parameters, $N_{\nu}$ and $\Sigma m_{\nu}$, are taken to be
   $N_{\nu} = 3.046$ and $\Sigma m_{\nu} = 0.06$~eV.}
   \label{fig:Nnu006_fsky0016}
 \end{center}
\end{figure*}
%%%%%%%%%%%%%%%%%%%%%%%%%%%%%%%%%%%%%%%%%

%%%%%%%%%%%%%%%%%FIGURE%%%%%%%%%%%%%%%%%%%%
\begin{figure*}[htbp]
 \begin{center}
   \includegraphics[bb=34 127 579 665,width=1\linewidth]{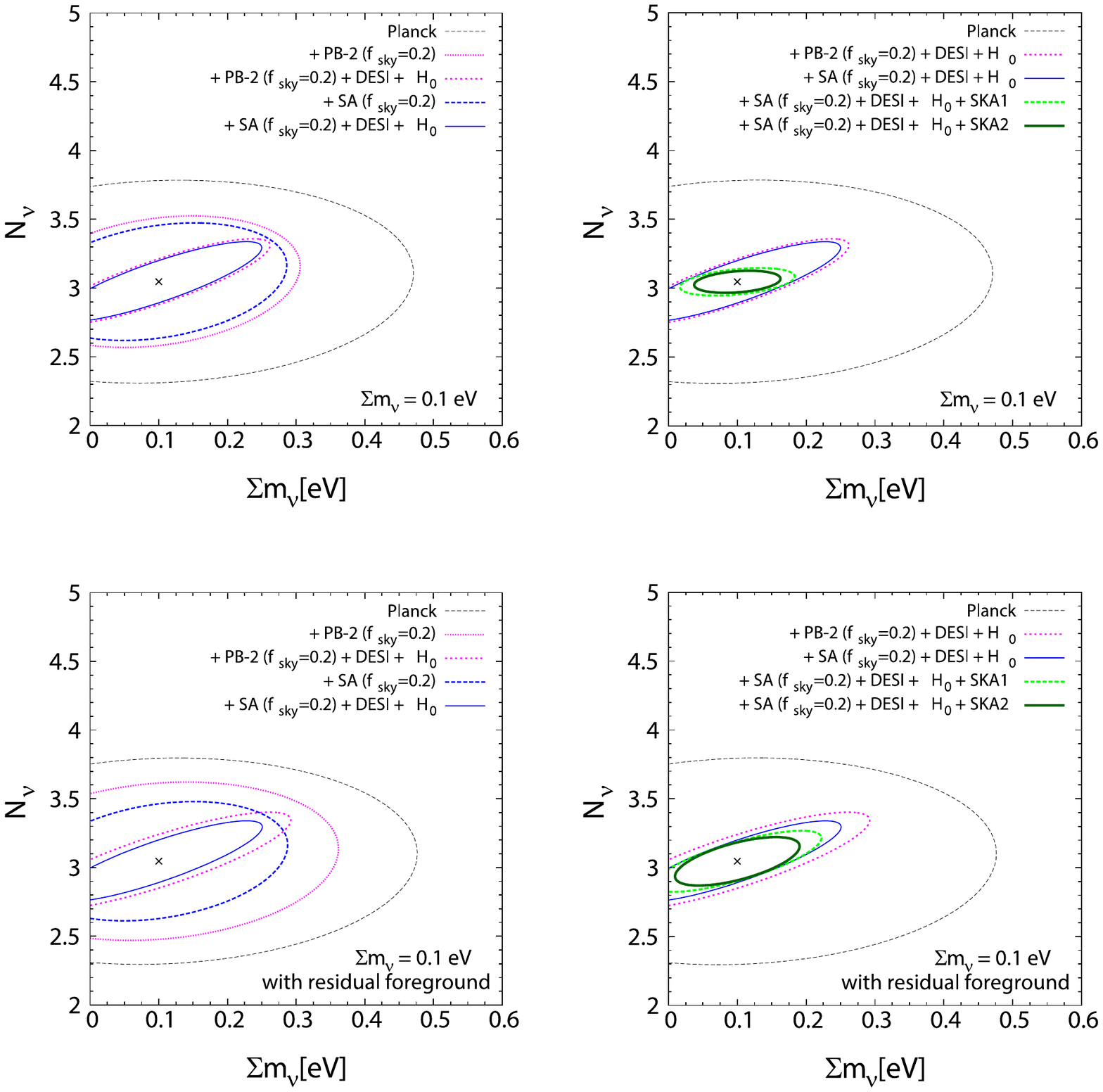}
   \caption{ 
   Same as Fig.\ref{fig:Nnu01_fsky0016}, 
   but the sky coverages of \textsc{Polarbear}-2 and Simons Array
   are $f_{{\rm sky}}=0.2$.
}
   \label{fig:Nnu01_fsky02}
 \end{center}
\end{figure*}
%%%%%%%%%%%%%%%%%%%%%%%%%%%%%%%%%%%%%%%%%

%%%%%%%%%%%%%%%%%FIGURE%%%%%%%%%%%%%%%%%%%%
\begin{figure*}[htbp]
 \begin{center}
   \includegraphics[bb= 34 127 579 665,width=1\linewidth]{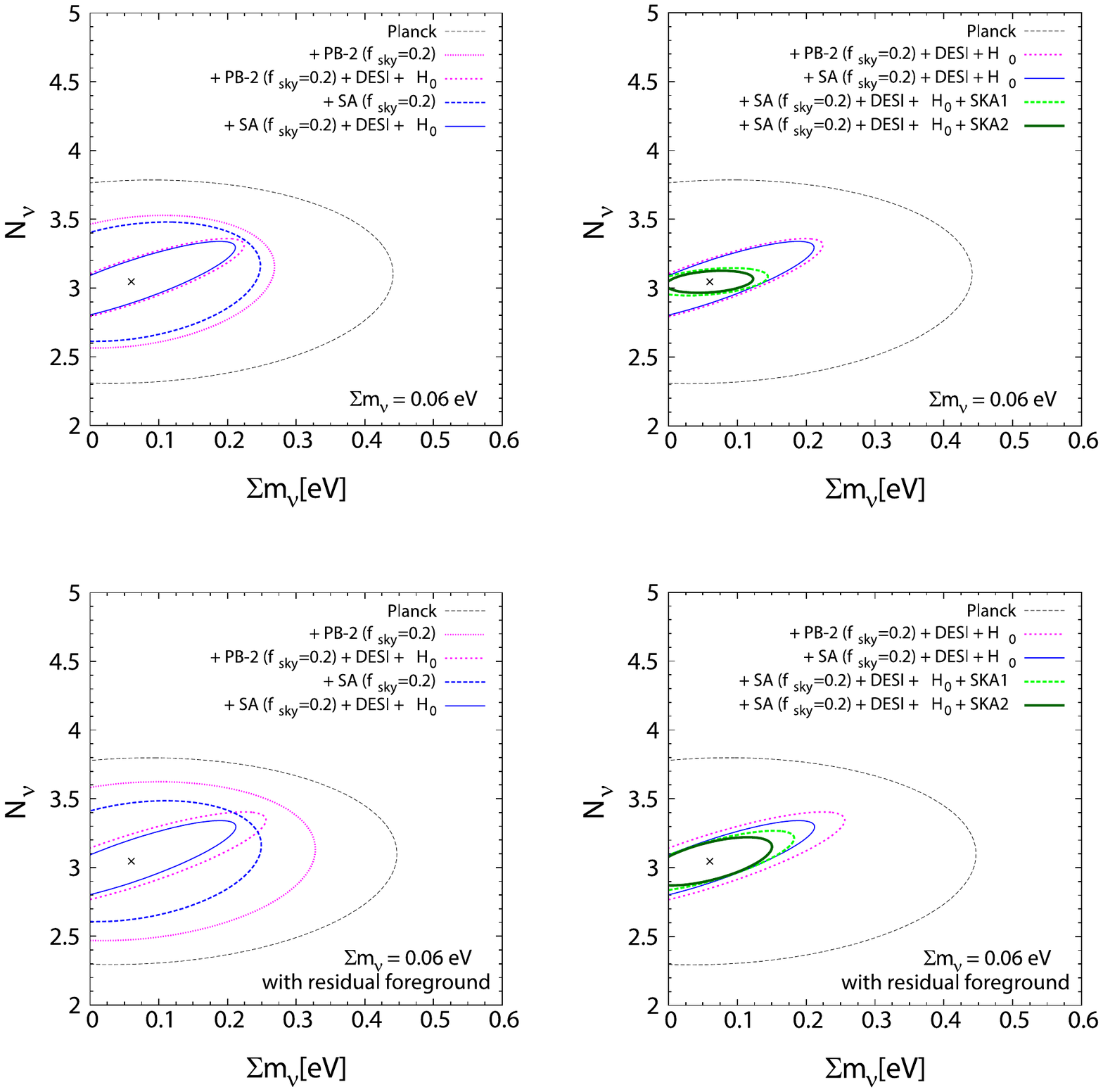}
   \caption{
   Same as Fig.\ref{fig:Nnu01_fsky02}, but 
   the fiducial values of neutrino parameters, $N_{\nu}$ and $\Sigma m_{\nu}$, are taken to be
   $N_{\nu} = 3.046$ and $\Sigma m_{\nu} = 0.06$~eV.
}
   \label{fig:Nnu006_fsky02}
 \end{center}
\end{figure*}
%%%%%%%%%%%%%%%%%%%%%%%%%%%%%%%%%%%%%%%%%

%%%%%%%%%%%%%%%%%FIGURE%%%%%%%%%%%%%%%%%%%%
\begin{figure*}[htbp]
 \begin{center}
   \includegraphics[bb= 34 127 579 665,width=1\linewidth]{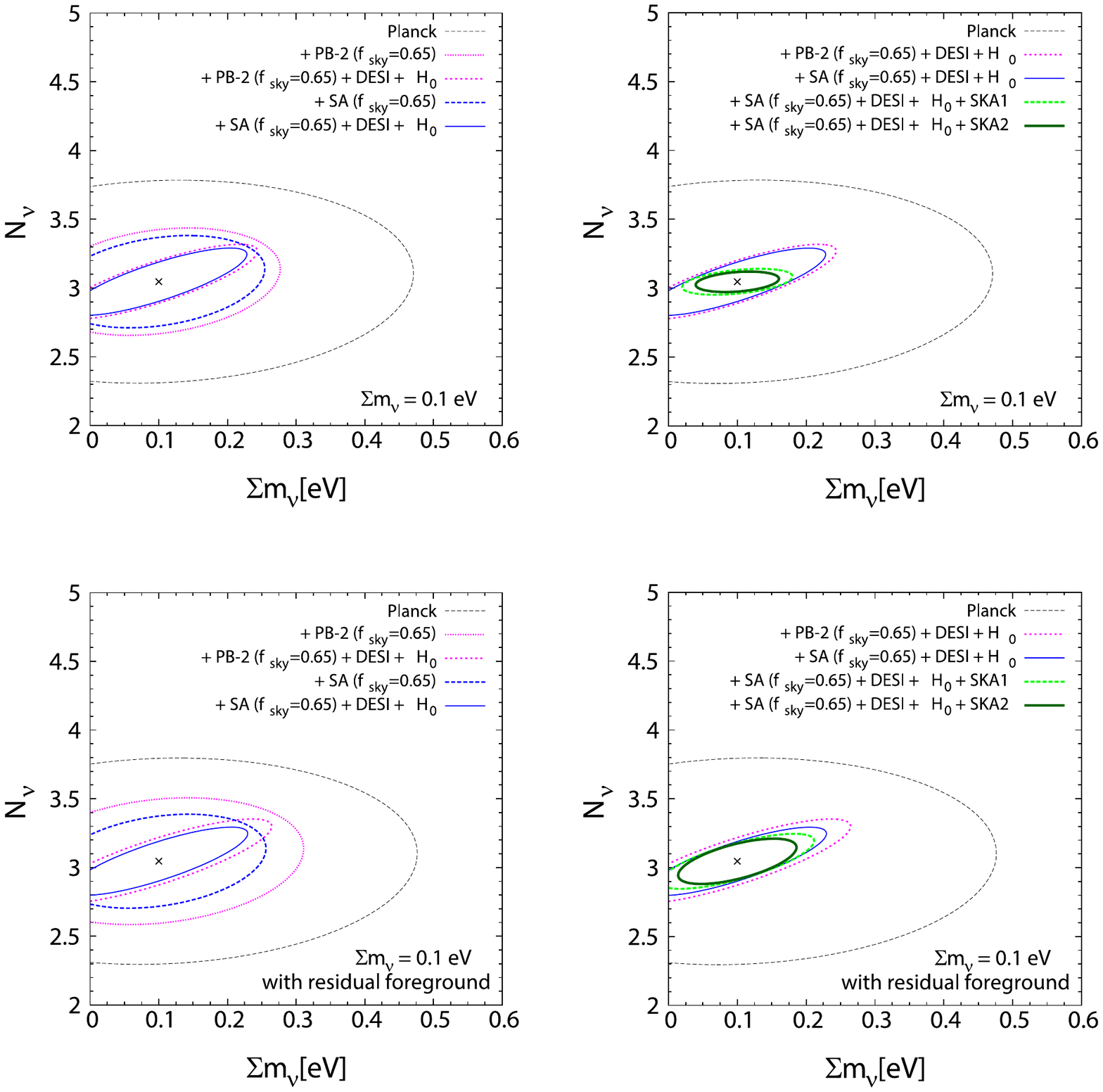}
   %\pdffile{Nnu.pdf ,width=1.0\hsize}
   \caption{ 
   Same as Fig.\ref{fig:Nnu01_fsky0016}, 
   but the sky coverages of \textsc{Polarbear}-2 and Simons Array
   are $f_{{\rm sky}}=0.65$.
}
   \label{fig:Nnu01_fsky065}
 \end{center}
\end{figure*}
%%%%%%%%%%%%%%%%%%%%%%%%%%%%%%%%%%%%%%%%%

%%%%%%%%%%%%%%%%%FIGURE%%%%%%%%%%%%%%%%%%%%
\begin{figure*}[htbp]
 \begin{center}
   \includegraphics[bb= 34 127 579 665,width=1\linewidth]{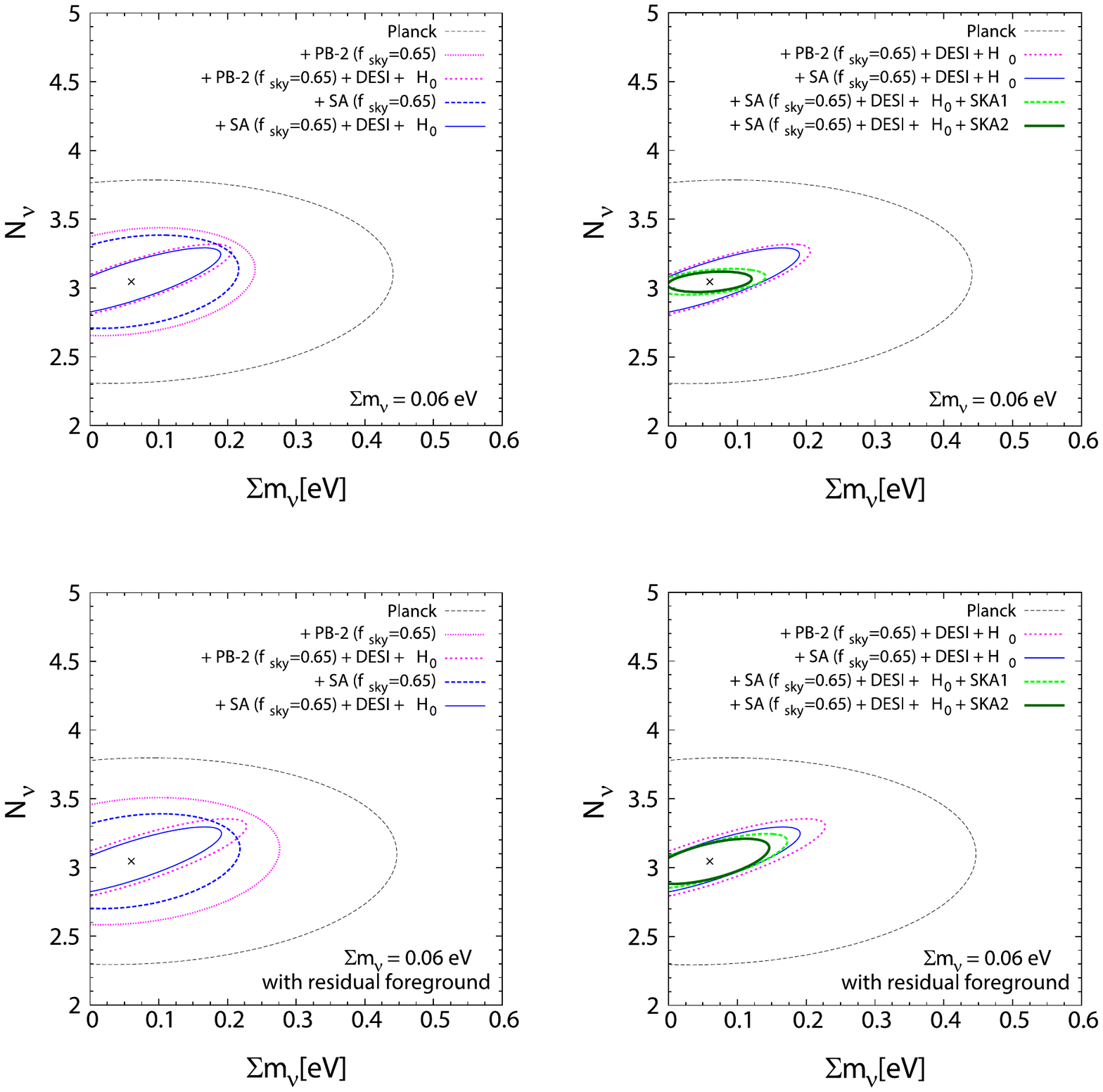}
   %\pdffile{Nnu.pdf ,width=1.0\hsize}
   \caption{ 
   Same as Fig.\ref{fig:Nnu01_fsky065}, but 
   the fiducial values of neutrino parameters, $N_{\nu}$ and $\Sigma m_{\nu}$, are taken to be
   $N_{\nu} = 3.046$ and $\Sigma m_{\nu} = 0.06$~eV.}
   \label{fig:Nnu006_fsky065}
 \end{center}
\end{figure*}
%%%%%%%%%%%%%%%%%%%%%%%%%%%%%%%%%%%%%%%%%

%%%%%%%%%%%%%%%%%FIGURE%%%%%%%%%%%%%%%%%%%%
\begin{figure*}[htbp]
 \begin{center}
   \includegraphics[bb= 41 23 572 767,width=0.85\linewidth]{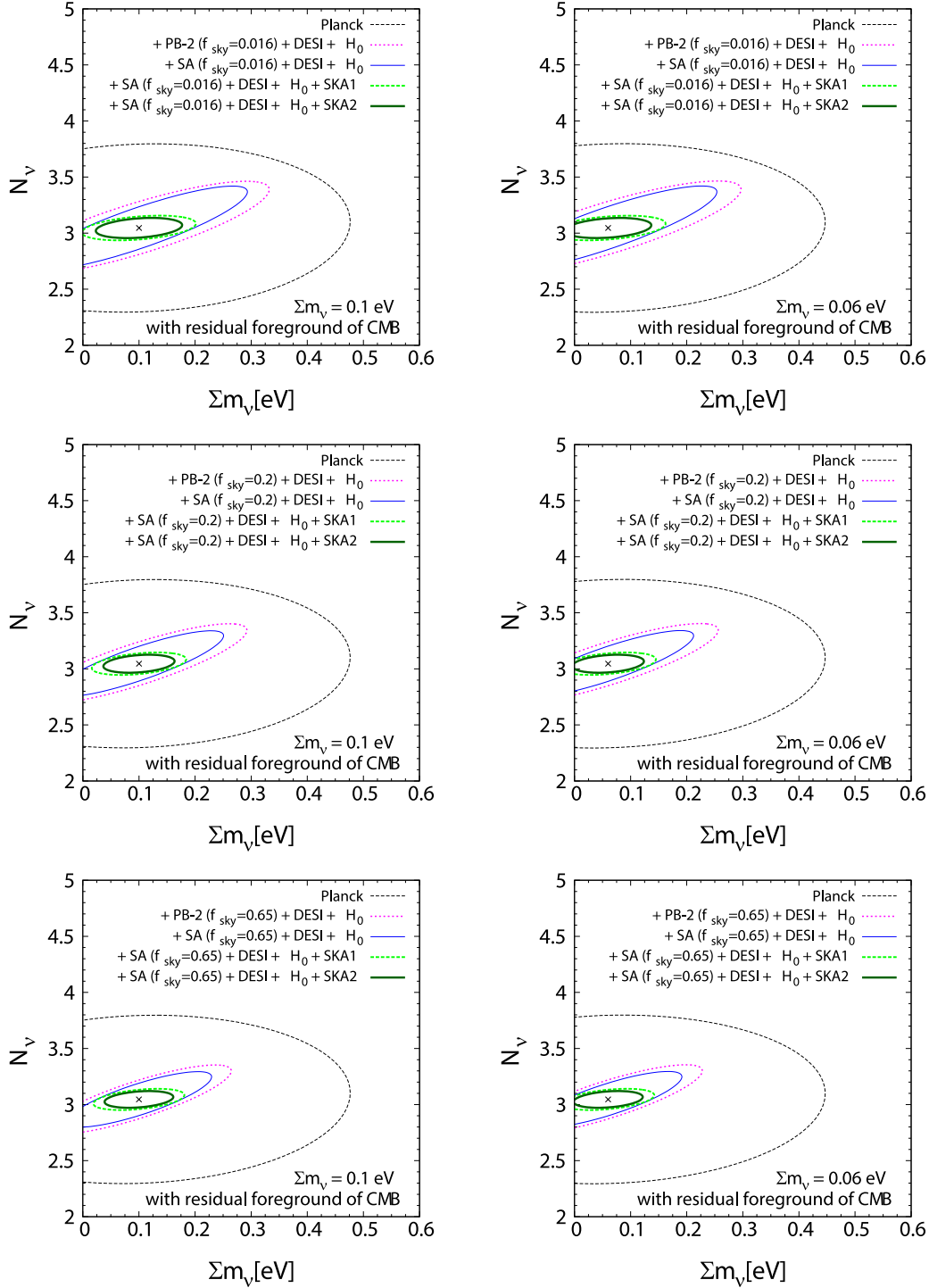}
   %\pdffile{Nnu.pdf ,width=1.0\hsize}
   \caption{
%   Contours of 95\% C.L. forecasts in $\Sigma m_{\nu}$-$N_{\nu}$ plane
%   when the foreground of 21 cm line is completely removed
   Contours of 95\% C.L. forecasts in $\Sigma m_{\nu}$-$N_{\nu}$ plane
   without residual foreground of 21 cm line
   (i.e. the residual foreground of 21 cm line is completely removed, 
   and that of CMB exists only).
   The fiducial values of neutrino parameters, $N_{\nu}$ and $\Sigma m_{\nu}$, are taken to be
   $N_{\nu} = 3.046$, $\Sigma m_{\nu} = 0.1$~eV in left panels,
   and  $\Sigma m_{\nu} = 0.06$~eV in right panels.
%   Same as Fig.\ref{fig:Nnu01_fsky0016}, but 
%   fiducial values of neutrino parameters, $N_{\nu}$ and $\Sigma m_{\nu}$, are taken to be
%   $N_{\nu} = 3.046$ and $\Sigma m_{\nu} = 0.06$~eV.}
   }
   \label{fig:Nnu_noSKAfg}
 \end{center}
\end{figure*}
%%%%%%%%%%%%%%%%%%%%%%%%%%%%%%%%%%%%%%%%%

%%%%%%%%%%%%%%%%%FIGURE%%%%%%%%%%%%%%%%%%%%
\begin{figure*}[htbp]
 \begin{center}
   \includegraphics[bb= 41 23 572 767,width=0.85\linewidth]{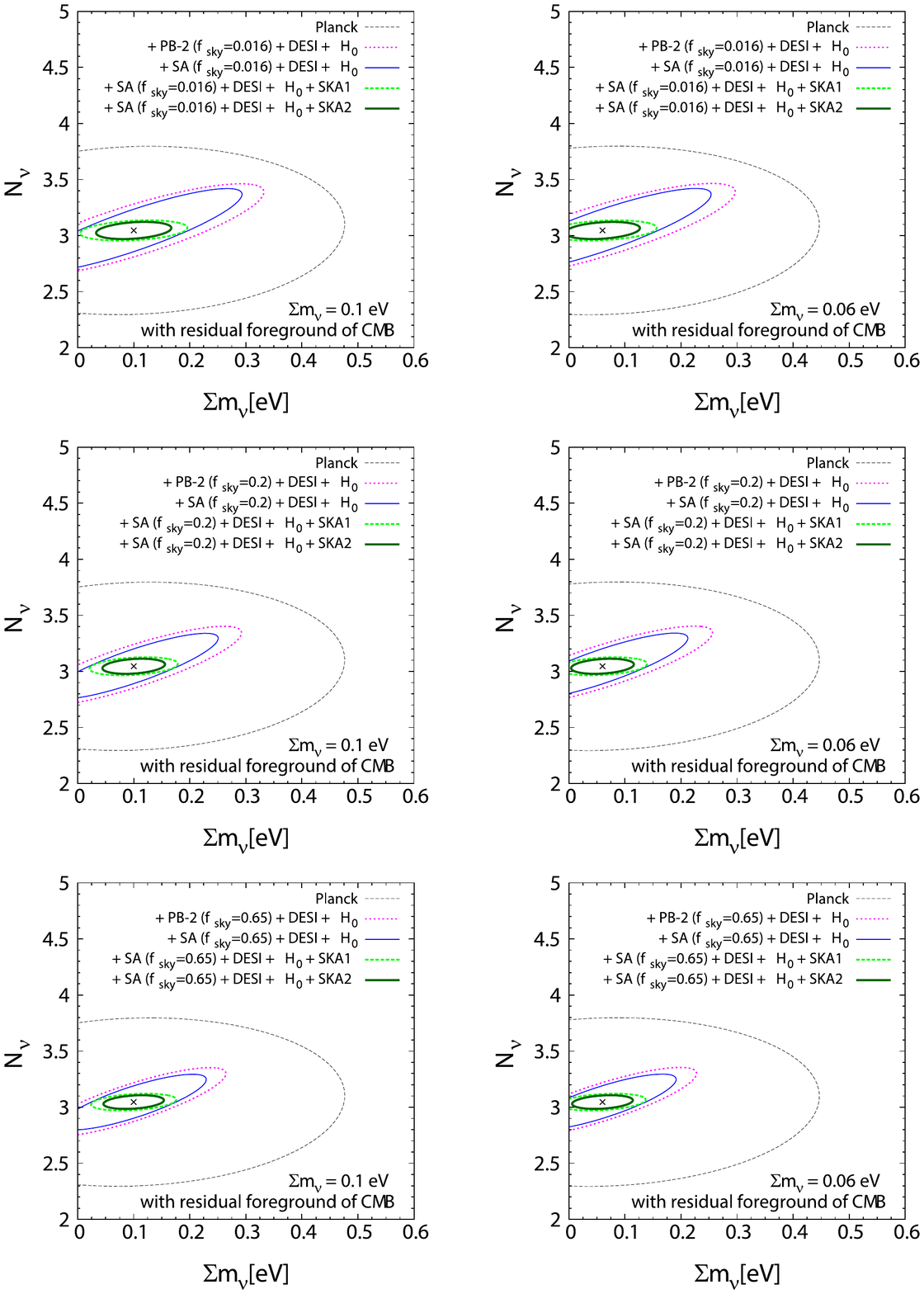}
   %\pdffile{Nnu.pdf ,width=1.0\hsize}
   \caption{ 
   Same as Fig.\ref{fig:Nnu_noSKAfg}, but 
   the observed fields of SKA are twice as large as that of Fig.\ref{fig:Nnu_noSKAfg}
   (i.e. $N_{{\rm field}} = 8$).
   %fiducial values of neutrino parameters, $N_{\nu}$ and $\Sigma m_{\nu}$, are taken to be
   %$N_{\nu} = 3.046$ and $\Sigma m_{\nu} = 0.06$~eV.}
   }
   \label{fig:Nnu_noSKAfg_m8}
 \end{center}
\end{figure*}
%%%%%%%%%%%%%%%%%%%%%%%%%%%%%%%%%%%%%%%%%

%%%%%%%%%%%%%%%%%%FIGURE%%%%%%%%%%%%%%%%%%%

\begin{figure*}[htbp]
 \begin{center}
   %\includegraphics
%   \resizebox{160mm}{!}{
   \includegraphics[bb=  178 270 435 520,width=0.5\linewidth]{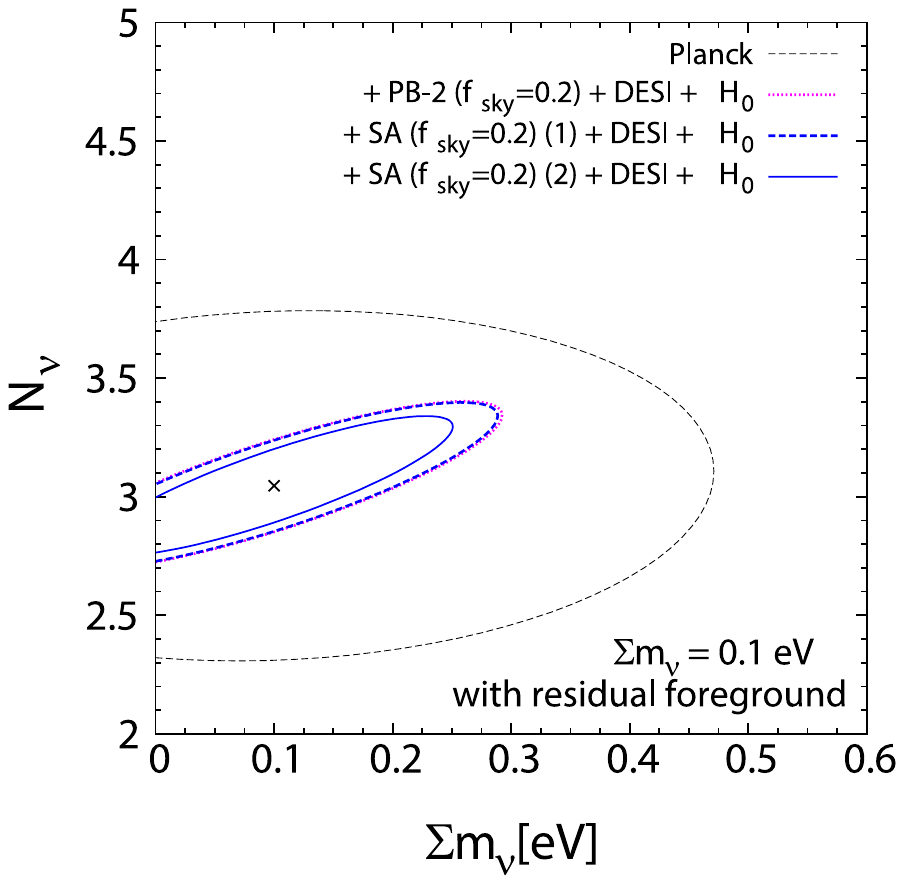}
%   }
   \caption{
   Contours of 95\% C.L. forecasts in $\Sigma m_{\nu}$-$N_{\nu}$ plane.
   The fiducial values of neutrino parameters, $N_{\nu}$ and $\Sigma m_{\nu}$, are taken to be
   $N_{\nu} = 3.046$ and $\Sigma m_{\nu} = 0.1$~eV.
   The contours are the constraints by adopting Planck (outer dashed line),
   Planck + BAO(DESI) + Hubble prior + \textsc{Polarbear}-2 (PB-2) ($f_{{\rm sky}}=0.2$) 
   (outer dotted line).
   For Simons Array, we plot results of two different cases.
   At first, we assume that the 220~GHz band of Simons Array 
   is used for only observation of CMB,
   and not used for the foreground removal (outer thick dashed line, 
   we call this situation Simons Array (1) or SA(1)).
   Secondary,  the 220~GHz band is used 
   for the foreground removal
   (inner solid line: we call this case Simons Array (2) or SA(2)).
   The constraint of Simons Array (1) almost laps over that of \textsc{Polarbear}-2.
}
   \label{fig:Nnu01_fsky02_Simons_fg}
 \end{center}
\end{figure*}

%%%%%%%%%%%%%%%%%%%%%%%%%%%%%%%%%%%%%%%%%

In Figs.\ref{fig:Nnu01_fsky0016}-\ref{fig:Nnu_noSKAfg_m8}, 
we plot contours of 95\% confidence levels (C.L.) forecasts 
of each combination of CMB, 21cm line and BAO observations
in $\Sigma m_{\nu}$-$N_{\nu}$ plane. 
Constraints on cosmological parameters are summarized
in Tables \ref{tab:fsk0016_Sigma_m_nu010} - \ref{tab:fsk065_fg_Sigma_m_nu006_SKAnoFG}.

The upper panels of Figs.\ref{fig:Nnu01_fsky0016}-\ref{fig:Nnu006_fsky065}
show forecasts without residual foregrounds 
of CMB and 21 cm line (i.e. they are completely removed),
and the lower ones show those
with given level residual foregrounds, respectively.
For Simons Array, we assume that 
its 220GHz band is used for the foreground removal
when we include the contribution of residual foreground,
and used for CMB observation when we do not include the residual foreground.
The fiducial value of the total neutrino mass is set to be $\Sigma m_{\nu} = 0.1$~eV 
(Figs.\ref{fig:Nnu01_fsky0016}, \ref{fig:Nnu01_fsky02} 
and \ref{fig:Nnu01_fsky065}) %and \ref{fig:Nnu01_fsky02_Simons_fg}), 
or $\Sigma m_{\nu} = 0.06$~eV (Figs.\ref{fig:Nnu006_fsky0016},
\ref{fig:Nnu006_fsky02} and \ref{fig:Nnu006_fsky065}).
%
%Additionally,
Sky coverages of \textsc{Polarbear}-2 and Simons Array are $f_{{\rm sky}}=0.016$
(Figs.\ref{fig:Nnu01_fsky0016} and \ref{fig:Nnu006_fsky0016}),
% and \ref{fig:Nnu01_fsky02_Simons_fg}),
$f_{{\rm sky}}=0.2$ (Figs.\ref{fig:Nnu01_fsky02} and \ref{fig:Nnu006_fsky02}),
or $f_{{\rm sky}}=0.65$ (Figs.\ref{fig:Nnu01_fsky065} and \ref{fig:Nnu006_fsky065}).
%
%
%The fiducial values of the total neutrino mass is $\Sigma m_{\nu} = 0.1$~eV 
%(Fig.\ref{fig:Nnu01_fsky02}, \ref{fig:Nnu01_fsky0016} and \ref{fig:Nnu01_fsky02_Simons_fg}), 
%or $\Sigma m_{\nu} = 0.06$~eV (Fig.\ref{fig:Nnu006_fsky02} and \ref{fig:Nnu006_fsky0016}).
%
%Additionally,
%sky coverages of \textsc{Polarbear}-2 and Simons Array are $f_{sky}=0.2$
%(Fig.\ref{fig:Nnu01_fsky02}, \ref{fig:Nnu006_fsky02} and \ref{fig:Nnu01_fsky02_Simons_fg})
%or $f_{sky}=0.016$ (Fig.\ref{fig:Nnu01_fsky0016} and \ref{fig:Nnu006_fsky0016}).
%
In the left two panels of Figs.\ref{fig:Nnu01_fsky0016}-\ref{fig:Nnu006_fsky065},
each contour represents constraints by CMB only or CMB + BAO~(DESI) + Hubble prior.
In the right two panels of them, each contour represents constraints
by Planck only or CMB + BAO~(DESI) + Hubble prior + 
21cm line~(SKA).
In these six figures, we assume that 
%the observed fields of SKA are 4 field ($N_{{\rm field}}=4$).
the number of the observed fields by SKA is four ($N_{{\rm field}}=4$).
%
%In Fig.\ref{fig:Nnu01_fsky02}-\ref{fig:Nnu006_fsky0016},
%In these four figures,
%upper two panels are results when we assume that residual foregrounds are completely removed.
%In contrast, lower panels are results 
%when the residual foregrounds are remaining
%and 

From these figures, adding the BAO experiments to the CMB ones,
we see that there is a strong improvement on the sensitivities to $\Sigma m_{\nu}$ and $N_{\nu}$
because several parameter degeneracies are broken by those combinations.
Besides, we find that 
%larger sky coverage is more effective than smaller one.
the larger sky coverage improves the constraints more. % more effective than smaller one.
However, it is difficult to detect the nonzero neutrino mass at 2$\sigma$ level
even by using the combination of Simons Array and DESI.
%except for the case of $f_{{\rm sky}}=0.65$ and $\Sigma m_{\nu}=0.1$~eV
%without residual foreground.

%
On the other hand, adding the 21 cm experiments (SKA phase~1) to the CMB observation, 
we see that there is a substantial improvement.
When we neglect the contribution from the residual foreground,
the combination of SKA phase~1 and Simons Array ($f_{\textrm{sky}}=0.2$ or $f_{\textrm{sk}y}=0.65$)
 has enough sensitivity to detect nonzero neutrino mass 
in the case of $\Sigma m_{\nu}=0.1$ eV to be fiducial.
%without residual foregrounds.
%when we neglect the contribution of residual foreground.
%
Of course, CMB + SKA phase~2 can obviously do the same job.
For the effective number of neutrino species,
its errors by the combination of CMB and BAO with SKA 
are several times as small as those of only CMB + BAO.
%
%In the case with the residual foregrounds,
%Planck + Simons Array ($f_{sky}=0.2$) + BAO + SKA phase 1 can
%detect the neutrino mass, however 
%Planck + Simons Array ($f_{sky}=0.016$) + BAO + SKA phase 1 
%does not have enough sensitivity, and
%only combination with SKA phase 2 can do it.
%
In the case with a given level of residual foregrounds,
%Planck + Simons Array ($f_{sky}=0.2$) + BAO + SKA phase 1 can
%detect the neutrino mass, 
Planck + Simons Array + BAO + SKA phase~1 
can not detect nonzero neutrino mass, and
%does not have enough sensitivity, and
only a combination of SKA phase~2 with
Simons Array ($f_{{\rm sky}}=0.2$ or $0.65$) can do it.
In the case of $\Sigma m_{\nu}=0.06$ eV to be fiducial
(this value corresponds to the lowest value for the normal hierarchy),
%
%without residual foregrounds, 
%only Planck + Simons Array ($f_{{\rm sky}} = 0.2$ or $0.65$) 
%+ BAO + SKA phase2 can detect the nonzero neutrino mass
%when we neglect the contribution of residual foreground.
%
%
%However, in the case with given level residual foregrounds,
%even combination with SKA phase 2 can not do it.
it is difficult to detect the nonzero neutrino mass
even by the combination with SKA phase~2.
For the effective number of neutrino species, its errors in this case
are a few times as large as those in the case without the residual foreground.
Hence, a stronger foreground removal of the 21 cm line observation is necessary
if we want to obtain a stringent constraint on the effective number of neutrino species.
%
%Therefore, stronger foreground removal of 21 cm line observation is necessary.

We can find the necessity of the foreground removal of the 21cm line
from Figs.\ref{fig:Nnu_noSKAfg} and \ref{fig:Nnu_noSKAfg_m8}.
These figures show forecasts without the residual foreground of 21 cm line
but with that of CMB (i.e. the foreground of the 21 cm line is completely removed).
We assume that the number of observed fields of SKA 
is $N_{{\rm field}}=4$ in Fig.\ref{fig:Nnu_noSKAfg}
and $N_{{\rm field}}=8$ in Fig.\ref{fig:Nnu_noSKAfg_m8}.
In these cases, constraints of $\Sigma m_{\nu}$ and $N_{\nu}$
are strongly improved in comparison with 
the lower panels of Figs.\ref{fig:Nnu01_fsky0016}-\ref{fig:Nnu006_fsky065},
and the errors of the effective number of neutrino species
%are about $\sigma (N_{\nu}) \sim 0.08$ ($N_{{\rm field}}=4$) 
%or $0.06$ ($N_{{\rm field}}=8$) at 2$\sigma$ by SKA.
are approximately $\sigma (N_{\nu}) \sim 0.06 - 0.09$
%or $\sigma (N_{\nu}) \sim 0.05 - 0.08$ at 2$\sigma$ 
%%%%%%2015/01/14 revised ver
or $\sigma (N_{\nu}) \sim 0.05 - 0.07$ at 2$\sigma$ 
%%%%%%%%%%
%
by Simons Array + SKA phase~1 or phase~2, respectively.
Besides, from Fig.\ref{fig:Nnu_noSKAfg},
by Planck + Simons Array ($f_{{\rm sky}}=0.2$ or $0.65$) + BAO + SKA phase~1
%and Planck + Simons Array ($f_{{\rm sky}}=0.016$) + BAO + SKA phase2
and Planck + Simons Array + BAO + SKA phase~2, 
we can detect the nonzero neutrino mass of $\Sigma m_{\nu} = 0.1$~eV
in this case.
Additionally, from Fig.\ref{fig:Nnu_noSKAfg_m8},
we find that more observed fields of 21 cm line observation 
($N_{{\rm filed}}=8$) are necessary
if we want to detect the nonzero neutrino mass of $\Sigma m_{\nu} = 0.06$~eV
by SKA phase~2.

In Fig.\ref{fig:Nnu01_fsky02_Simons_fg}, 
we show two different cases about observations of Simons Array.
At first, we assume that the 220 GHz band of Simons Array is
used for the foreground removal (inner solid line),
and this assumption is used for the analysis
of Figs.\ref{fig:Nnu01_fsky0016}-\ref{fig:Nnu_noSKAfg_m8}.
Secondary, the band is used 
for only the observation of CMB, and not used for the foreground removal,
which is done by only Planck (outer thick dashed line).
%
%We plot the both results in the Fig.\ref{fig:Nnu01_fsky02_Simons_fg}.
%
This figure shows that 
the latter constraint is almost the same level as that of \textsc{Polarbear}-2.
The reason is that the strength of the residual foreground depends 
on only the Planck sensitivity.
Therefore, it is better to use the 220 GHz band of Simon Array 
for the foreground removal.
Constraints on cosmological parameters in this situation
are summarized in Tables \ref{tab:fsk02_fg_Sigma_m_nu010_Simons} and \ref{tab:fsk02_fg_Sigma_m_nu006_Simons}.

\begin{table}[htbp]
\centering \scalebox{0.85}[0.85]{
\begin{tabular}{l|ccccc}
\hline \hline \\
& $\sigma( \Omega_{m}h^{2} )$ & $\sigma( \Omega_{b}h^{2} )$ &  $\sigma( \Omega_{\Lambda} )$ & $ \sigma( n_s ) $ & $\sigma( A_{s} \times 10^{10} )$ 
\\
\hline
Planck 
 & $ 5.31 \times 10^{  -3} $ & $ 2.44 \times 10^{  -4} $ & $ 2.38 \times 10^{  -2} $ & $ 8.61 \times 10^{  -3} $ & $ 2.83 \times 10^{  -1} $ 
\\
+ PB-2 (0.016)
 & $ 4.57 \times 10^{  -3} $ & $ 2.16 \times 10^{  -4} $ & $ 1.91 \times 10^{  -2} $ & $ 7.69 \times 10^{  -3} $ & $ 2.62 \times 10^{  -1} $ 
\\
+ PB-2 + BAO + $H_{0}$
 & $ 3.64 \times 10^{  -3} $ & $ 1.56 \times 10^{  -4} $ & $ 9.77 \times 10^{  -3} $ & $ 5.37 \times 10^{  -3} $ & $ 2.34 \times 10^{  -1} $ 
\\
+ PB-2 + BAO + $H_{0}$ + SKA1
 & $ 9.38 \times 10^{  -4} $ & $ 1.46 \times 10^{  -4} $ & $ 4.45 \times 10^{  -3} $ & $ 4.80 \times 10^{  -3} $ & $ 2.07 \times 10^{  -1} $ 
\\
+ PB-2 + BAO + $H_{0}$ + SKA2
 & $ 5.76 \times 10^{  -4} $ & $ 1.32 \times 10^{  -4} $ & $ 2.70 \times 10^{  -3} $ & $ 2.89 \times 10^{  -3} $ & $ 1.91 \times 10^{  -1} $ 
\\
+ SA (0.016)
 & $ 4.40 \times 10^{  -3} $ & $ 2.13 \times 10^{  -4} $ & $ 1.84 \times 10^{  -2} $ & $ 7.57 \times 10^{  -3} $ & $ 2.58 \times 10^{  -1} $ 
\\
\hline
\end{tabular} }
\caption{$f_{\textrm{sky}}=0.016$, $N_{\textrm{field}}=4$, $\Sigma m_{\nu} = 0.1$~eV, without the residual foregrounds.}
\label{tab:fsk0016_Sigma_m_nu010}

\vspace{10pt}

\centering \scalebox{0.85}[0.85]{
\begin{tabular}{l|ccccc}
\hline \hline \\
& $\sigma( \Omega_{m}h^{2} )$ & $\sigma( \Omega_{b}h^{2} )$ &  $\sigma( \Omega_{\Lambda} )$ & $ \sigma( n_s ) $ & $\sigma( A_{s} \times 10^{10} )$ 
\\
\hline
Planck 
 & $ 5.35 \times 10^{  -3} $ & $ 2.45 \times 10^{  -4} $ & $ 2.41 \times 10^{  -2} $ & $ 8.61 \times 10^{  -3} $ & $ 2.83 \times 10^{  -1} $ 
\\
+ PB-2 (0.016)
 & $ 4.62 \times 10^{  -3} $ & $ 2.16 \times 10^{  -4} $ & $ 1.91 \times 10^{  -2} $ & $ 7.69 \times 10^{  -3} $ & $ 2.64 \times 10^{  -1} $ 
\\
+ PB-2 + BAO + $H_{0}$
 & $ 3.68 \times 10^{  -3} $ & $ 1.56 \times 10^{  -4} $ & $ 9.87 \times 10^{  -3} $ & $ 5.37 \times 10^{  -3} $ & $ 2.34 \times 10^{  -1} $ 
\\
+ PB-2 + BAO + $H_{0}$ + SKA1
 & $ 9.43 \times 10^{  -4} $ & $ 1.46 \times 10^{  -4} $ & $ 4.46 \times 10^{  -3} $ & $ 4.80 \times 10^{  -3} $ & $ 2.07 \times 10^{  -1} $ 
\\
+ PB-2 + BAO + $H_{0}$ + SKA2
 & $ 5.78 \times 10^{  -4} $ & $ 1.31 \times 10^{  -4} $ & $ 2.69 \times 10^{  -3} $ & $ 2.86 \times 10^{  -3} $ & $ 1.91 \times 10^{  -1} $ 
\\
+ SA (0.016)
 & $ 4.46 \times 10^{  -3} $ & $ 2.13 \times 10^{  -4} $ & $ 1.84 \times 10^{  -2} $ & $ 7.58 \times 10^{  -3} $ & $ 2.59 \times 10^{  -1} $ 
\\
\hline
\end{tabular} }
\caption{$f_{\textrm{sky}}=0.016$, $N_{\textrm{field}}=4$, $\Sigma m_{\nu} = 0.06$~eV, without the residual foregrounds.}
\label{tab:fsk0016_Sigma_m_nu006}

\end{table}

\begin{table}[htbp]
\centering \scalebox{0.85}[0.85]{
\begin{tabular}{l|ccccc}
\hline \hline \\
& $\sigma( \Omega_{m}h^{2} )$ & $\sigma( \Omega_{b}h^{2} )$ &  $\sigma( \Omega_{\Lambda} )$ & $ \sigma( n_s ) $ & $\sigma( A_{s} \times 10^{10} )$ 
\\
\hline
Planck 
 & $ 5.37 \times 10^{  -3} $ & $ 2.49 \times 10^{  -4} $ & $ 2.43 \times 10^{  -2} $ & $ 8.82 \times 10^{  -3} $ & $ 2.86 \times 10^{  -1} $ 
\\
+ PB-2 (0.016)
 & $ 5.14 \times 10^{  -3} $ & $ 2.33 \times 10^{  -4} $ & $ 2.26 \times 10^{  -2} $ & $ 8.24 \times 10^{  -3} $ & $ 2.76 \times 10^{  -1} $ 
\\
+ PB-2 + BAO + $H_{0}$
 & $ 3.92 \times 10^{  -3} $ & $ 1.70 \times 10^{  -4} $ & $ 1.05 \times 10^{  -2} $ & $ 5.75 \times 10^{  -3} $ & $ 2.34 \times 10^{  -1} $ 
\\
+ PB-2 + BAO + $H_{0}$ + SKA1
 & $ 2.49 \times 10^{  -3} $ & $ 1.67 \times 10^{  -4} $ & $ 7.27 \times 10^{  -3} $ & $ 5.63 \times 10^{  -3} $ & $ 2.18 \times 10^{  -1} $ 
\\
+ PB-2 + BAO + $H_{0}$ + SKA2
 & $ 1.65 \times 10^{  -3} $ & $ 1.57 \times 10^{  -4} $ & $ 4.59 \times 10^{  -3} $ & $ 4.58 \times 10^{  -3} $ & $ 2.05 \times 10^{  -1} $ 
\\
+ SA (0.016)
 & $ 4.42 \times 10^{  -3} $ & $ 2.14 \times 10^{  -4} $ & $ 1.85 \times 10^{  -2} $ & $ 7.59 \times 10^{  -3} $ & $ 2.55 \times 10^{  -1} $ 
\\
\hline
\end{tabular} }
\caption{$f_{\textrm{sky}}=0.016$, $N_{\textrm{field}}=4$, $\Sigma m_{\nu} = 0.1$~eV, with the residual foregrounds.}
\label{tab:fsk0016_fg_Sigma_m_nu010}

\vspace{10pt}

\centering \scalebox{0.85}[0.85]{
\begin{tabular}{l|ccccc}
\hline \hline \\
& $\sigma( \Omega_{m}h^{2} )$ & $\sigma( \Omega_{b}h^{2} )$ &  $\sigma( \Omega_{\Lambda} )$ & $ \sigma( n_s ) $ & $\sigma( A_{s} \times 10^{10} )$ 
\\
\hline
Planck 
 & $ 5.42 \times 10^{  -3} $ & $ 2.50 \times 10^{  -4} $ & $ 2.46 \times 10^{  -2} $ & $ 8.82 \times 10^{  -3} $ & $ 2.86 \times 10^{  -1} $ 
\\
+ PB-2 (0.016)
 & $ 5.18 \times 10^{  -3} $ & $ 2.33 \times 10^{  -4} $ & $ 2.28 \times 10^{  -2} $ & $ 8.23 \times 10^{  -3} $ & $ 2.76 \times 10^{  -1} $ 
\\
+ PB-2 + BAO + $H_{0}$
 & $ 3.97 \times 10^{  -3} $ & $ 1.70 \times 10^{  -4} $ & $ 1.06 \times 10^{  -2} $ & $ 5.75 \times 10^{  -3} $ & $ 2.34 \times 10^{  -1} $ 
\\
+ PB-2 + BAO + $H_{0}$ + SKA1
 & $ 2.50 \times 10^{  -3} $ & $ 1.67 \times 10^{  -4} $ & $ 7.28 \times 10^{  -3} $ & $ 5.62 \times 10^{  -3} $ & $ 2.18 \times 10^{  -1} $ 
\\
+ PB-2 + BAO + $H_{0}$ + SKA2
 & $ 1.64 \times 10^{  -3} $ & $ 1.57 \times 10^{  -4} $ & $ 4.56 \times 10^{  -3} $ & $ 4.55 \times 10^{  -3} $ & $ 2.04 \times 10^{  -1} $ 
\\
+ SA (0.016)
 & $ 4.48 \times 10^{  -3} $ & $ 2.14 \times 10^{  -4} $ & $ 1.84 \times 10^{  -2} $ & $ 7.59 \times 10^{  -3} $ & $ 2.57 \times 10^{  -1} $ 
\\
\hline
\end{tabular} }
\caption{$f_{\textrm{sky}}=0.016$, $N_{\textrm{field}}=4$, $\Sigma m_{\nu} = 0.06$~eV, with the residual foregrounds.}
\label{tab:fsk0016_fg_Sigma_m_nu006}

\end{table}

\begin{table}[htbp]
\centering \scalebox{0.85}[0.85]{
\begin{tabular}{l|ccccc}
\hline \hline \\
& $\sigma( \Omega_{m}h^{2} )$ & $\sigma( \Omega_{b}h^{2} )$ &  $\sigma( \Omega_{\Lambda} )$ & $ \sigma( n_s ) $ & $\sigma( A_{s} \times 10^{10} )$ 
\\
\hline
Planck 
 & $ 5.31 \times 10^{  -3} $ & $ 2.44 \times 10^{  -4} $ & $ 2.38 \times 10^{  -2} $ & $ 8.61 \times 10^{  -3} $ & $ 2.83 \times 10^{  -1} $ 
\\
+ PB-2 (0.2)
 & $ 3.58 \times 10^{  -3} $ & $ 1.50 \times 10^{  -4} $ & $ 1.40 \times 10^{  -2} $ & $ 5.36 \times 10^{  -3} $ & $ 2.38 \times 10^{  -1} $ 
\\
+ PB-2 + BAO + $H_{0}$
 & $ 3.03 \times 10^{  -3} $ & $ 1.19 \times 10^{  -4} $ & $ 8.23 \times 10^{  -3} $ & $ 4.13 \times 10^{  -3} $ & $ 2.25 \times 10^{  -1} $ 
\\
+ PB-2 + BAO + $H_{0}$ + SKA1
 & $ 9.18 \times 10^{  -4} $ & $ 1.15 \times 10^{  -4} $ & $ 4.35 \times 10^{  -3} $ & $ 3.84 \times 10^{  -3} $ & $ 1.99 \times 10^{  -1} $ 
\\
+ PB-2 + BAO + $H_{0}$ + SKA2
 & $ 5.67 \times 10^{  -4} $ & $ 1.07 \times 10^{  -4} $ & $ 2.67 \times 10^{  -3} $ & $ 2.58 \times 10^{  -3} $ & $ 1.85 \times 10^{  -1} $ 
\\
+ SA (0.2)
 & $ 3.26 \times 10^{  -3} $ & $ 1.42 \times 10^{  -4} $ & $ 1.30 \times 10^{  -2} $ & $ 5.04 \times 10^{  -3} $ & $ 2.31 \times 10^{  -1} $ 
\\
\hline
\end{tabular} }
\caption{$f_{\textrm{sky}}=0.2$, $N_{\textrm{field}}=4$, $\Sigma m_{\nu} = 0.1$~eV, without the residual foregrounds.}
\label{tab:fsk02_Sigma_m_nu010}

\vspace{10pt}

\centering \scalebox{0.85}[0.85]{
\begin{tabular}{l|ccccc}
\hline \hline \\
& $\sigma( \Omega_{m}h^{2} )$ & $\sigma( \Omega_{b}h^{2} )$ &  $\sigma( \Omega_{\Lambda} )$ & $ \sigma( n_s ) $ & $\sigma( A_{s} \times 10^{10} )$ 
\\
\hline
Planck 
 & $ 5.35 \times 10^{  -3} $ & $ 2.45 \times 10^{  -4} $ & $ 2.41 \times 10^{  -2} $ & $ 8.61 \times 10^{  -3} $ & $ 2.83 \times 10^{  -1} $ 
\\
+ PB-2 (0.2)
 & $ 3.63 \times 10^{  -3} $ & $ 1.50 \times 10^{  -4} $ & $ 1.40 \times 10^{  -2} $ & $ 5.36 \times 10^{  -3} $ & $ 2.38 \times 10^{  -1} $ 
\\
+ PB-2 + BAO + $H_{0}$
 & $ 3.07 \times 10^{  -3} $ & $ 1.19 \times 10^{  -4} $ & $ 8.32 \times 10^{  -3} $ & $ 4.13 \times 10^{  -3} $ & $ 2.25 \times 10^{  -1} $ 
\\
+ PB-2 + BAO + $H_{0}$ + SKA1
 & $ 9.23 \times 10^{  -4} $ & $ 1.15 \times 10^{  -4} $ & $ 4.36 \times 10^{  -3} $ & $ 3.83 \times 10^{  -3} $ & $ 1.99 \times 10^{  -1} $ 
\\
+ PB-2 + BAO + $H_{0}$ + SKA2
 & $ 5.69 \times 10^{  -4} $ & $ 1.07 \times 10^{  -4} $ & $ 2.66 \times 10^{  -3} $ & $ 2.56 \times 10^{  -3} $ & $ 1.85 \times 10^{  -1} $ 
\\
+ SA (0.2)
 & $ 3.32 \times 10^{  -3} $ & $ 1.42 \times 10^{  -4} $ & $ 1.30 \times 10^{  -2} $ & $ 5.04 \times 10^{  -3} $ & $ 2.32 \times 10^{  -1} $ 
\\
\hline
\end{tabular} }
\caption{$f_{\textrm{sky}}=0.2$, $N_{\textrm{field}}=4$, $\Sigma m_{\nu} = 0.06$~eV, without the residual foregrounds.}
\label{tab:fsk02_Sigma_m_nu006}

\end{table}

\begin{table}[htbp]
\centering \scalebox{0.85}[0.85]{
\begin{tabular}{l|ccccc}
\hline \hline \\
& $\sigma( \Omega_{m}h^{2} )$ & $\sigma( \Omega_{b}h^{2} )$ &  $\sigma( \Omega_{\Lambda} )$ & $ \sigma( n_s ) $ & $\sigma( A_{s} \times 10^{10} )$ 
\\
\hline
Planck 
 & $ 5.37 \times 10^{  -3} $ & $ 2.49 \times 10^{  -4} $ & $ 2.43 \times 10^{  -2} $ & $ 8.82 \times 10^{  -3} $ & $ 2.86 \times 10^{  -1} $ 
\\
+ PB-2 (0.2)
 & $ 4.22 \times 10^{  -3} $ & $ 1.75 \times 10^{  -4} $ & $ 1.69 \times 10^{  -2} $ & $ 6.20 \times 10^{  -3} $ & $ 2.51 \times 10^{  -1} $ 
\\
+ PB-2 + BAO + $H_{0}$
 & $ 3.41 \times 10^{  -3} $ & $ 1.35 \times 10^{  -4} $ & $ 9.18 \times 10^{  -3} $ & $ 4.60 \times 10^{  -3} $ & $ 2.28 \times 10^{  -1} $ 
\\
+ PB-2 + BAO + $H_{0}$ + SKA1
 & $ 2.34 \times 10^{  -3} $ & $ 1.34 \times 10^{  -4} $ & $ 6.85 \times 10^{  -3} $ & $ 4.53 \times 10^{  -3} $ & $ 2.13 \times 10^{  -1} $ 
\\
+ PB-2 + BAO + $H_{0}$ + SKA2
 & $ 1.60 \times 10^{  -3} $ & $ 1.29 \times 10^{  -4} $ & $ 4.47 \times 10^{  -3} $ & $ 3.91 \times 10^{  -3} $ & $ 2.01 \times 10^{  -1} $ 
\\
+ SA (0.2)
 & $ 3.30 \times 10^{  -3} $ & $ 1.43 \times 10^{  -4} $ & $ 1.31 \times 10^{  -2} $ & $ 5.08 \times 10^{  -3} $ & $ 2.29 \times 10^{  -1} $ 
\\
\hline
\end{tabular} }
\caption{$f_{\textrm{sky}}=0.2$, $N_{\textrm{field}}=4$, $\Sigma m_{\nu} = 0.1$~eV, with the residual foregrounds.}
\label{tab:fsk02_fg_Sigma_m_nu010}

\vspace{10pt}

\centering \scalebox{0.85}[0.85]{
\begin{tabular}{l|ccccc}
\hline \hline \\
& $\sigma( \Omega_{m}h^{2} )$ & $\sigma( \Omega_{b}h^{2} )$ &  $\sigma( \Omega_{\Lambda} )$ & $ \sigma( n_s ) $ & $\sigma( A_{s} \times 10^{10} )$ 
\\
\hline
Planck 
 & $ 5.42 \times 10^{  -3} $ & $ 2.50 \times 10^{  -4} $ & $ 2.46 \times 10^{  -2} $ & $ 8.82 \times 10^{  -3} $ & $ 2.86 \times 10^{  -1} $ 
\\
+ PB-2 (0.2)
 & $ 4.26 \times 10^{  -3} $ & $ 1.75 \times 10^{  -4} $ & $ 1.70 \times 10^{  -2} $ & $ 6.19 \times 10^{  -3} $ & $ 2.51 \times 10^{  -1} $ 
\\
+ PB-2 + BAO + $H_{0}$
 & $ 3.45 \times 10^{  -3} $ & $ 1.35 \times 10^{  -4} $ & $ 9.29 \times 10^{  -3} $ & $ 4.60 \times 10^{  -3} $ & $ 2.28 \times 10^{  -1} $ 
\\
+ PB-2 + BAO + $H_{0}$ + SKA1
 & $ 2.35 \times 10^{  -3} $ & $ 1.34 \times 10^{  -4} $ & $ 6.87 \times 10^{  -3} $ & $ 4.53 \times 10^{  -3} $ & $ 2.13 \times 10^{  -1} $ 
\\
+ PB-2 + BAO + $H_{0}$ + SKA2
 & $ 1.59 \times 10^{  -3} $ & $ 1.29 \times 10^{  -4} $ & $ 4.45 \times 10^{  -3} $ & $ 3.89 \times 10^{  -3} $ & $ 2.00 \times 10^{  -1} $ 
\\
+ SA (0.2)
 & $ 3.35 \times 10^{  -3} $ & $ 1.43 \times 10^{  -4} $ & $ 1.31 \times 10^{  -2} $ & $ 5.08 \times 10^{  -3} $ & $ 2.30 \times 10^{  -1} $ 
\\
\hline
\end{tabular} }
\caption{$f_{\textrm{sky}}=0.2$, $N_{\textrm{field}}=4$, $\Sigma m_{\nu} = 0.06$~eV, with the residual foregrounds.}
\label{tab:fsk02_fg_Sigma_m_nu006}

\end{table}

\begin{table}[htbp]
\centering \scalebox{0.85}[0.85]{
\begin{tabular}{l|ccccc}
\hline \hline \\
& $\sigma( \Omega_{m}h^{2} )$ & $\sigma( \Omega_{b}h^{2} )$ &  $\sigma( \Omega_{\Lambda} )$ & $ \sigma( n_s ) $ & $\sigma( A_{s} \times 10^{10} )$ 
\\
\hline
Planck 
 & $ 5.31 \times 10^{  -3} $ & $ 2.44 \times 10^{  -4} $ & $ 2.38 \times 10^{  -2} $ & $ 8.61 \times 10^{  -3} $ & $ 2.83 \times 10^{  -1} $ 
\\
+ PB-2 (0.65)
 & $ 2.98 \times 10^{  -3} $ & $ 1.17 \times 10^{  -4} $ & $ 1.13 \times 10^{  -2} $ & $ 4.21 \times 10^{  -3} $ & $ 2.23 \times 10^{  -1} $ 
\\
+ PB-2 + BAO + $H_{0}$
 & $ 2.62 \times 10^{  -3} $ & $ 9.60 \times 10^{  -5} $ & $ 7.20 \times 10^{  -3} $ & $ 3.44 \times 10^{  -3} $ & $ 2.17 \times 10^{  -1} $ 
\\
+ PB-2 + BAO + $H_{0}$ + SKA1
 & $ 8.99 \times 10^{  -4} $ & $ 9.36 \times 10^{  -5} $ & $ 4.25 \times 10^{  -3} $ & $ 3.25 \times 10^{  -3} $ & $ 1.93 \times 10^{  -1} $ 
\\
+ PB-2 + BAO + $H_{0}$ + SKA2
 & $ 5.60 \times 10^{  -4} $ & $ 8.88 \times 10^{  -5} $ & $ 2.65 \times 10^{  -3} $ & $ 2.36 \times 10^{  -3} $ & $ 1.80 \times 10^{  -1} $ 
\\
+ SA (0.65)
 & $ 2.61 \times 10^{  -3} $ & $ 1.05 \times 10^{  -4} $ & $ 1.01 \times 10^{  -2} $ & $ 3.82 \times 10^{  -3} $ & $ 2.16 \times 10^{  -1} $ 
\\
\hline
\end{tabular} }
\caption{$f_{\textrm{sky}}=0.65$, $N_{\textrm{field}}=4$, $\Sigma m_{\nu} = 0.1$~eV, without the residual foregrounds.}
\label{tab:fsk065_Sigma_m_nu010}

\vspace{10pt}

\centering \scalebox{0.85}[0.85]{
\begin{tabular}{l|ccccc}
\hline \hline \\
& $\sigma( \Omega_{m}h^{2} )$ & $\sigma( \Omega_{b}h^{2} )$ &  $\sigma( \Omega_{\Lambda} )$ & $ \sigma( n_s ) $ & $\sigma( A_{s} \times 10^{10} )$ 
\\
\hline
Planck 
 & $ 5.35 \times 10^{  -3} $ & $ 2.45 \times 10^{  -4} $ & $ 2.41 \times 10^{  -2} $ & $ 8.61 \times 10^{  -3} $ & $ 2.83 \times 10^{  -1} $ 
\\
+ PB-2 (0.65)
 & $ 3.02 \times 10^{  -3} $ & $ 1.17 \times 10^{  -4} $ & $ 1.14 \times 10^{  -2} $ & $ 4.21 \times 10^{  -3} $ & $ 2.23 \times 10^{  -1} $ 
\\
+ PB-2 + BAO + $H_{0}$
 & $ 2.65 \times 10^{  -3} $ & $ 9.60 \times 10^{  -5} $ & $ 7.28 \times 10^{  -3} $ & $ 3.44 \times 10^{  -3} $ & $ 2.17 \times 10^{  -1} $ 
\\
+ PB-2 + BAO + $H_{0}$ + SKA1
 & $ 9.05 \times 10^{  -4} $ & $ 9.36 \times 10^{  -5} $ & $ 4.26 \times 10^{  -3} $ & $ 3.25 \times 10^{  -3} $ & $ 1.92 \times 10^{  -1} $ 
\\
+ PB-2 + BAO + $H_{0}$ + SKA2
 & $ 5.63 \times 10^{  -4} $ & $ 8.87 \times 10^{  -5} $ & $ 2.64 \times 10^{  -3} $ & $ 2.34 \times 10^{  -3} $ & $ 1.80 \times 10^{  -1} $ 
\\
+ SA (0.65)
 & $ 2.65 \times 10^{  -3} $ & $ 1.05 \times 10^{  -4} $ & $ 1.01 \times 10^{  -2} $ & $ 3.82 \times 10^{  -3} $ & $ 2.17 \times 10^{  -1} $ 
\\
\hline
\end{tabular} }
\caption{$f_{\textrm{sky}}=0.65$, $N_{\textrm{field}}=4$, $\Sigma m_{\nu} = 0.06$~eV, without the residual foregrounds.}
\label{tab:fsk065_Sigma_m_nu006}

\end{table}

\begin{table}[htbp]
\centering \scalebox{0.85}[0.85]{
\begin{tabular}{l|ccccc}
\hline \hline \\
& $\sigma( \Omega_{m}h^{2} )$ & $\sigma( \Omega_{b}h^{2} )$ &  $\sigma( \Omega_{\Lambda} )$ & $ \sigma( n_s ) $ & $\sigma( A_{s} \times 10^{10} )$ 
\\
\hline
Planck 
 & $ 5.37 \times 10^{  -3} $ & $ 2.49 \times 10^{  -4} $ & $ 2.43 \times 10^{  -2} $ & $ 8.82 \times 10^{  -3} $ & $ 2.86 \times 10^{  -1} $ 
\\
+ PB-2 (0.65)
 & $ 3.46 \times 10^{  -3} $ & $ 1.36 \times 10^{  -4} $ & $ 1.32 \times 10^{  -2} $ & $ 4.82 \times 10^{  -3} $ & $ 2.34 \times 10^{  -1} $ 
\\
+ PB-2 + BAO + $H_{0}$
 & $ 2.94 \times 10^{  -3} $ & $ 1.09 \times 10^{  -4} $ & $ 7.99 \times 10^{  -3} $ & $ 3.81 \times 10^{  -3} $ & $ 2.23 \times 10^{  -1} $ 
\\
+ PB-2 + BAO + $H_{0}$ + SKA1
 & $ 2.17 \times 10^{  -3} $ & $ 1.09 \times 10^{  -4} $ & $ 6.38 \times 10^{  -3} $ & $ 3.77 \times 10^{  -3} $ & $ 2.11 \times 10^{  -1} $ 
\\
+ PB-2 + BAO + $H_{0}$ + SKA2
 & $ 1.54 \times 10^{  -3} $ & $ 1.06 \times 10^{  -4} $ & $ 4.34 \times 10^{  -3} $ & $ 3.37 \times 10^{  -3} $ & $ 1.99 \times 10^{  -1} $ 
\\
+ SA (0.65)
 & $ 2.65 \times 10^{  -3} $ & $ 1.06 \times 10^{  -4} $ & $ 1.02 \times 10^{  -2} $ & $ 3.86 \times 10^{  -3} $ & $ 2.15 \times 10^{  -1} $ 
\\
\hline
\end{tabular} }
\caption{$f_{\textrm{sky}}=0.65$, $N_{\textrm{field}}=4$, $\Sigma m_{\nu} = 0.1$~eV, with the residual foregrounds.}
\label{tab:fsk065_fg_Sigma_m_nu010}

\vspace{10pt}

\centering \scalebox{0.85}[0.85]{
\begin{tabular}{l|ccccc}
\hline \hline \\
& $\sigma( \Omega_{m}h^{2} )$ & $\sigma( \Omega_{b}h^{2} )$ &  $\sigma( \Omega_{\Lambda} )$ & $ \sigma( n_s ) $ & $\sigma( A_{s} \times 10^{10} )$ 
\\
\hline
Planck 
 & $ 5.42 \times 10^{  -3} $ & $ 2.50 \times 10^{  -4} $ & $ 2.46 \times 10^{  -2} $ & $ 8.82 \times 10^{  -3} $ & $ 2.86 \times 10^{  -1} $ 
\\
+ PB-2 (0.65)
 & $ 3.49 \times 10^{  -3} $ & $ 1.36 \times 10^{  -4} $ & $ 1.33 \times 10^{  -2} $ & $ 4.81 \times 10^{  -3} $ & $ 2.34 \times 10^{  -1} $ 
\\
+ PB-2 + BAO + $H_{0}$
 & $ 2.98 \times 10^{  -3} $ & $ 1.09 \times 10^{  -4} $ & $ 8.09 \times 10^{  -3} $ & $ 3.81 \times 10^{  -3} $ & $ 2.23 \times 10^{  -1} $ 
\\
+ PB-2 + BAO + $H_{0}$ + SKA1
 & $ 2.18 \times 10^{  -3} $ & $ 1.09 \times 10^{  -4} $ & $ 6.41 \times 10^{  -3} $ & $ 3.77 \times 10^{  -3} $ & $ 2.11 \times 10^{  -1} $ 
\\
+ PB-2 + BAO + $H_{0}$ + SKA2
 & $ 1.54 \times 10^{  -3} $ & $ 1.06 \times 10^{  -4} $ & $ 4.32 \times 10^{  -3} $ & $ 3.36 \times 10^{  -3} $ & $ 1.98 \times 10^{  -1} $ 
\\
+ SA (0.65)
 & $ 2.69 \times 10^{  -3} $ & $ 1.06 \times 10^{  -4} $ & $ 1.02 \times 10^{  -2} $ & $ 3.86 \times 10^{  -3} $ & $ 2.15 \times 10^{  -1} $ 
\\
\hline
\end{tabular} }
\caption{$f_{\textrm{sky}}=0.65$, $N_{\textrm{field}}=4$, $\Sigma m_{\nu} = 0.06$~eV, with the residual foregrounds.}
\label{tab:fsk065_fg_Sigma_m_nu006}

\end{table}

\begin{table}[htbp]
\centering \scalebox{0.8}[0.77]{
\begin{tabular}{l|ccccc}
\hline \hline \\
& $\sigma( \Omega_{m}h^{2} )$ & $\sigma( \Omega_{b}h^{2} )$ &  $\sigma( \Omega_{\Lambda} )$ & $ \sigma( n_s ) $ & $\sigma( A_{s} \times 10^{10} )$ 
\\
\hline
Planck + PB-2(0.016) + BAO + $H_{0}$ 
 & $ 3.92 \times 10^{  -3} $ & $ 1.70 \times 10^{  -4} $ & $ 1.05 \times 10^{  -2} $ & $ 5.75 \times 10^{  -3} $ & $ 2.34 \times 10^{  -1} $ 
\\
+ SKA1 ($N_{\textrm{field}}=4$)
 & $ 9.47 \times 10^{  -4} $ & $ 1.55 \times 10^{  -4} $ & $ 4.51 \times 10^{  -3} $ & $ 5.12 \times 10^{  -3} $ & $ 2.09 \times 10^{  -1} $ 
\\
+ SKA1 ($N_{\textrm{field}}=8$)
 & $ 7.23 \times 10^{  -4} $ & $ 1.46 \times 10^{  -4} $ & $ 4.12 \times 10^{  -3} $ & $ 4.66 \times 10^{  -3} $ & $ 2.06 \times 10^{  -1} $ 
\\
+ SKA2 ($N_{\textrm{field}}=4$)
 & $ 5.79 \times 10^{  -4} $ & $ 1.37 \times 10^{  -4} $ & $ 2.72 \times 10^{  -3} $ & $ 2.99 \times 10^{  -3} $ & $ 1.96 \times 10^{  -1} $ 
\\
+ SKA2 ($N_{\textrm{field}}=8$)
 & $ 4.20 \times 10^{  -4} $ & $ 1.26 \times 10^{  -4} $ & $ 2.12 \times 10^{  -3} $ & $ 2.30 \times 10^{  -3} $ & $ 1.91 \times 10^{  -1} $ 
\\
Planck + SA(0.016) + BAO + $H_{0}$ 
 & $ 3.57 \times 10^{  -3} $ & $ 1.54 \times 10^{  -4} $ & $ 9.61 \times 10^{  -3} $ & $ 5.23 \times 10^{  -3} $ & $ 2.29 \times 10^{  -1} $ 
\\
\hline
\end{tabular} }
\caption{$f_{\textrm{sky}}=0.016$, $\Sigma m_{\nu} = 0.1$~eV, with the residual foregrounds of CMB, without those of 21 cm line.}
\label{tab:fsk0016_fg_Sigma_m_nu010_SKAnoFG}

\vspace{10pt}

\centering \scalebox{0.8}[0.77]{
\begin{tabular}{l|ccccc}
\hline \hline \\
& $\sigma( \Omega_{m}h^{2} )$ & $\sigma( \Omega_{b}h^{2} )$ &  $\sigma( \Omega_{\Lambda} )$ & $ \sigma( n_s ) $ & $\sigma( A_{s} \times 10^{10} )$ 
\\
\hline
Planck + PB-2(0.016) + BAO + $H_{0}$ 
 & $ 3.97 \times 10^{  -3} $ & $ 1.70 \times 10^{  -4} $ & $ 1.06 \times 10^{  -2} $ & $ 5.75 \times 10^{  -3} $ & $ 2.34 \times 10^{  -1} $ 
\\
+ SKA1 ($N_{\textrm{field}}=4$)
 & $ 9.54 \times 10^{  -4} $ & $ 1.55 \times 10^{  -4} $ & $ 4.52 \times 10^{  -3} $ & $ 5.11 \times 10^{  -3} $ & $ 2.09 \times 10^{  -1} $ 
\\
+ SKA1 ($N_{\textrm{field}}=8$)
 & $ 7.29 \times 10^{  -4} $ & $ 1.46 \times 10^{  -4} $ & $ 4.13 \times 10^{  -3} $ & $ 4.65 \times 10^{  -3} $ & $ 2.06 \times 10^{  -1} $ 
\\
+ SKA2 ($N_{\textrm{field}}=4$)
 & $ 5.82 \times 10^{  -4} $ & $ 1.36 \times 10^{  -4} $ & $ 2.71 \times 10^{  -3} $ & $ 2.95 \times 10^{  -3} $ & $ 1.95 \times 10^{  -1} $ 
\\
+ SKA2 ($N_{\textrm{field}}=8$)
 & $ 4.21 \times 10^{  -4} $ & $ 1.26 \times 10^{  -4} $ & $ 2.10 \times 10^{  -3} $ & $ 2.27 \times 10^{  -3} $ & $ 1.91 \times 10^{  -1} $ 
\\
Planck + SA(0.016) + BAO + $H_{0}$ 
 & $ 3.61 \times 10^{  -3} $ & $ 1.54 \times 10^{  -4} $ & $ 9.71 \times 10^{  -3} $ & $ 5.23 \times 10^{  -3} $ & $ 2.29 \times 10^{  -1} $ 
\\
\hline
\end{tabular} }
\caption{$f_{\textrm{sky}}=0.016$, $\Sigma m_{\nu} = 0.06$~eV, with the residual foregrounds of CMB, without those of 21 cm line.}
\label{tab:fsk0016_fg_Sigma_m_nu006_SKAnoFG}

\end{table}

\begin{table}[htbp]
\centering \scalebox{0.8}[0.77]{
\begin{tabular}{l|ccccc}
\hline \hline \\
& $\sigma( \Omega_{m}h^{2} )$ & $\sigma( \Omega_{b}h^{2} )$ &  $\sigma( \Omega_{\Lambda} )$ & $ \sigma( n_s ) $ & $\sigma( A_{s} \times 10^{10} )$ 
\\
\hline
Planck + PB-2(0.2) + BAO + $H_{0}$ 
 & $ 3.41 \times 10^{  -3} $ & $ 1.35 \times 10^{  -4} $ & $ 9.18 \times 10^{  -3} $ & $ 4.60 \times 10^{  -3} $ & $ 2.28 \times 10^{  -1} $ 
\\
+ SKA1 ($N_{\textrm{field}}=4$)
 & $ 9.32 \times 10^{  -4} $ & $ 1.28 \times 10^{  -4} $ & $ 4.43 \times 10^{  -3} $ & $ 4.24 \times 10^{  -3} $ & $ 2.03 \times 10^{  -1} $ 
\\
+ SKA1 ($N_{\textrm{field}}=8$)
 & $ 7.11 \times 10^{  -4} $ & $ 1.23 \times 10^{  -4} $ & $ 4.05 \times 10^{  -3} $ & $ 3.95 \times 10^{  -3} $ & $ 2.01 \times 10^{  -1} $ 
\\
+ SKA2 ($N_{\textrm{field}}=4$)
 & $ 5.72 \times 10^{  -4} $ & $ 1.18 \times 10^{  -4} $ & $ 2.70 \times 10^{  -3} $ & $ 2.74 \times 10^{  -3} $ & $ 1.91 \times 10^{  -1} $ 
\\
+ SKA2 ($N_{\textrm{field}}=8$)
 & $ 4.14 \times 10^{  -4} $ & $ 1.10 \times 10^{  -4} $ & $ 2.11 \times 10^{  -3} $ & $ 2.17 \times 10^{  -3} $ & $ 1.85 \times 10^{  -1} $ 
\\
Planck + SA(0.2) + BAO + $H_{0}$ 
 & $ 2.86 \times 10^{  -3} $ & $ 1.14 \times 10^{  -4} $ & $ 7.82 \times 10^{  -3} $ & $ 3.96 \times 10^{  -3} $ & $ 2.19 \times 10^{  -1} $ 
\\
\hline
\end{tabular} }
\caption{$f_{\textrm{sky}}=0.2$, $\Sigma m_{\nu} = 0.1$~eV, with the residual foregrounds of CMB, without those of 21 cm line.}
\label{tab:fsk02_fg_Sigma_m_nu010_SKAnoFG}

\vspace{10pt}

\centering \scalebox{0.8}[0.77]{
\begin{tabular}{l|ccccc}
\hline \hline \\
& $\sigma( \Omega_{m}h^{2} )$ & $\sigma( \Omega_{b}h^{2} )$ &  $\sigma( \Omega_{\Lambda} )$ & $ \sigma( n_s ) $ & $\sigma( A_{s} \times 10^{10} )$ 
\\
\hline
Planck + PB-2(0.2) + BAO + $H_{0}$ 
 & $ 3.45 \times 10^{  -3} $ & $ 1.35 \times 10^{  -4} $ & $ 9.29 \times 10^{  -3} $ & $ 4.60 \times 10^{  -3} $ & $ 2.28 \times 10^{  -1} $ 
\\
+ SKA1 ($N_{\textrm{field}}=4$)
 & $ 9.38 \times 10^{  -4} $ & $ 1.28 \times 10^{  -4} $ & $ 4.44 \times 10^{  -3} $ & $ 4.23 \times 10^{  -3} $ & $ 2.03 \times 10^{  -1} $ 
\\
+ SKA1 ($N_{\textrm{field}}=8$)
 & $ 7.16 \times 10^{  -4} $ & $ 1.23 \times 10^{  -4} $ & $ 4.06 \times 10^{  -3} $ & $ 3.94 \times 10^{  -3} $ & $ 2.00 \times 10^{  -1} $ 
\\
+ SKA2 ($N_{\textrm{field}}=4$)
 & $ 5.74 \times 10^{  -4} $ & $ 1.18 \times 10^{  -4} $ & $ 2.69 \times 10^{  -3} $ & $ 2.71 \times 10^{  -3} $ & $ 1.90 \times 10^{  -1} $ 
\\
+ SKA2 ($N_{\textrm{field}}=8$)
 & $ 4.15 \times 10^{  -4} $ & $ 1.10 \times 10^{  -4} $ & $ 2.09 \times 10^{  -3} $ & $ 2.14 \times 10^{  -3} $ & $ 1.84 \times 10^{  -1} $ 
\\
Planck + SA(0.2) + BAO + $H_{0}$ 
 & $ 2.90 \times 10^{  -3} $ & $ 1.14 \times 10^{  -4} $ & $ 7.91 \times 10^{  -3} $ & $ 3.95 \times 10^{  -3} $ & $ 2.19 \times 10^{  -1} $ 
\\
\hline
\end{tabular} }
\caption{$f_{\textrm{sky}}=0.2$, $\Sigma m_{\nu} = 0.06$~eV, with the residual foregrounds of CMB, without those of 21 cm line.}
\label{tab:fsk02_fg_Sigma_m_nu006_SKAnoFG}

\end{table}

\begin{table}[htbp]
\centering \scalebox{0.8}[0.77]{
\begin{tabular}{l|ccccc}
\hline \hline \\
& $\sigma( \Omega_{m}h^{2} )$ & $\sigma( \Omega_{b}h^{2} )$ &  $\sigma( \Omega_{\Lambda} )$ & $ \sigma( n_s ) $ & $\sigma( A_{s} \times 10^{10} )$ 
\\
\hline
Planck + PB-2(0.65) + BAO + $H_{0}$ 
 & $ 2.94 \times 10^{  -3} $ & $ 1.09 \times 10^{  -4} $ & $ 7.99 \times 10^{  -3} $ & $ 3.81 \times 10^{  -3} $ & $ 2.23 \times 10^{  -1} $ 
\\
+ SKA1 ($N_{\textrm{field}}=4$)
 & $ 9.14 \times 10^{  -4} $ & $ 1.06 \times 10^{  -4} $ & $ 4.34 \times 10^{  -3} $ & $ 3.58 \times 10^{  -3} $ & $ 2.00 \times 10^{  -1} $ 
\\
+ SKA1 ($N_{\textrm{field}}=8$)
 & $ 7.00 \times 10^{  -4} $ & $ 1.03 \times 10^{  -4} $ & $ 3.98 \times 10^{  -3} $ & $ 3.40 \times 10^{  -3} $ & $ 1.98 \times 10^{  -1} $ 
\\
+ SKA2 ($N_{\textrm{field}}=4$)
 & $ 5.65 \times 10^{  -4} $ & $ 9.88 \times 10^{  -5} $ & $ 2.68 \times 10^{  -3} $ & $ 2.50 \times 10^{  -3} $ & $ 1.88 \times 10^{  -1} $ 
\\
+ SKA2 ($N_{\textrm{field}}=8$)
 & $ 4.09 \times 10^{  -4} $ & $ 9.38 \times 10^{  -5} $ & $ 2.10 \times 10^{  -3} $ & $ 2.03 \times 10^{  -3} $ & $ 1.81 \times 10^{  -1} $ 
\\
Planck + SA(0.65) + BAO + $H_{0}$ 
 & $ 2.39 \times 10^{  -3} $ & $ 8.90 \times 10^{  -5} $ & $ 6.66 \times 10^{  -3} $ & $ 3.22 \times 10^{  -3} $ & $ 2.11 \times 10^{  -1} $ 
\\
\hline
\end{tabular} }
\caption{$f_{\textrm{sky}}=0.65$, $\Sigma m_{\nu} = 0.1$~eV, with the residual foregrounds of CMB, without those of 21 cm line.}
\label{tab:fsk065_fg_Sigma_m_nu010_SKAnoFG}

\vspace{10pt}

\centering \scalebox{0.8}[0.77]{
\begin{tabular}{l|ccccc}
\hline \hline \\
& $\sigma( \Omega_{m}h^{2} )$ & $\sigma( \Omega_{b}h^{2} )$ &  $\sigma( \Omega_{\Lambda} )$ & $ \sigma( n_s ) $ & $\sigma( A_{s} \times 10^{10} )$ 
\\
\hline
Planck + PB-2(0.65) + BAO + $H_{0}$ 
 & $ 2.98 \times 10^{  -3} $ & $ 1.09 \times 10^{  -4} $ & $ 8.09 \times 10^{  -3} $ & $ 3.81 \times 10^{  -3} $ & $ 2.23 \times 10^{  -1} $ 
\\
+ SKA1 ($N_{\textrm{field}}=4$)
 & $ 9.21 \times 10^{  -4} $ & $ 1.06 \times 10^{  -4} $ & $ 4.35 \times 10^{  -3} $ & $ 3.58 \times 10^{  -3} $ & $ 2.00 \times 10^{  -1} $ 
\\
+ SKA1 ($N_{\textrm{field}}=8$)
 & $ 7.05 \times 10^{  -4} $ & $ 1.03 \times 10^{  -4} $ & $ 3.99 \times 10^{  -3} $ & $ 3.39 \times 10^{  -3} $ & $ 1.97 \times 10^{  -1} $ 
\\
+ SKA2 ($N_{\textrm{field}}=4$)
 & $ 5.68 \times 10^{  -4} $ & $ 9.87 \times 10^{  -5} $ & $ 2.66 \times 10^{  -3} $ & $ 2.49 \times 10^{  -3} $ & $ 1.87 \times 10^{  -1} $ 
\\
+ SKA2 ($N_{\textrm{field}}=8$)
 & $ 4.11 \times 10^{  -4} $ & $ 9.37 \times 10^{  -5} $ & $ 2.08 \times 10^{  -3} $ & $ 2.01 \times 10^{  -3} $ & $ 1.80 \times 10^{  -1} $ 
\\
Planck + SA(0.65) + BAO + $H_{0}$ 
 & $ 2.42 \times 10^{  -3} $ & $ 8.89 \times 10^{  -5} $ & $ 6.73 \times 10^{  -3} $ & $ 3.22 \times 10^{  -3} $ & $ 2.11 \times 10^{  -1} $ 
\\
\hline
\end{tabular} }
\caption{$f_{\textrm{sky}}=0.65$, $\Sigma m_{\nu} = 0.06$~eV, with the residual foregrounds of CMB, without those of 21 cm line.}
\label{tab:fsk065_fg_Sigma_m_nu006_SKAnoFG}

\end{table}

\begin{table}[htbp]
\centering \scalebox{0.85}[0.85]{
\begin{tabular}{l|ccccc}
\hline \hline \\
& $\sigma( \Omega_{m}h^{2} )$ & $\sigma( \Omega_{b}h^{2} )$ &  $\sigma( \Omega_{\Lambda} )$ & $ \sigma( n_s ) $ & $\sigma( A_{s} \times 10^{10} )$ 
\\
\hline
Planck 
 & $ 5.37 \times 10^{  -3} $ & $ 2.49 \times 10^{  -4} $ & $ 2.43 \times 10^{  -2} $ & $ 8.82 \times 10^{  -3} $ & $ 2.86 \times 10^{  -1} $ 
\\
+ PB-2 (0.2)
 & $ 4.22 \times 10^{  -3} $ & $ 1.75 \times 10^{  -4} $ & $ 1.69 \times 10^{  -2} $ & $ 6.20 \times 10^{  -3} $ & $ 2.51 \times 10^{  -1} $ 
\\
+ PB-2 + BAO + $H_{0}$
 & $ 3.41 \times 10^{  -3} $ & $ 1.35 \times 10^{  -4} $ & $ 9.18 \times 10^{  -3} $ & $ 4.60 \times 10^{  -3} $ & $ 2.28 \times 10^{  -1} $ 
\\
+ SA (1) (0.2)
 & $ 4.15 \times 10^{  -3} $ & $ 1.72 \times 10^{  -4} $ & $ 1.65 \times 10^{  -2} $ & $ 6.08 \times 10^{  -3} $ & $ 2.49 \times 10^{  -1} $ 
\\
+ SA (2) (0.2)
 & $ 3.30 \times 10^{  -3} $ & $ 1.43 \times 10^{  -4} $ & $ 1.31 \times 10^{  -2} $ & $ 5.08 \times 10^{  -3} $ & $ 2.29 \times 10^{  -1} $ 
\\
+ SA (1) + BAO + $H_{0}$
 & $ 3.37 \times 10^{  -3} $ & $ 1.33 \times 10^{  -4} $ & $ 9.07 \times 10^{  -3} $ & $ 4.53 \times 10^{  -3} $ & $ 2.27 \times 10^{  -1} $ 
\\
\hline
\end{tabular} }
\caption{$f_{\textrm{sky}}=0.2$, $\Sigma m_{\nu} = 0.1$~eV, with the residual foregrounds. SA (1) and SA (2) mean that the 220GHz band is used for CMB observation or foreground removal, respectively (see Fig.\ref{fig:Nnu01_fsky02_Simons_fg}).}
\label{tab:fsk02_fg_Sigma_m_nu010_Simons}

\vspace{10pt}

\centering \scalebox{0.85}[0.85]{
\begin{tabular}{l|ccccc}
\hline \hline \\
& $\sigma( \Omega_{m}h^{2} )$ & $\sigma( \Omega_{b}h^{2} )$ &  $\sigma( \Omega_{\Lambda} )$ & $ \sigma( n_s ) $ & $\sigma( A_{s} \times 10^{10} )$ 
\\
\hline
Planck 
 & $ 5.42 \times 10^{  -3} $ & $ 2.50 \times 10^{  -4} $ & $ 2.46 \times 10^{  -2} $ & $ 8.82 \times 10^{  -3} $ & $ 2.86 \times 10^{  -1} $ 
\\
+ PB-2 (0.2)
 & $ 4.26 \times 10^{  -3} $ & $ 1.75 \times 10^{  -4} $ & $ 1.70 \times 10^{  -2} $ & $ 6.19 \times 10^{  -3} $ & $ 2.51 \times 10^{  -1} $ 
\\
+ PB-2 + BAO + $H_{0}$
 & $ 3.45 \times 10^{  -3} $ & $ 1.35 \times 10^{  -4} $ & $ 9.29 \times 10^{  -3} $ & $ 4.60 \times 10^{  -3} $ & $ 2.28 \times 10^{  -1} $ 
\\
+ SA (1) (0.2)
 & $ 4.19 \times 10^{  -3} $ & $ 1.72 \times 10^{  -4} $ & $ 1.66 \times 10^{  -2} $ & $ 6.07 \times 10^{  -3} $ & $ 2.49 \times 10^{  -1} $ 
\\
+ SA (2) (0.2)
 & $ 3.35 \times 10^{  -3} $ & $ 1.43 \times 10^{  -4} $ & $ 1.31 \times 10^{  -2} $ & $ 5.08 \times 10^{  -3} $ & $ 2.30 \times 10^{  -1} $ 
\\
+ SA (1) + BAO + $H_{0}$
 & $ 3.41 \times 10^{  -3} $ & $ 1.33 \times 10^{  -4} $ & $ 9.18 \times 10^{  -3} $ & $ 4.53 \times 10^{  -3} $ & $ 2.27 \times 10^{  -1} $ 
\\
\hline
\end{tabular} }
\caption{$f_{\textrm{sky}}=0.2$, $\Sigma m_{\nu} = 0.06$~eV, with the residual foregrounds.}
\label{tab:fsk02_fg_Sigma_m_nu006_Simons}

\end{table}

%%%%%%%%%%%%%%%%%%%%%%%%%%%%%%%%%%%%%%%%%%%%%%%%%%%%%%%%%%%%%%%%%%%%%
\subsection{Constraints on neutrino mass hierarchy}
%\section{Constraints on neutrino mass hierarchy}
\label{subsec:const_hie}
%%%%%%%%%%%%%%%%%%%%%%%%%%%%%%%%%%%%%%%%%%%%%%%%%%%%%%%%%%%%%%%%%%%%%

\begin{figure*}[tbp]
 \begin{center}
   \resizebox{150mm}{!}{
   \includegraphics[bb= 93 236 523 553,width=0.95\linewidth]{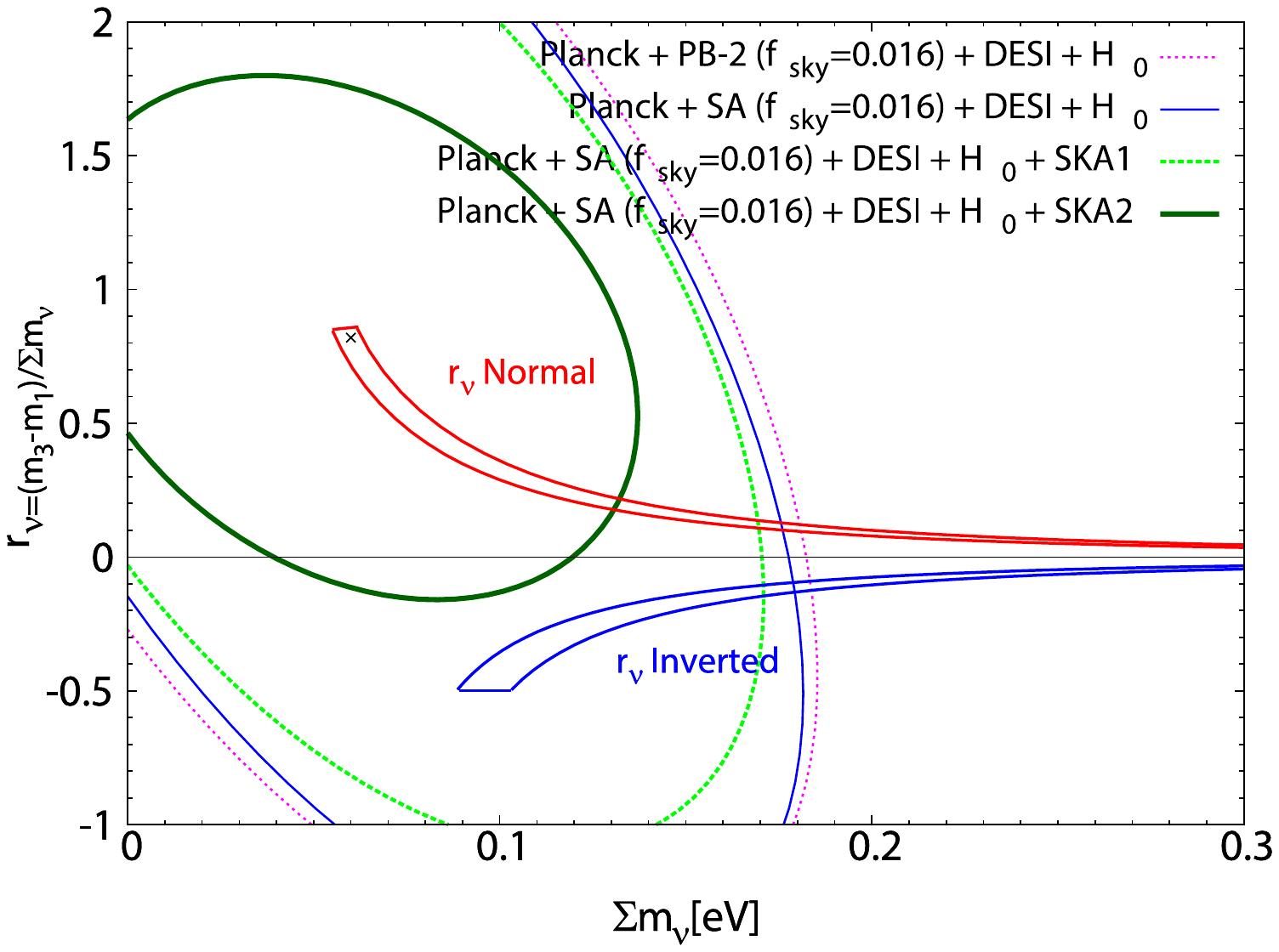} 
   \includegraphics[bb= 93 236 523 553,width=0.95\linewidth]{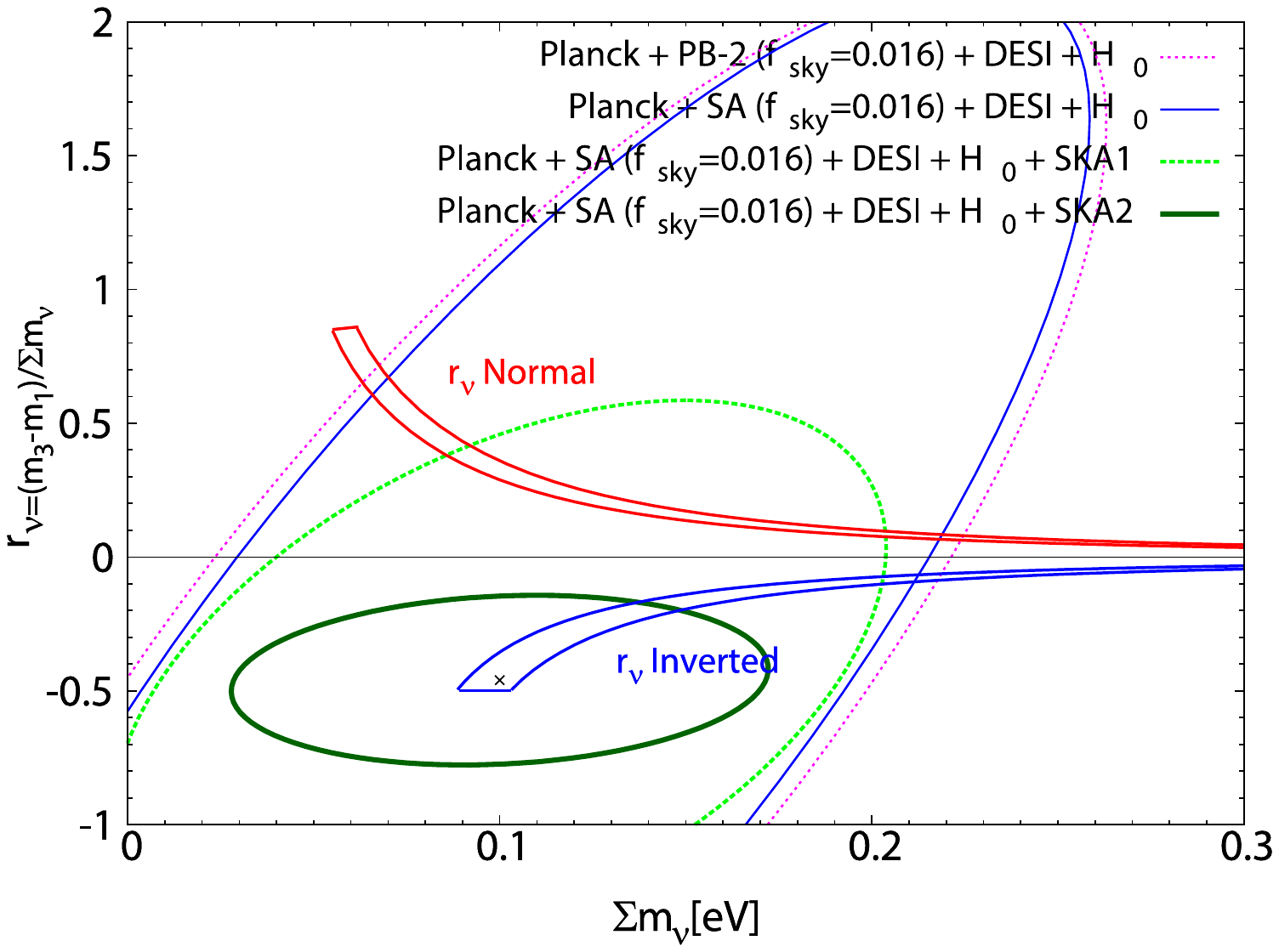} 
   }
   \resizebox{150mm}{!}{
   \includegraphics[bb= 93 236 523 553,width=0.95\linewidth]{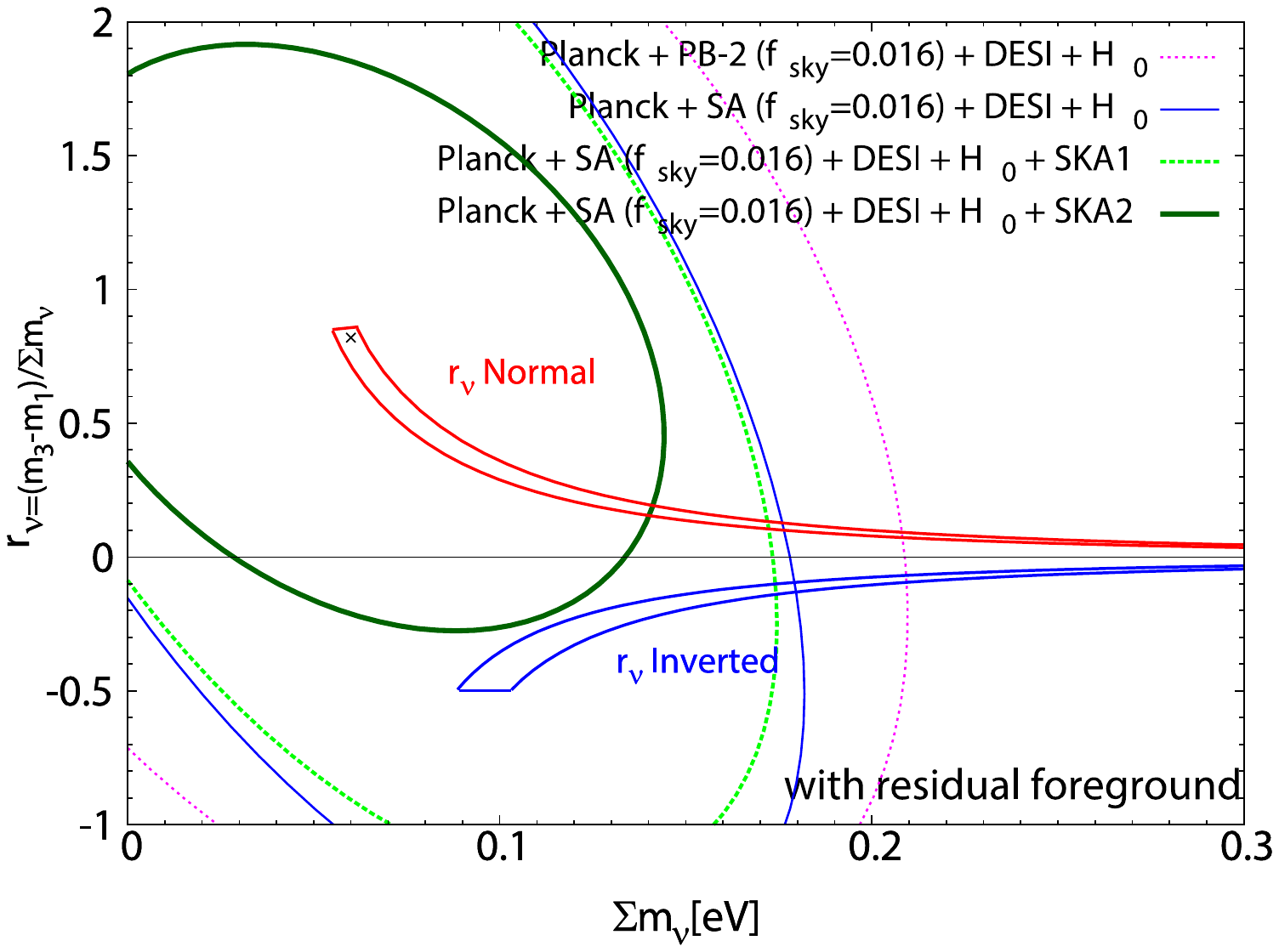} 
   \includegraphics[bb= 93 236 523 553,width=0.95\linewidth]{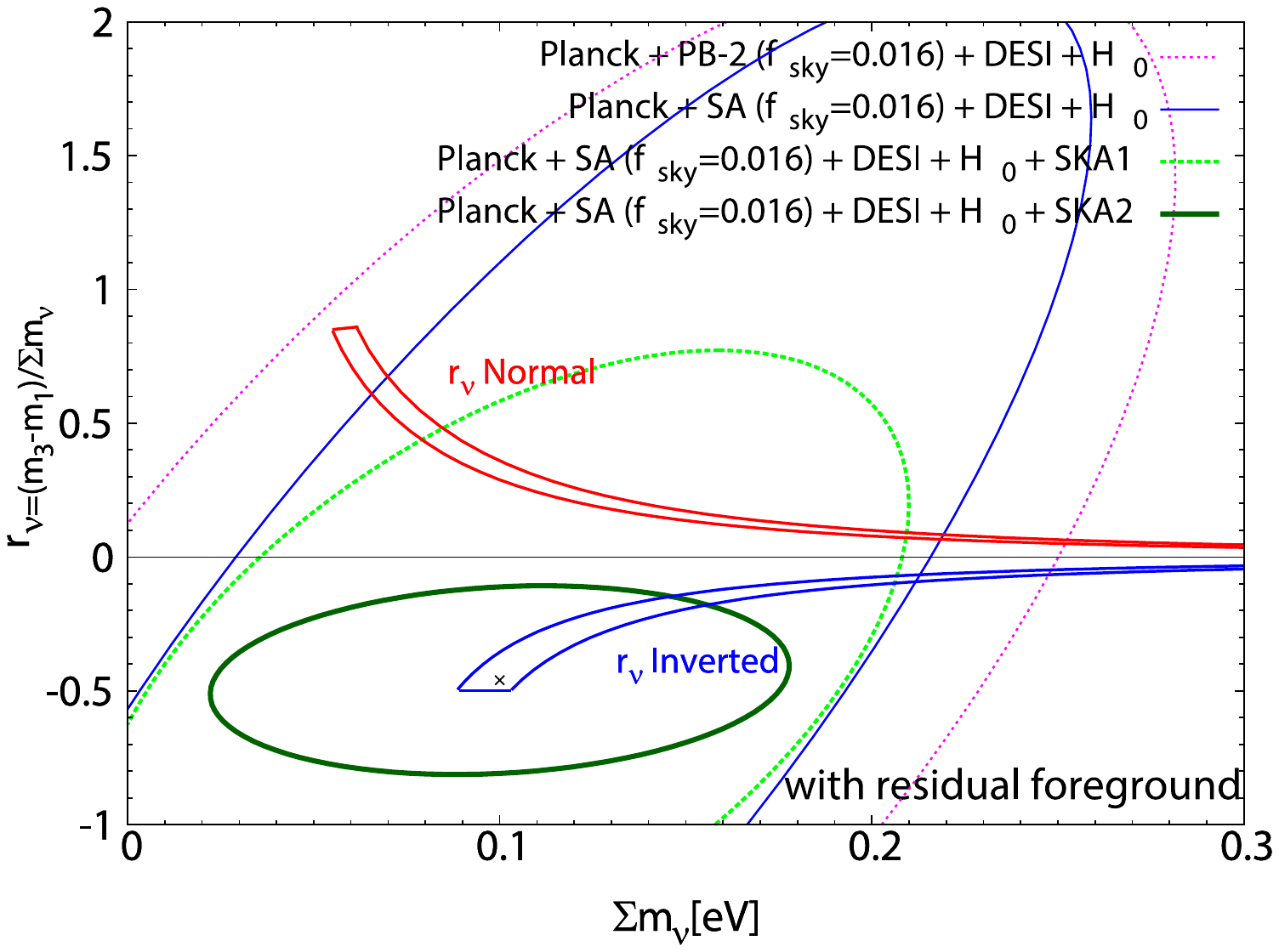} 
   }
   \resizebox{150mm}{!}{
   \includegraphics[bb= 93 236 523 553,width=0.95\linewidth]{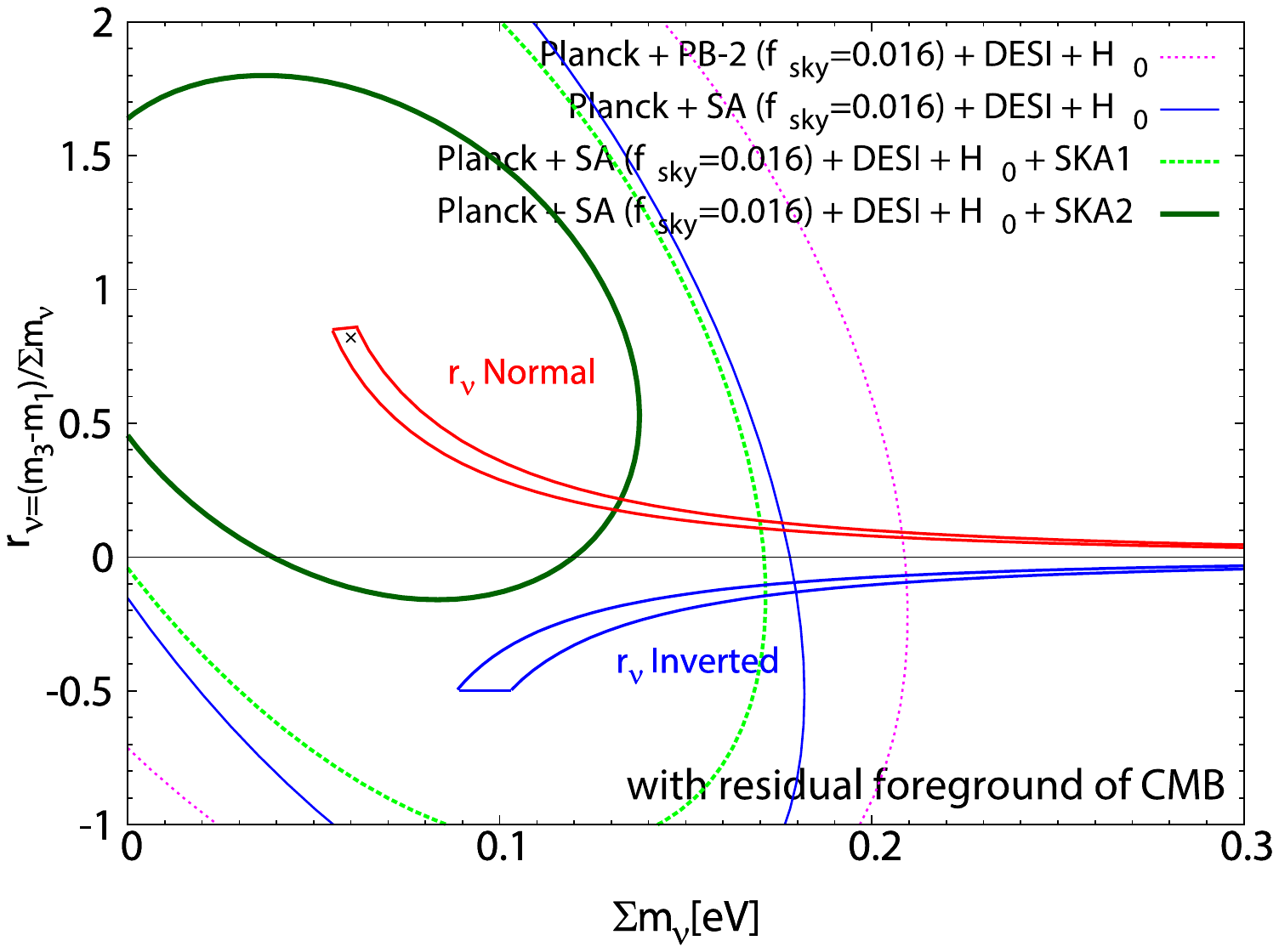} 
   \includegraphics[bb= 93 236 523 553,width=0.95\linewidth]{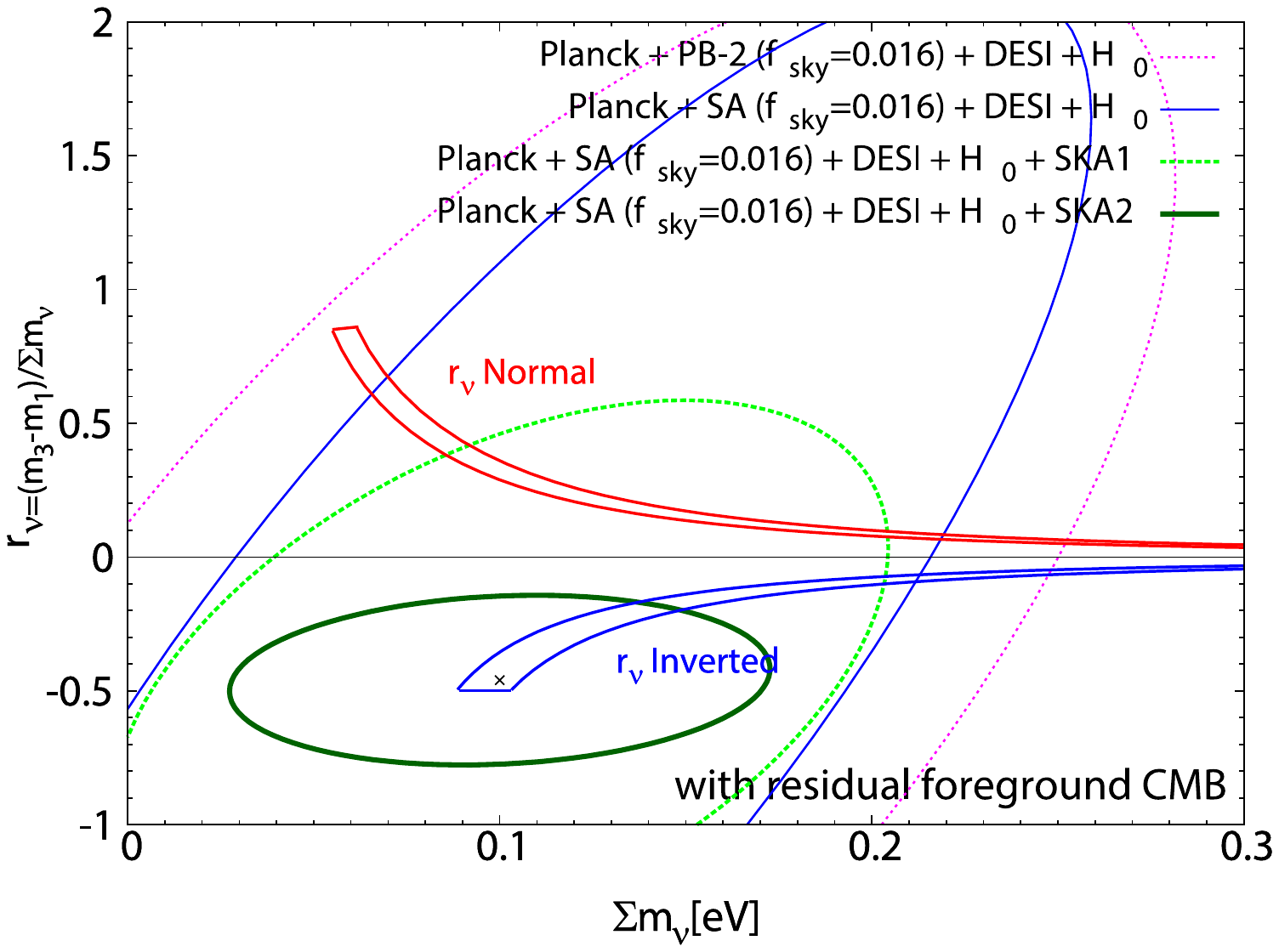} 
   }
   \caption{
   Contours of 95\% C.L. forecasts in $\Sigma m_{\nu}$-$r_{\nu}$ plane. 
   We plot the cases of 
   Planck + BAO(DESI) + Hubble prior combined with 
   \textsc{Polarbear}-2 (PB-2) ($f_{{\rm sky}}=0.016$) (dotted line) or 
   Simons Array (SA) ($f_{{\rm sky}}=0.016$) (outer thin solid line),
   Planck + BAO(DESI) + Hubble prior + Simons Array 
   combined with SKA phase~1 ($N_{{\rm field}} = 4$) (inner thick dashed line) or 
   phase~2 ($N_{{\rm field}} = 4$) (inner thick solid line), respectively.
   Allowed parameters on $r_{\nu}$ by neutrino oscillation experiments 
   are also plotted as two bands for the inverted and the normal hierarchies,
   respectively (the name of each hierarchy is written in the close vicinity of the line).
%   The solid line inside the band is the fiducial value of $r_{\nu}$ as 
%   a function of $\Sigma m_{\nu}$.
%and the solid line inside the band is the central value of $r_{\nu}$.
The lowest two panels are forecasts when the foreground of 21 cm line is completely removed
(i.e. only the residual foreground of CMB exists).
%as 
%   a function of $\Sigma m_{\nu}$
}
   \label{fig:hie_ellipse_fsky0016}
 \end{center}
\end{figure*}

\begin{figure*}[tbp]
 \begin{center}
   \resizebox{150mm}{!}{
   \includegraphics[bb= 93 236 523 553,width=0.95\linewidth]{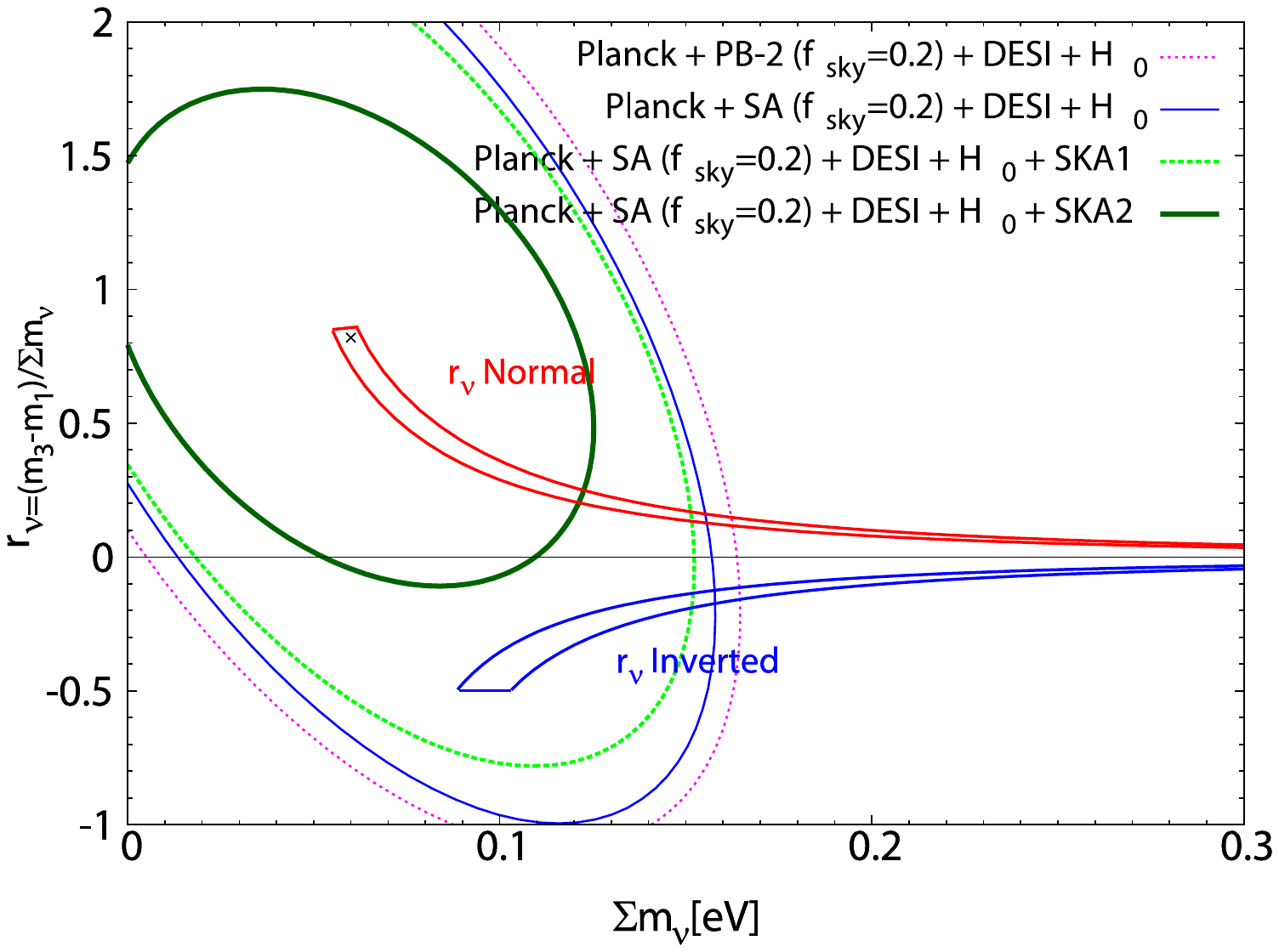} 
   \includegraphics[bb= 93 236 523 553,width=0.95\linewidth]{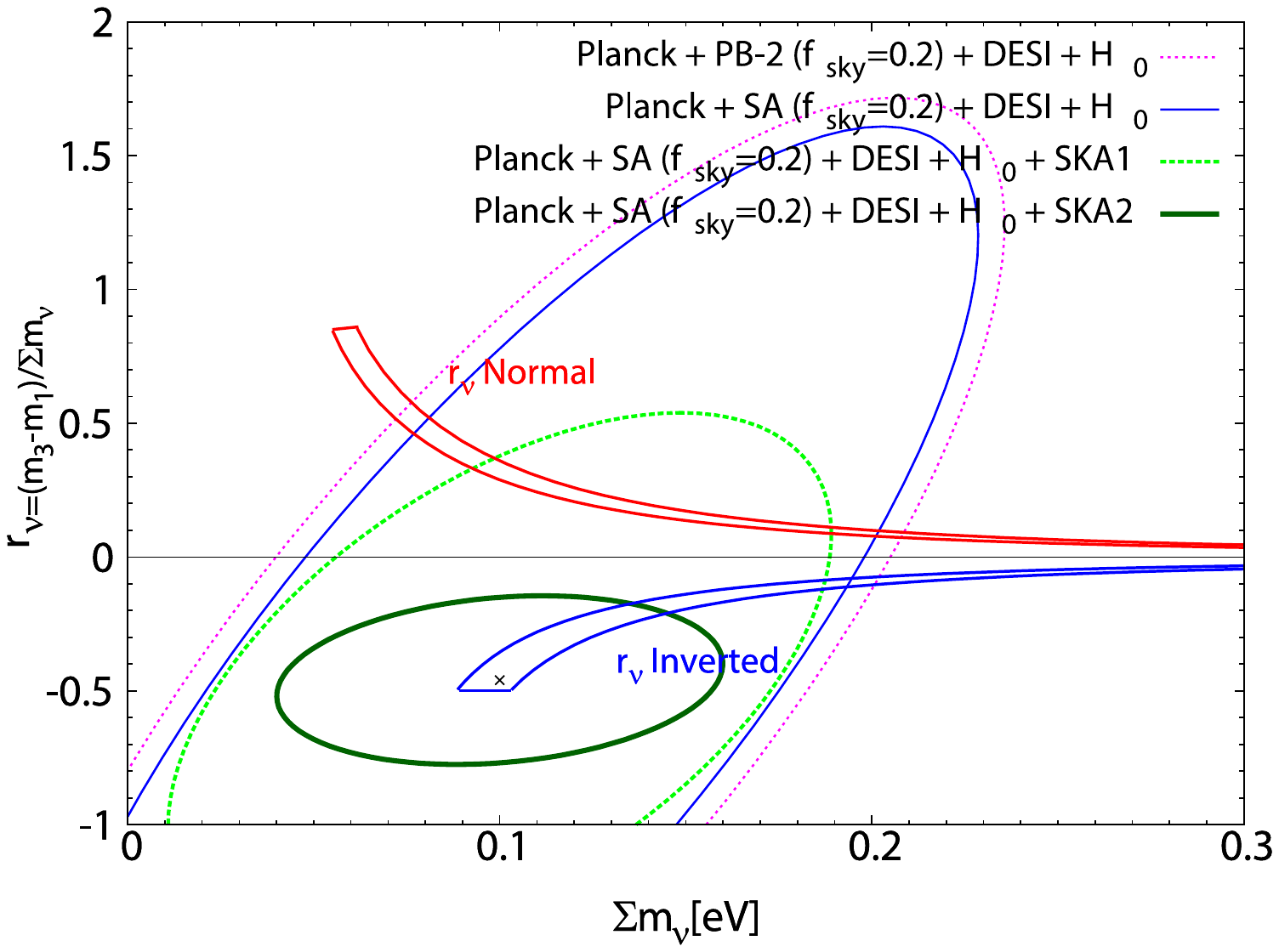} 
   }
   \resizebox{150mm}{!}{
   \includegraphics[bb= 93 236 523 553,width=0.95\linewidth]{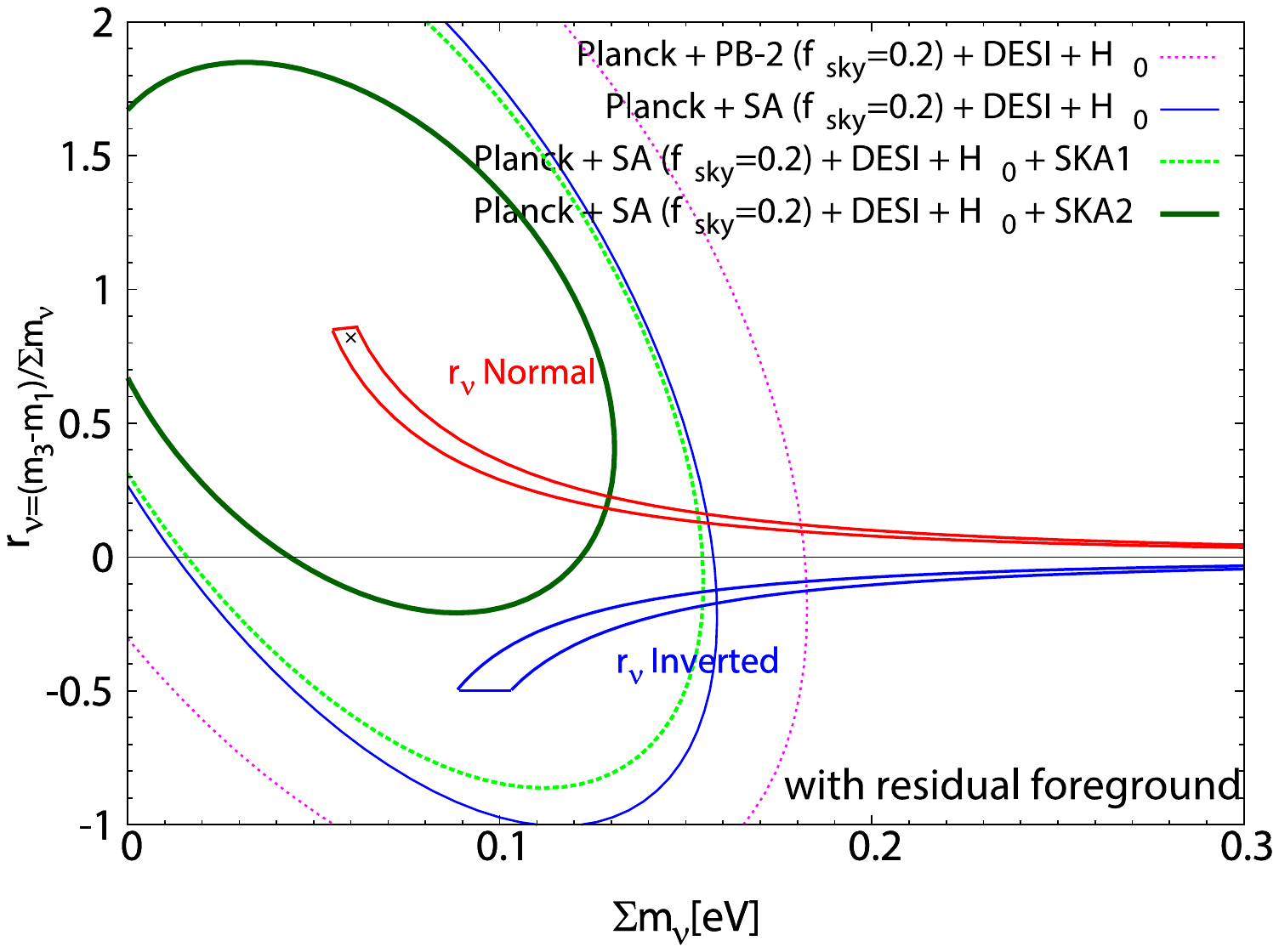} 
   \includegraphics[bb= 93 236 523 553,width=0.95\linewidth]{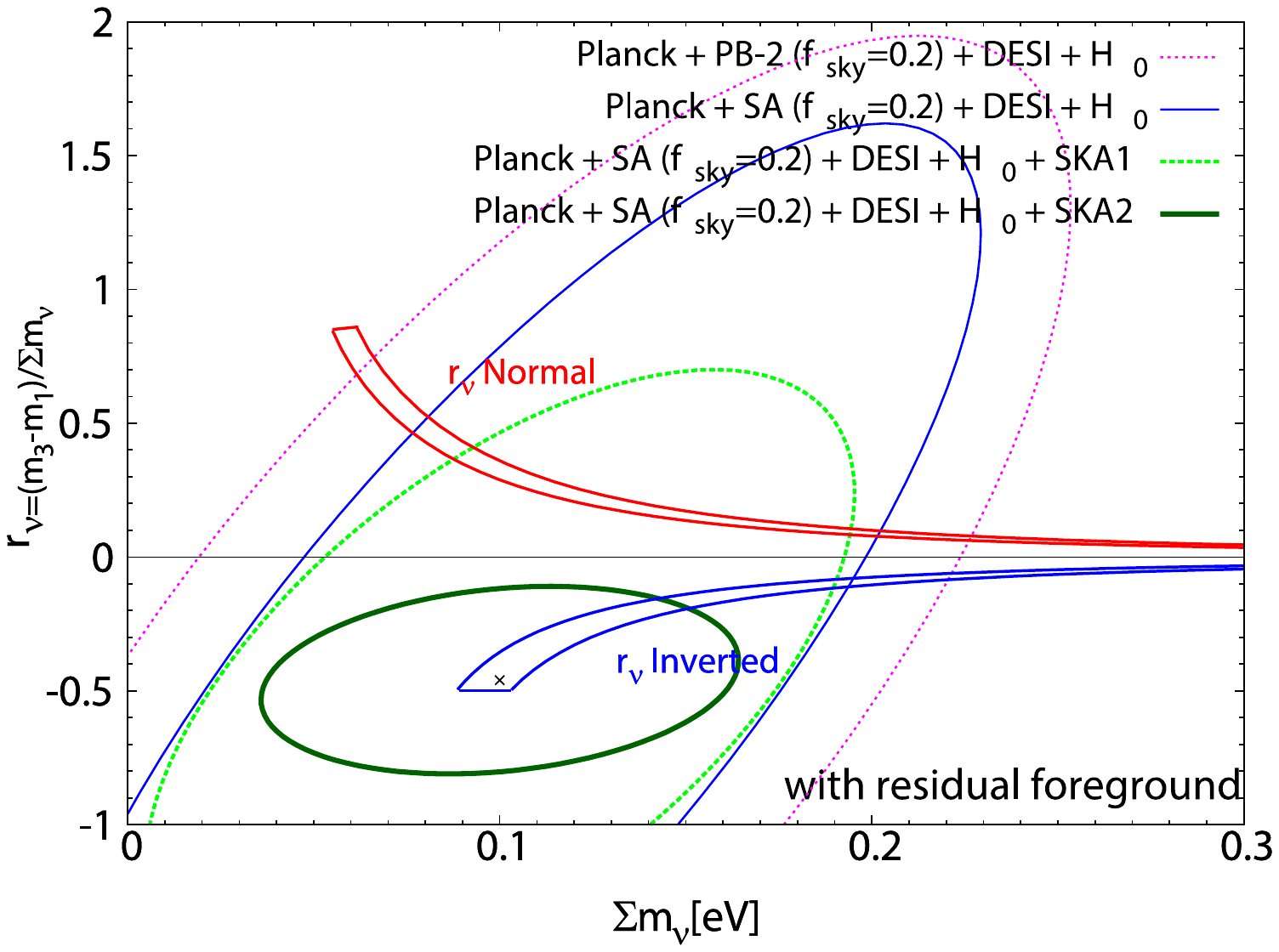} 
   }
   \resizebox{150mm}{!}{
   \includegraphics[bb= 93 236 523 553,width=0.95\linewidth]{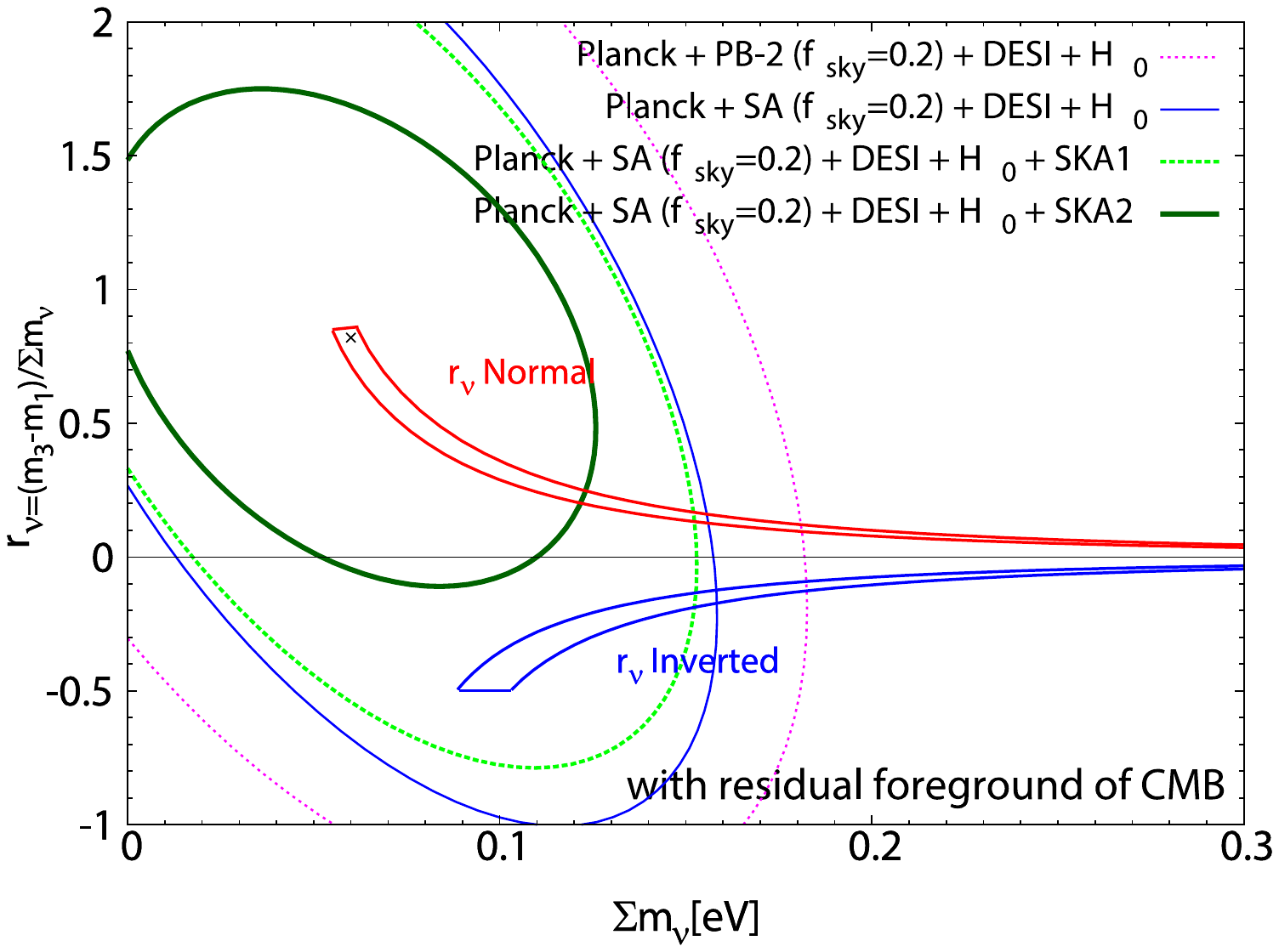} 
   \includegraphics[bb= 93 236 523 553,width=0.95\linewidth]{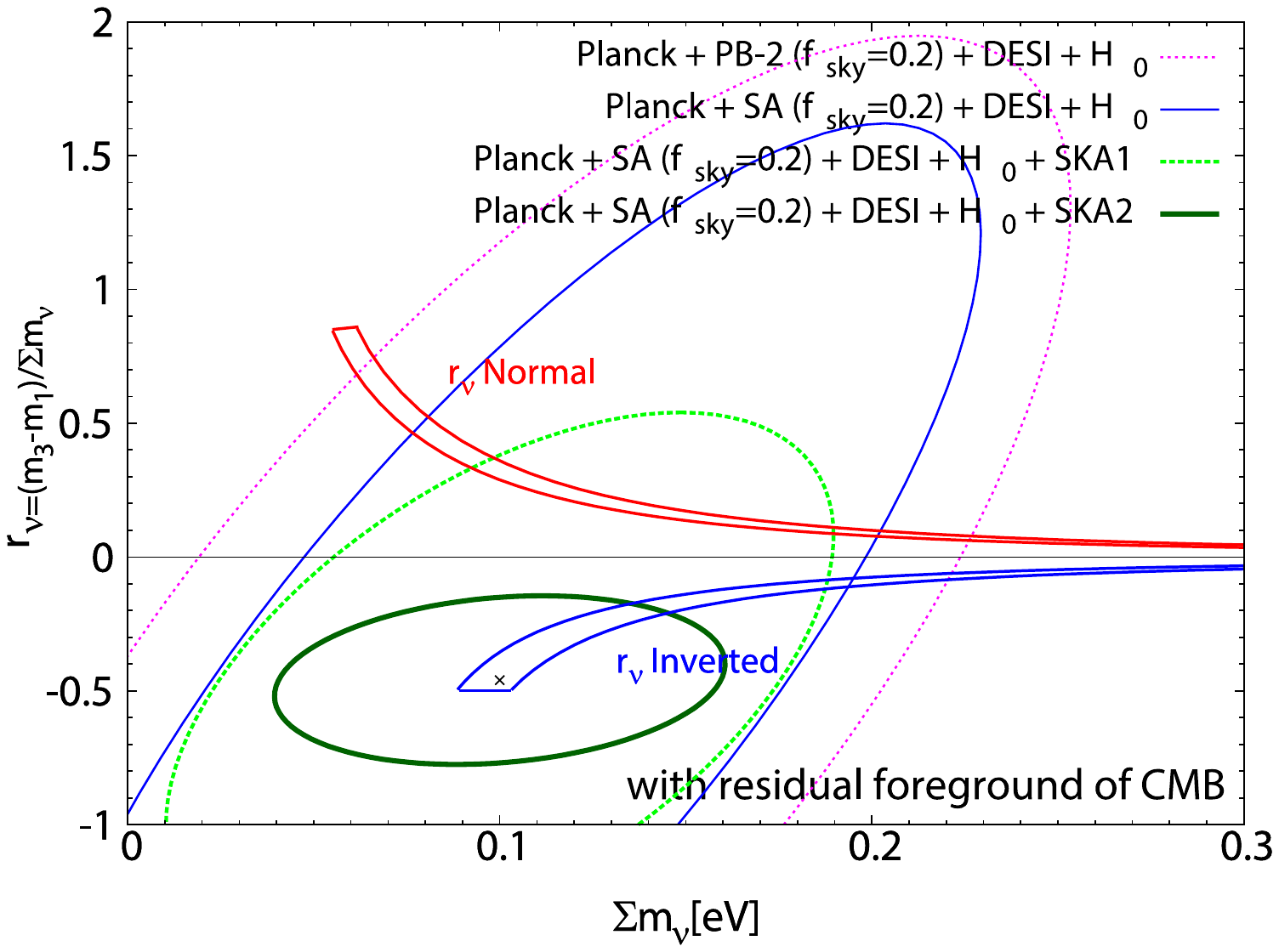} 
   }
   \caption{
   Same as Fig.\ref{fig:hie_ellipse_fsky0016}, 
   but the sky coverages of \textsc{Polarbear}-2 and Simons Array
   are $f_{{\rm sky}}=0.2$.
}
   \label{fig:hie_ellipse_fsky02}
 \end{center}
\end{figure*}

\begin{figure*}[tbp]
 \begin{center}
   \resizebox{150mm}{!}{
   \includegraphics[bb= 93 236 523 553,width=0.95\linewidth]{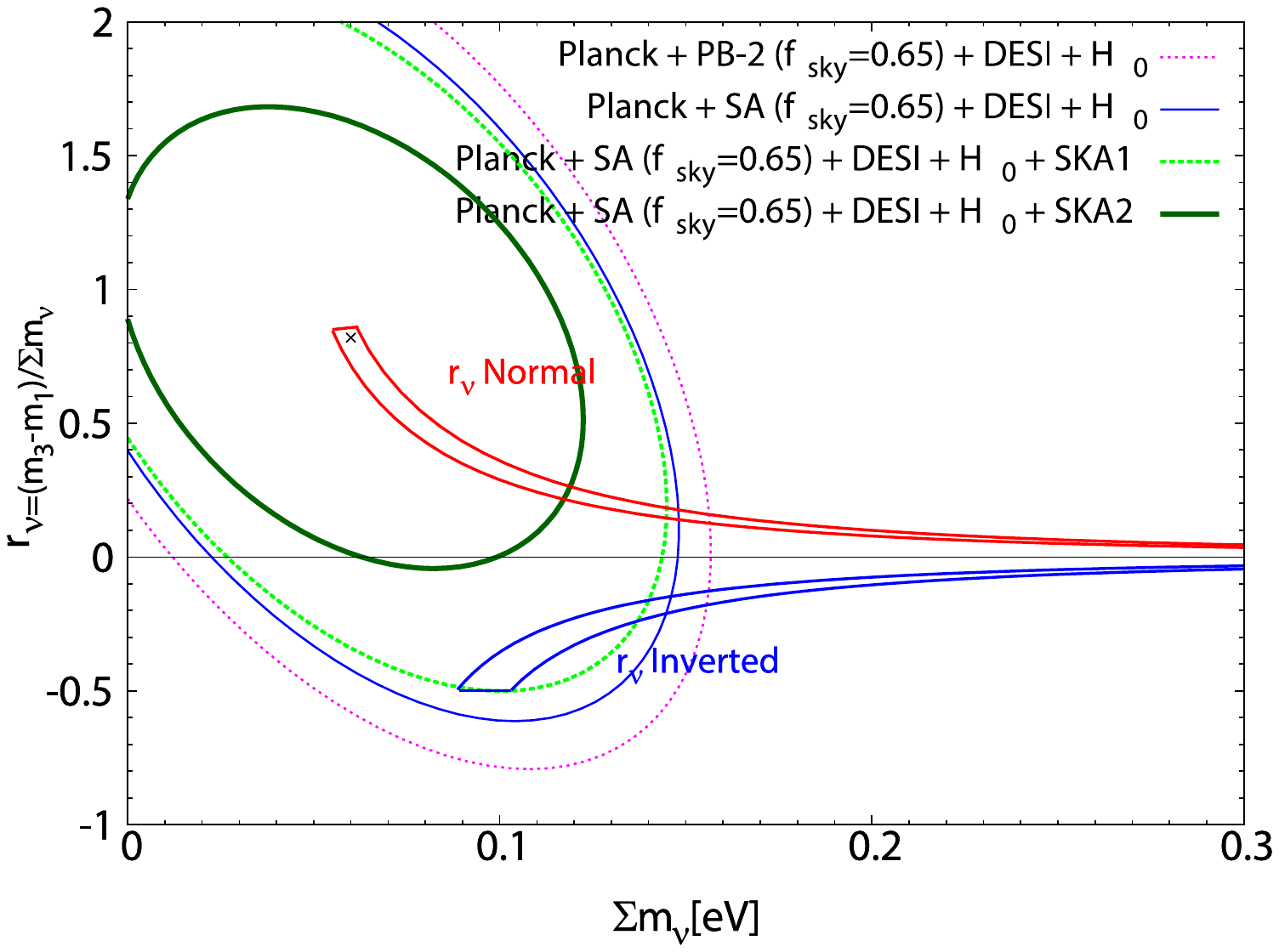} 
   \includegraphics[bb= 93 236 523 553,width=0.95\linewidth]{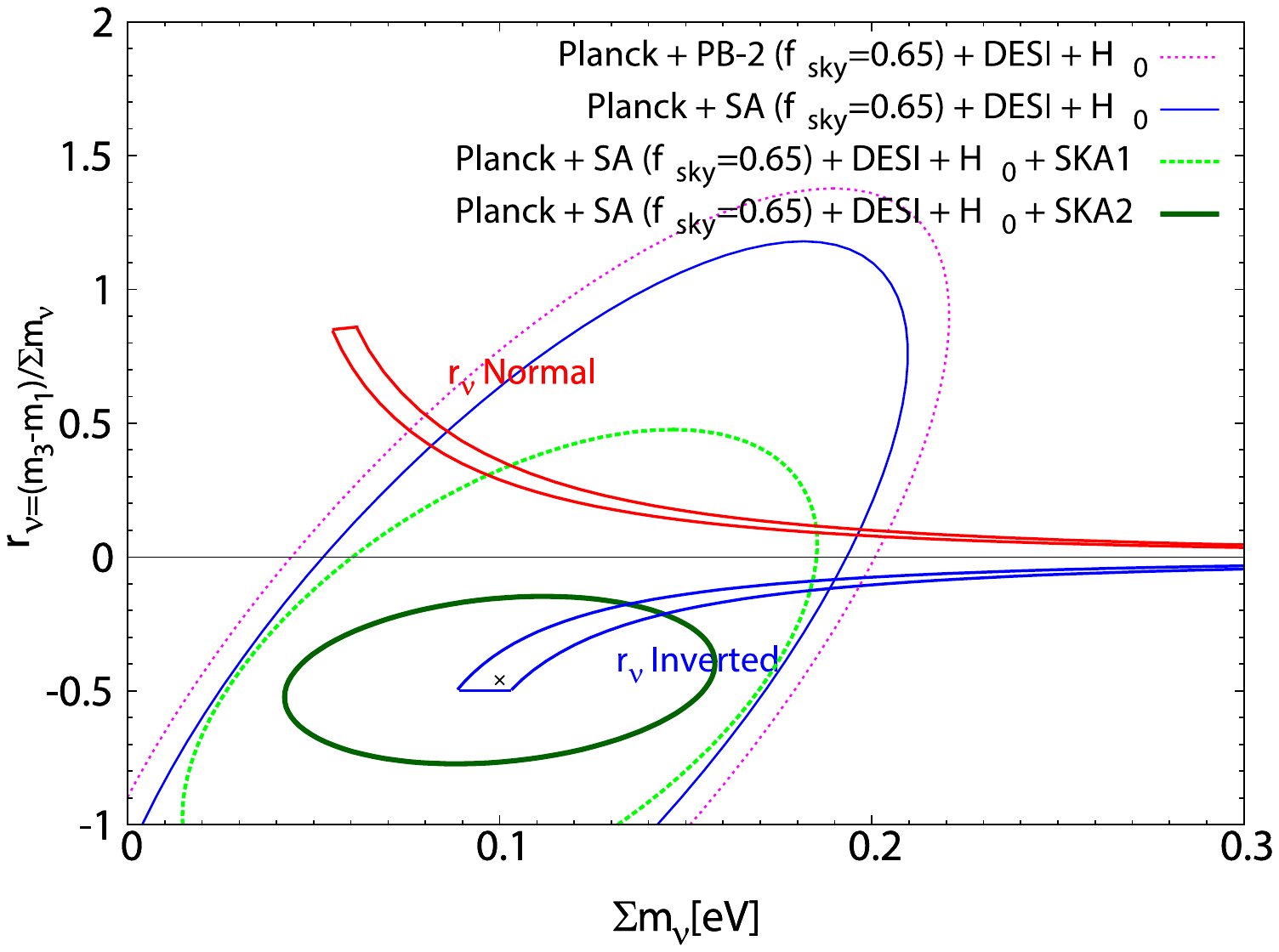} 
   }
   \resizebox{150mm}{!}{
   \includegraphics[bb= 93 236 523 553,width=0.95\linewidth]{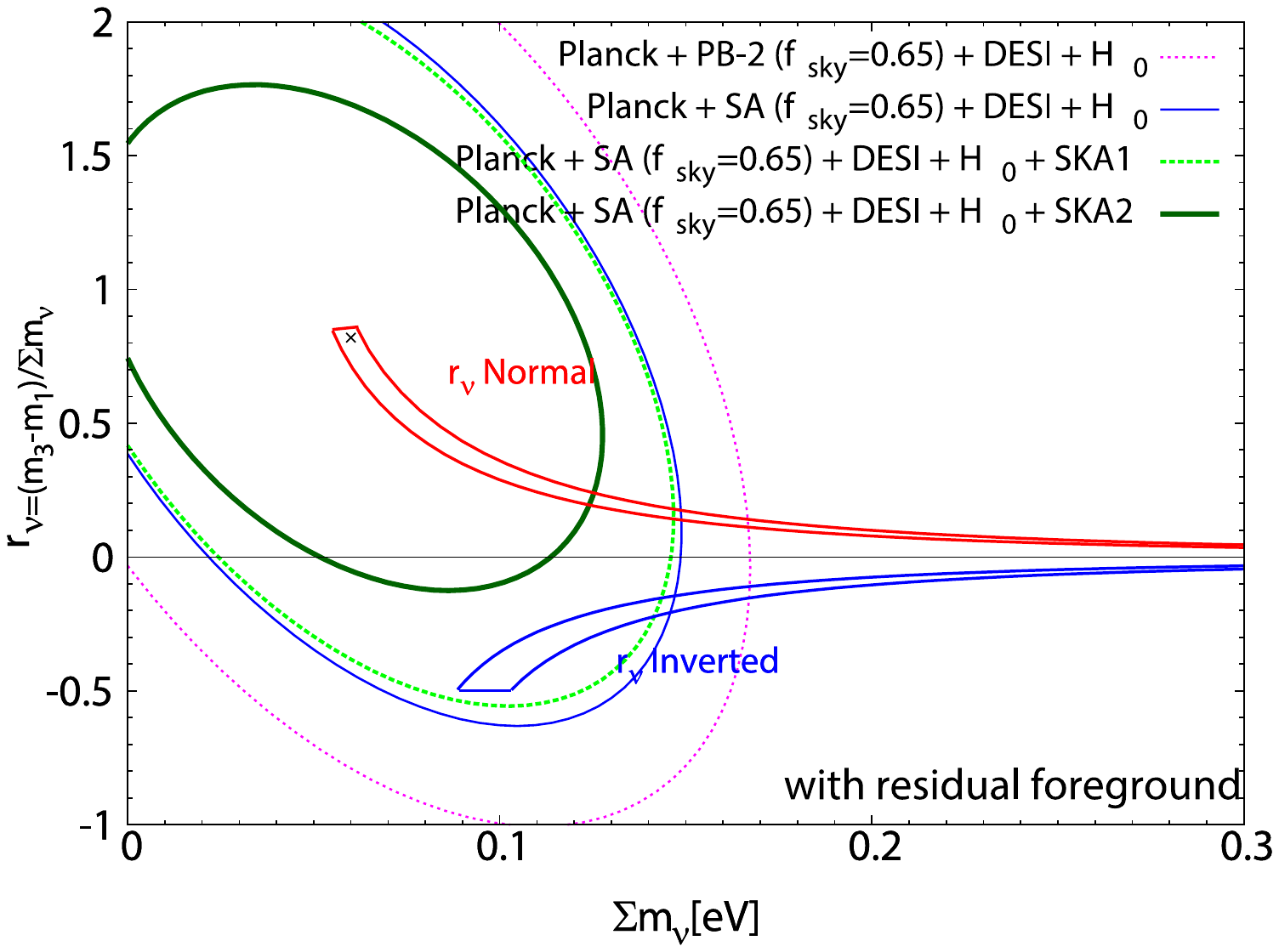} 
   \includegraphics[bb= 93 236 523 553,width=0.95\linewidth]{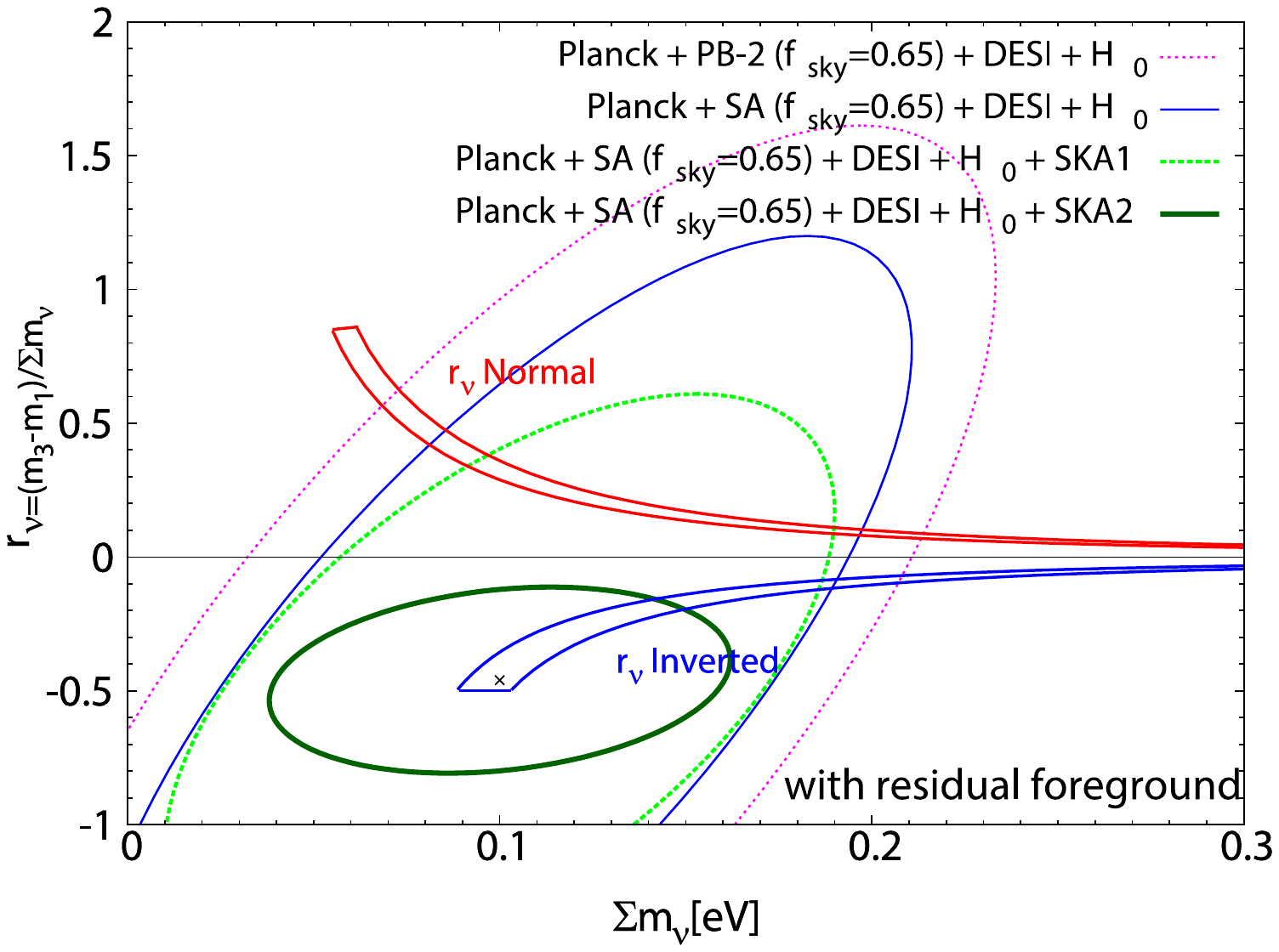} 
   }
   \resizebox{150mm}{!}{
   \includegraphics[bb= 93 236 523 553,width=0.95\linewidth]{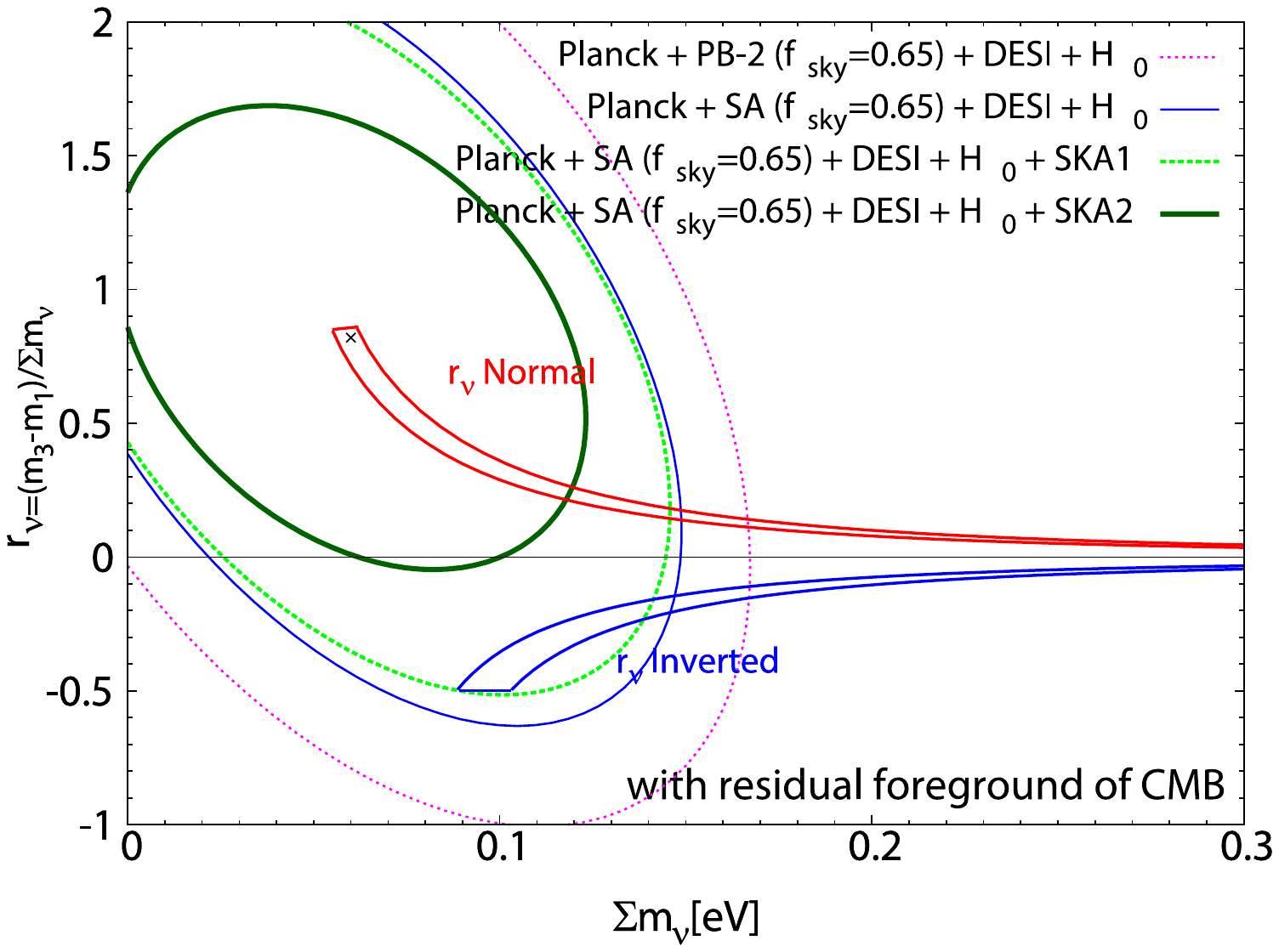} 
   \includegraphics[bb= 93 236 523 553,width=0.95\linewidth]{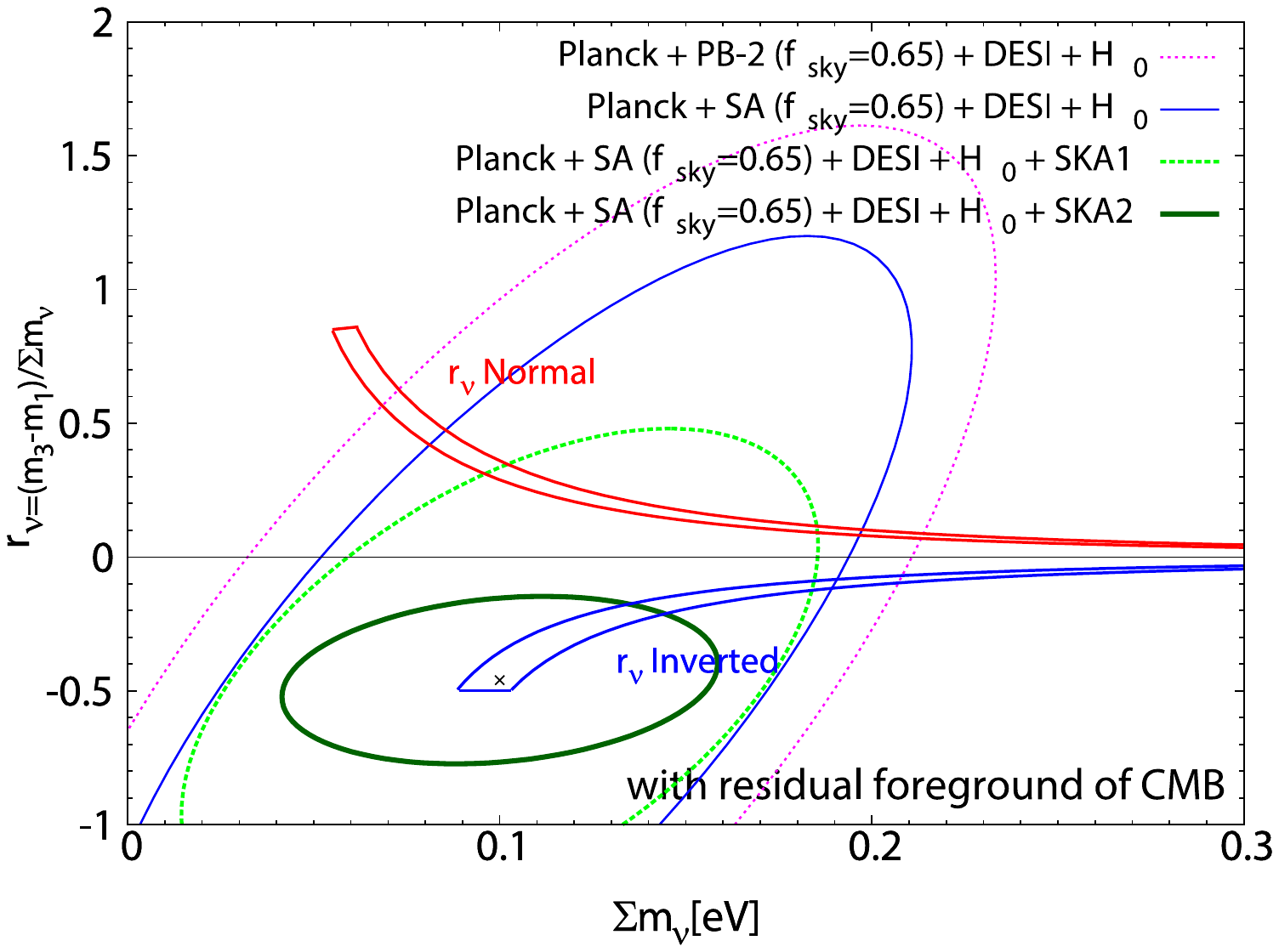} 
   }
   \caption{
   Same as Fig.\ref{fig:hie_ellipse_fsky0016}, 
   but the sky coverages of \textsc{Polarbear}-2 and Simons Array
   are $f_{{\rm sky}}=0.65$.
}
   \label{fig:hie_ellipse_fsky065}
 \end{center}
\end{figure*}

\begin{figure*}[tbp]
 \begin{center}
   \resizebox{150mm}{!}{
   \includegraphics[bb= 93 236 523 553,width=0.95\linewidth]{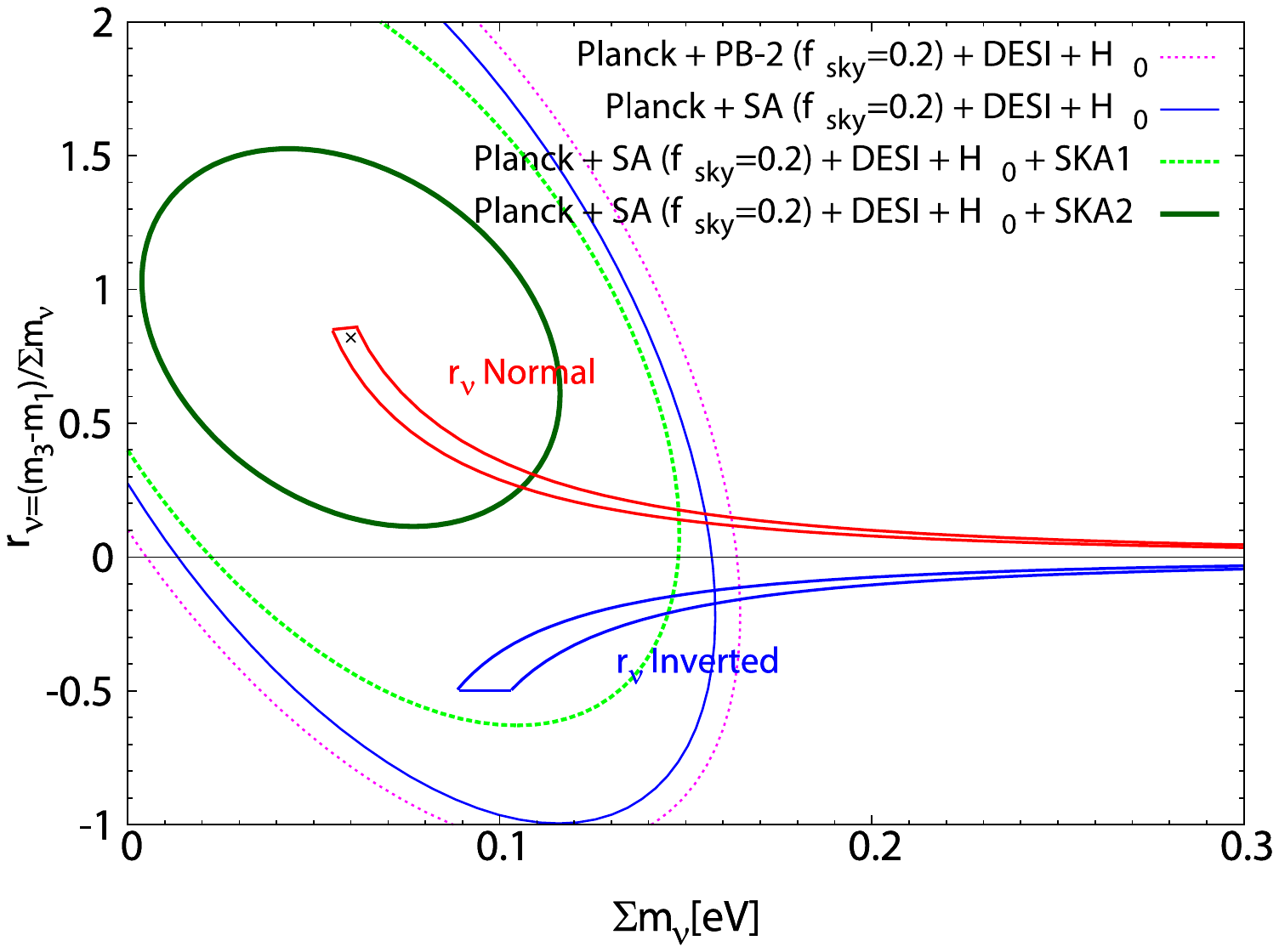} 
   \includegraphics[bb= 93 236 523 553,width=0.95\linewidth]{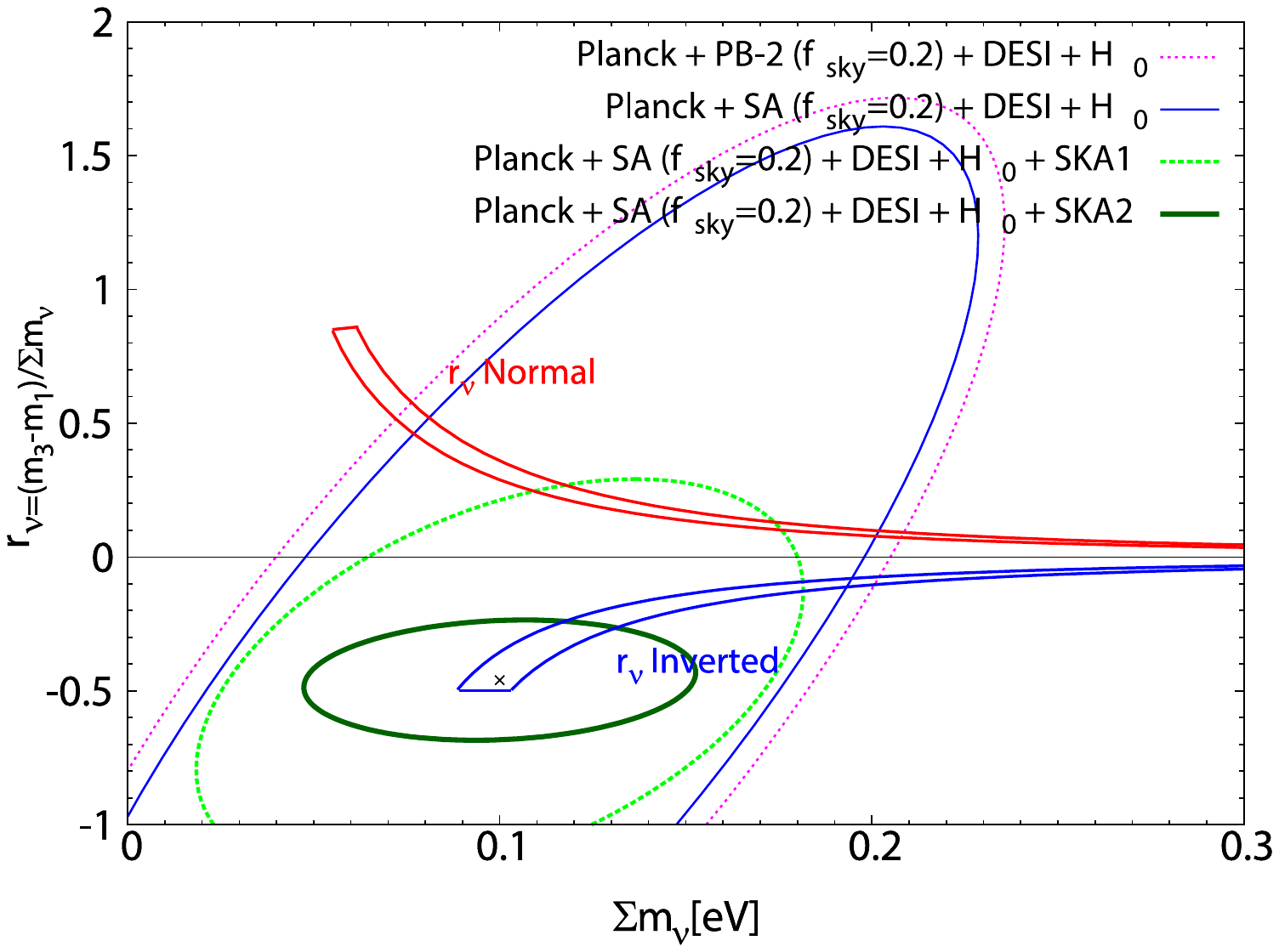} 
   }
   \resizebox{150mm}{!}{
   \includegraphics[bb= 93 236 523 553,width=0.95\linewidth]{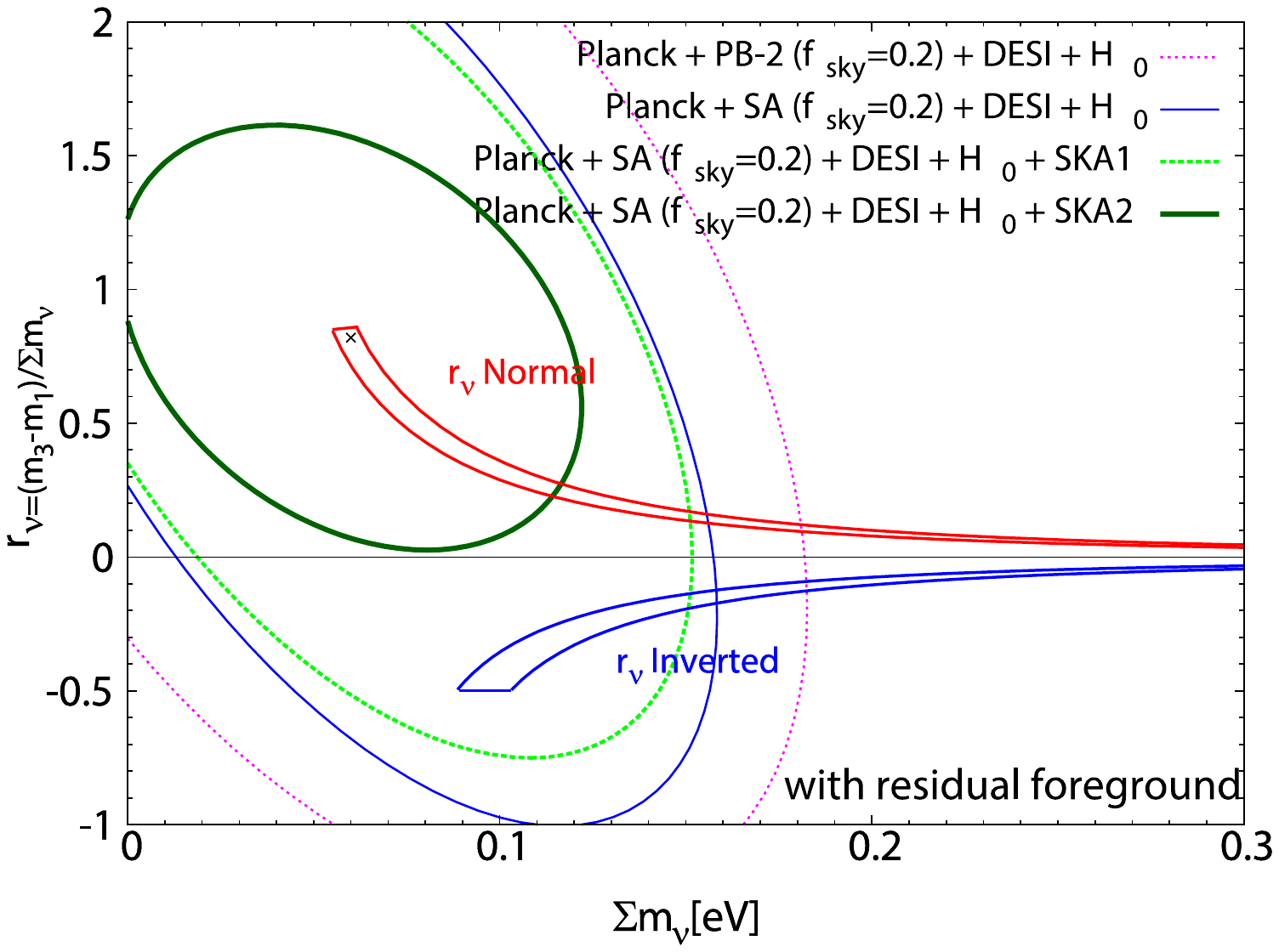} 
   \includegraphics[bb= 93 236 523 553,width=0.95\linewidth]{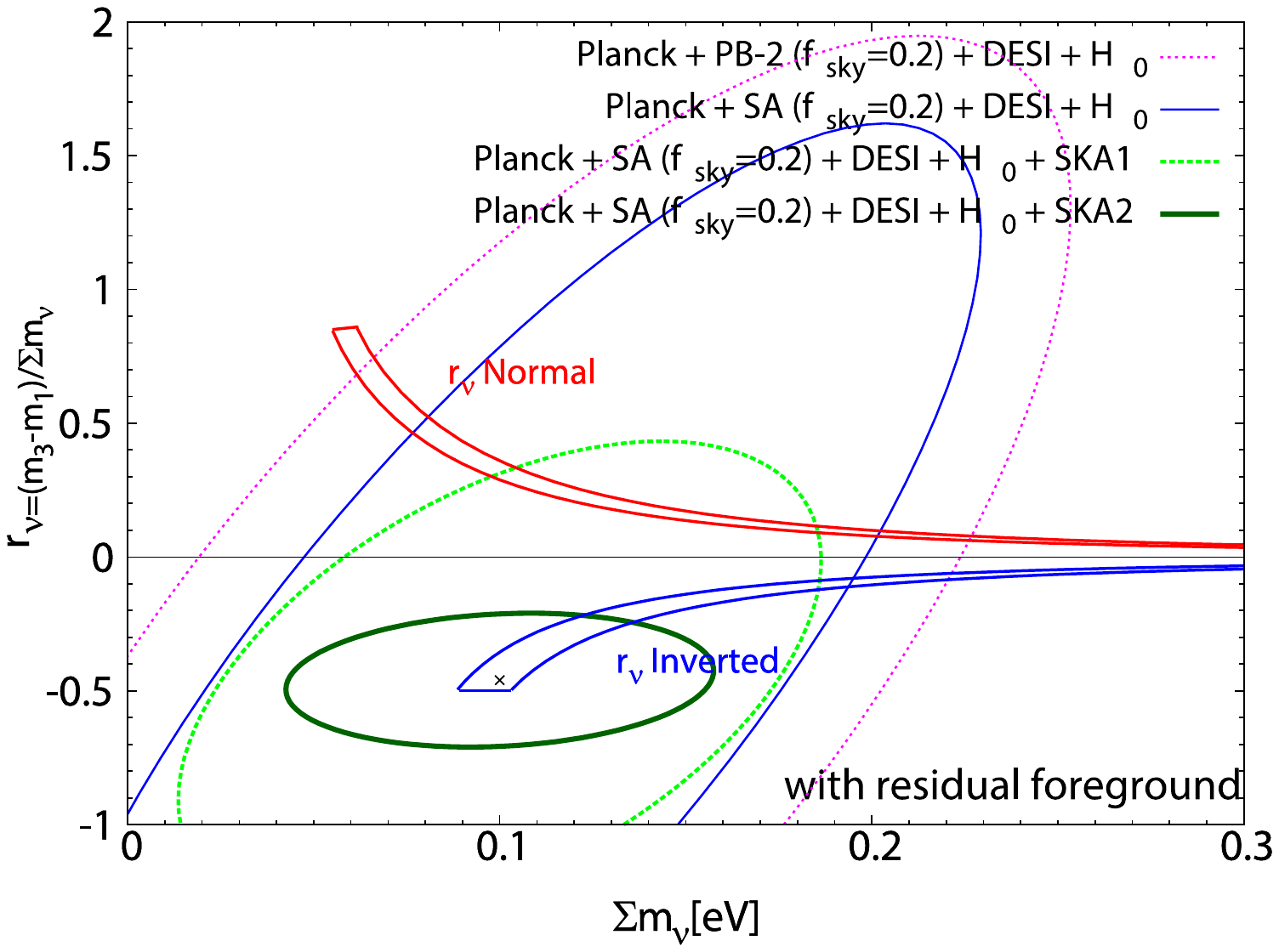} 
   }
   \resizebox{150mm}{!}{
   \includegraphics[bb= 93 236 523 553,width=0.95\linewidth]{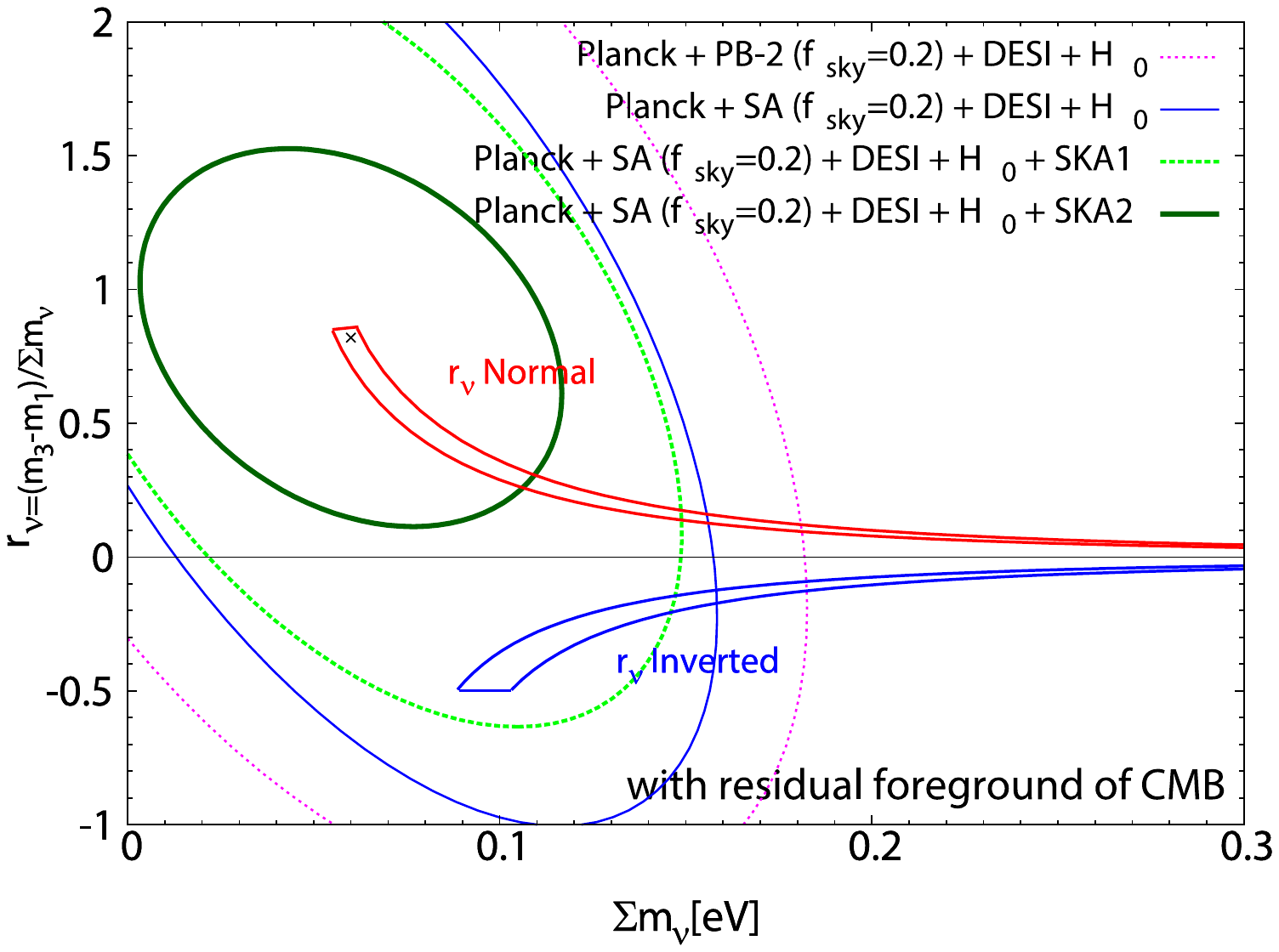} 
   \includegraphics[bb= 93 236 523 553,width=0.95\linewidth]{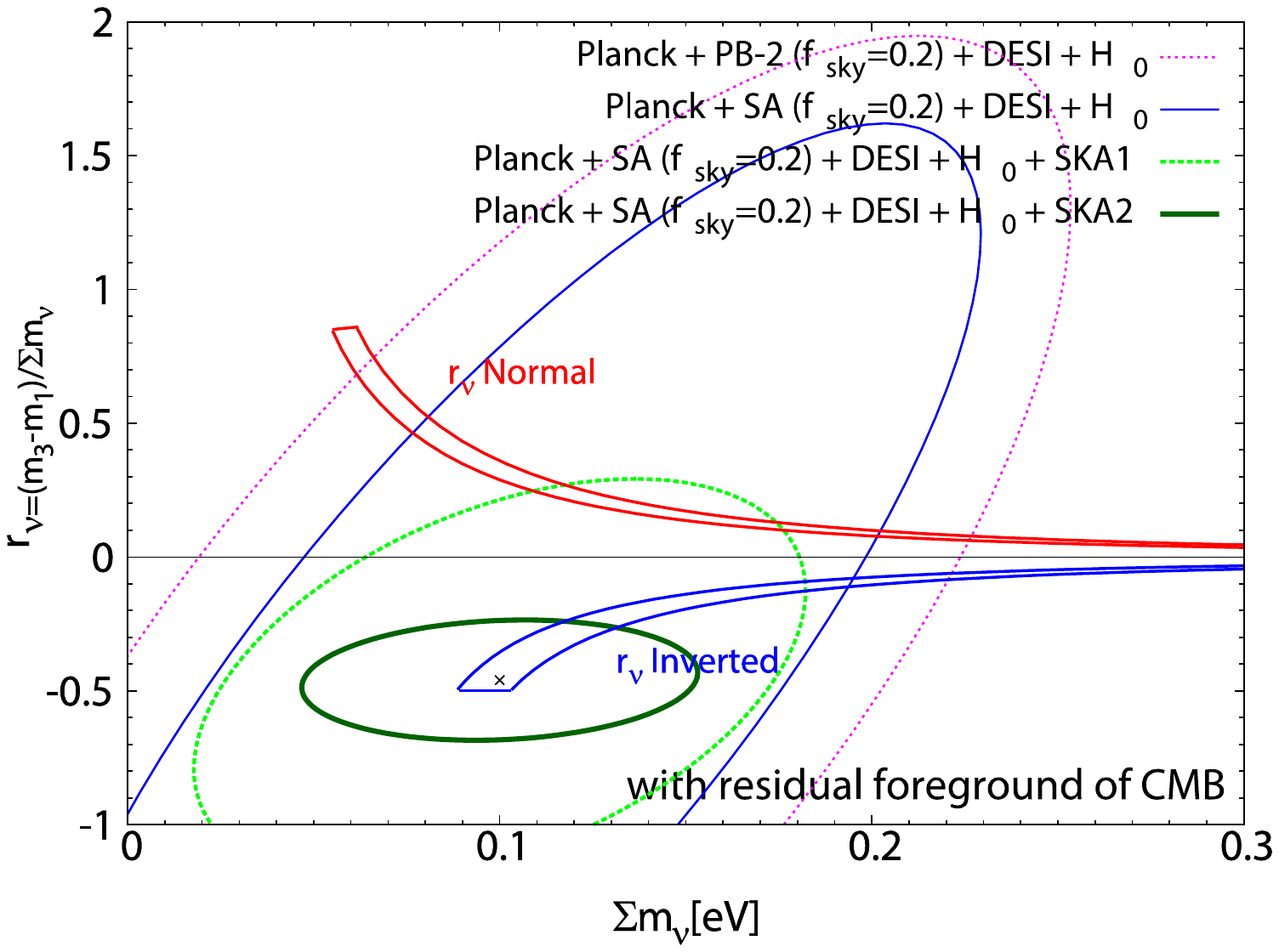} 
   }
   \caption{
   Same as Fig.\ref{fig:hie_ellipse_fsky02}, 
   but the observed fields of SKA are twice as large as that of 
   Fig.\ref{fig:hie_ellipse_fsky02} (i.e. $N_{{\rm field}}=8$).
}
   \label{fig:hie_ellipse_m8}
 \end{center}
\end{figure*}

Next, we discuss whether we will be able to determine the neutrino mass
hierarchies by using the future 21 cm line and the CMB observations. 
Constraints on cosmological parameters are summarized
in Tables \ref{tab:fsk0016_HieN} - \ref{tab:fsk065_fg_HieI_SKAnoFG}.
%
%In Fig.~\ref{fig:hie} we plot 95\% C.L. on a parameter
%%%%%%%%revised
In Figs.~\ref{fig:hie_ellipse_fsky0016}-\ref{fig:hie_ellipse_m8}, 
we plot $2{\bf \sigma}$ errors of the parameter 
%%%%%%%%%%%%%%
$r_{\nu}\equiv (m_{3} - m_{1} )/\Sigma m_{\nu}$ constrained by both
the 21 cm line and the CMB observations in case of the inverted hierarchy
to be fiducial (left panels), and the normal hierarchy to be fiducial (right panels). 
In these figure, upper two panels represent results 
without residual foregrounds, i.e. they are completely removed.
%On the other hand, lower panels show the results
%when the residual foregrounds are remaining
%and the 220 GHz band of Simons Array are used for foreground removal.
On the other hand, middle two panels show the results
with given level residual foregrounds.
Here, we assume that the amplitude of the residual foreground of the 21 cm line 
is 10 times as small as that of the previous section, % are remaining
and the 220 GHz band of Simons Array is used for the foreground removal.
Lower two panels show the results with only the residual foreground of CMB,
i.e. that of the 21 cm line is completely removed.
We also assume that the 220 GHz band of Simons Array is 
used for the foreground removal in these cases.
The number of observed field of SKA is $N_{{\rm field}}=4$ 
in Figs.\ref{fig:hie_ellipse_fsky0016}-\ref{fig:hie_ellipse_fsky065}
and $N_{{\rm field}}=8$ in Fig.\ref{fig:hie_ellipse_m8}.

It is notable that the difference between $r_{\nu}$ of these two hierarchies 
becomes larger as the total mass $\Sigma m_{\nu}$ becomes smaller. 
Therefore, $r_{\nu}$ is quite useful to distinguish a true mass hierarchy from the other.  
Allowed parameters on $r_{\nu}$ by neutrino oscillation experiments are plotted as two
bands for the inverted and the normal hierarchies, respectively.  
%The thin solid lines inside the bands are the experimental central values by oscillations 
The thin solid lines inside the bands are the central values of the oscillation experiments.
%one of which is taken to be a corresponding fiducial
%value of $r_{\nu}$ as a function of $\Sigma m_{\nu}$ in each analysis. 

%
%As is clearly shown in Fig.~\ref{fig:hie_ellipse}, 
%actually those combinations of the observations will be able to determine 
%the neutrino mass hierarchy to be inverted  or normal for 
%$\Sigma m_{\nu} \sim 0.06$ eV or 
%$\Sigma m_{\nu} \sim 0.1 \ {\rm eV} $ at 95 \% C.L.,
%%%%%%%%%%%%%%%%%%%%%%%%%%%%%%%%%%%%%%%
%
%respectively.  Although the determination is possible only at around
%$\Sigma m_{\nu}\lesssim {\cal O}(0.1) \ {\rm eV}$, those results should be
%reasonable. That is  because a precise discrimination of the mass
%hierarchy itself may have no meaning if the masses are highly
%degenerate, i.e., if $\Sigma m_{\nu} \gg 0.1 - 0.3$~eV.

From these figures,
we find that it is difficult to determine the neutrino mass hierarchy
by the combination of CMB and BAO with SKA phase~1.
%except for the case of Simons Array ($f_{{\rm sky}}=0.65$)
%and without the contribution of residual foreground.
%
However, if we use SKA phase~2, by the combinations of them,
we will be able to determine the neutrino mass hierarchy to be inverted 
or normal for $\Sigma m_{\nu} \sim 0.06$ eV or 
$\Sigma m_{\nu} \sim 0.1 \ {\rm eV} $ at 95 \% C.L., respectively.
In particular, the detectability becomes larger
as the sky coverage of Simons Array and the number of observed field 
of SKA become larger.
This improvement on the observed field of SKA is more effective
than that of the large integrated time per one field.
%%%%%%%%%%%%%%%%%%%%%%%%%%%%%%%%%%%%%%%
%
%Although the determination is possible only at around
%$\Sigma m_{\nu}\lesssim {\cal O}(0.1) \ {\rm eV}$, 
%those results should be reasonable. 
%That is  
Those results should be reasonable 
because a precise discrimination of the mass
hierarchy itself may have no meaning if the masses are highly
degenerate, i.e., % if $\Sigma m_{\nu} \gg 0.1 - 0.3$~eV.
if $ 0.1 $~eV $\ll \Sigma m_{\nu}$.

Once a clear signature $\Sigma m_{\nu} \ll 0.1 \ {\rm eV}$ were
determined by observations or experiments, it should be obvious that
the hierarchy must be normal without any ambiguities. 
On the other hand, if the hierarchy were inverted, we could not determine it 
by using information of only $\Sigma m_{\nu}$. 
However, it is remarkable that our method is quite useful 
because we can discriminate the hierarchy from the other
even if the fiducial values were $\Sigma m_{\nu} \gtrsim 0.1$~eV for
both the normal and inverted cases.
%%
%%%%%%%%%%%%%%%%%
This is clearly shown in these figures.  
In case that a fiducial value of $\Sigma m_{\nu}$ is taken to be the lowest values 
in neutrino oscillation experiments, 
these figures indicate that by Planck + Simons Array + BAO + SKA phase~2, we can discriminate the 
inverted (normal) mass hierarchy from the normal (inverted) one.
%%%%%%%%%%%%%%%%%
%%
By using Omniscope, we can discriminate any mass hierarchies
up to $\Sigma m_{\nu}\sim 0.1 $ eV \cite{Oyama:2012tq}.

%%%%%%%%%%%%%%%%%%FIGURE%%%%%%%%%%%%%%%%%%%

%%%%%%%%%%%%%%%%%%%%%%%%%%%%%%%%%%%%%%%%%

%%%%%%%%%%%%%%%%%%FIGURE%%%%%%%%%%%%%%%%%%%

%%%%%%%%%%%%%%%%%%%%%%%%%%%%%%%%%%%%%%%%%

%%%%%%%%%%%%%%%%%%FIGURE%%%%%%%%%%%%%%%%%%%

%%%%%%%%%%%%%%%%%%%%%%%%%%%%%%%%%%%%%%%%%

%%%%%%%%%%%%%%%%%%FIGURE%%%%%%%%%%%%%%%%%%%

%%%%%%%%%%%%%%%%%%%%%%%%%%%%%%%%%%%%%%%%%

\begin{table}[htbp]
\centering \scalebox{0.8}[0.77]{
\begin{tabular}{l|ccccc}
\hline \hline \\
& $\sigma( \Omega_{m}h^{2} )$ & $\sigma( \Omega_{b}h^{2} )$ &  $\sigma( \Omega_{\Lambda} )$ & $ \sigma( n_s ) $ & $\sigma( A_{s} \times 10^{10} )$ 
\\
\hline
Planck + PB-2(0.016) + BAO + $H_{0}$
 & $ 7.90 \times 10^{  -4} $ & $ 1.52 \times 10^{  -4} $ & $ 4.68 \times 10^{  -3} $ & $ 5.35 \times 10^{  -3} $ & $ 2.12 \times 10^{  -1} $ 
\\
+ SKA1 ($N_{\textrm{field}}=4$)
 & $ 6.71 \times 10^{  -4} $ & $ 1.32 \times 10^{  -4} $ & $ 4.32 \times 10^{  -3} $ & $ 4.69 \times 10^{  -3} $ & $ 2.07 \times 10^{  -1} $ 
\\
+ SKA1 ($N_{\textrm{field}}=8$)
 & $ 6.02 \times 10^{  -4} $ & $ 1.20 \times 10^{  -4} $ & $ 4.06 \times 10^{  -3} $ & $ 4.27 \times 10^{  -3} $ & $ 2.03 \times 10^{  -1} $ 
\\
+ SKA2 ($N_{\textrm{field}}=4$)
 & $ 4.28 \times 10^{  -4} $ & $ 1.12 \times 10^{  -4} $ & $ 2.68 \times 10^{  -3} $ & $ 2.75 \times 10^{  -3} $ & $ 1.91 \times 10^{  -1} $ 
\\
+ SKA2 ($N_{\textrm{field}}=8$)
 & $ 3.36 \times 10^{  -4} $ & $ 9.59 \times 10^{  -5} $ & $ 2.09 \times 10^{  -3} $ & $ 2.13 \times 10^{  -3} $ & $ 1.84 \times 10^{  -1} $ 
\\
Planck + SA(0.016) + BAO + $H_{0}$
 & $ 7.76 \times 10^{  -4} $ & $ 1.51 \times 10^{  -4} $ & $ 4.66 \times 10^{  -3} $ & $ 5.21 \times 10^{  -3} $ & $ 2.11 \times 10^{  -1} $ 
\\
\hline
\end{tabular} }
\caption{$f_{\textrm{sky}}=0.016$, normal hierarchy, without the residual foregrounds.}
\label{tab:fsk0016_HieN}

\vspace{10pt}

\centering \scalebox{0.8}[0.77]{
\begin{tabular}{l|ccccc}
\hline \hline \\
& $\sigma( \Omega_{m}h^{2} )$ & $\sigma( \Omega_{b}h^{2} )$ &  $\sigma( \Omega_{\Lambda} )$ & $ \sigma( n_s ) $ & $\sigma( A_{s} \times 10^{10} )$ 
\\
\hline
Planck + PB-2(0.016) + BAO + $H_{0}$
 & $ 8.64 \times 10^{  -4} $ & $ 1.52 \times 10^{  -4} $ & $ 4.77 \times 10^{  -3} $ & $ 5.36 \times 10^{  -3} $ & $ 2.12 \times 10^{  -1} $ 
\\
+ SKA1 ($N_{\textrm{field}}=4$)
 & $ 6.59 \times 10^{  -4} $ & $ 1.32 \times 10^{  -4} $ & $ 4.28 \times 10^{  -3} $ & $ 4.69 \times 10^{  -3} $ & $ 2.07 \times 10^{  -1} $ 
\\
+ SKA1 ($N_{\textrm{field}}=8$)
 & $ 5.83 \times 10^{  -4} $ & $ 1.19 \times 10^{  -4} $ & $ 4.02 \times 10^{  -3} $ & $ 4.28 \times 10^{  -3} $ & $ 2.04 \times 10^{  -1} $ 
\\
+ SKA2 ($N_{\textrm{field}}=4$)
 & $ 4.22 \times 10^{  -4} $ & $ 1.12 \times 10^{  -4} $ & $ 2.68 \times 10^{  -3} $ & $ 2.79 \times 10^{  -3} $ & $ 1.92 \times 10^{  -1} $ 
\\
+ SKA2 ($N_{\textrm{field}}=8$)
 & $ 3.31 \times 10^{  -4} $ & $ 9.59 \times 10^{  -5} $ & $ 2.10 \times 10^{  -3} $ & $ 2.17 \times 10^{  -3} $ & $ 1.85 \times 10^{  -1} $ 
\\
Planck + SA(0.016) + BAO + $H_{0}$
 & $ 8.51 \times 10^{  -4} $ & $ 1.51 \times 10^{  -4} $ & $ 4.75 \times 10^{  -3} $ & $ 5.22 \times 10^{  -3} $ & $ 2.11 \times 10^{  -1} $ 
\\
\hline
\end{tabular} }
\caption{$f_{\textrm{sky}}=0.016$, inverted hierarchy, without the residual foregrounds.}
\label{tab:fsk0016_HieI}

\end{table}

\begin{table}[htbp]
\centering \scalebox{0.8}[0.77]{
\begin{tabular}{l|ccccc}
\hline \hline \\
& $\sigma( \Omega_{m}h^{2} )$ & $\sigma( \Omega_{b}h^{2} )$ &  $\sigma( \Omega_{\Lambda} )$ & $ \sigma( n_s ) $ & $\sigma( A_{s} \times 10^{10} )$ 
\\
\hline
Planck + PB-2(0.016) + BAO + $H_{0}$
 & $ 8.28 \times 10^{  -4} $ & $ 1.65 \times 10^{  -4} $ & $ 4.72 \times 10^{  -3} $ & $ 5.74 \times 10^{  -3} $ & $ 2.12 \times 10^{  -1} $ 
\\
+ SKA1 ($N_{\textrm{field}}=4$)
 & $ 7.41 \times 10^{  -4} $ & $ 1.54 \times 10^{  -4} $ & $ 4.47 \times 10^{  -3} $ & $ 5.33 \times 10^{  -3} $ & $ 2.08 \times 10^{  -1} $ 
\\
+ SKA1 ($N_{\textrm{field}}=8$)
 & $ 6.81 \times 10^{  -4} $ & $ 1.46 \times 10^{  -4} $ & $ 4.28 \times 10^{  -3} $ & $ 5.01 \times 10^{  -3} $ & $ 2.06 \times 10^{  -1} $ 
\\
+ SKA2 ($N_{\textrm{field}}=4$)
 & $ 5.13 \times 10^{  -4} $ & $ 1.34 \times 10^{  -4} $ & $ 3.01 \times 10^{  -3} $ & $ 3.38 \times 10^{  -3} $ & $ 1.96 \times 10^{  -1} $ 
\\
+ SKA2 ($N_{\textrm{field}}=8$)
 & $ 4.15 \times 10^{  -4} $ & $ 1.20 \times 10^{  -4} $ & $ 2.41 \times 10^{  -3} $ & $ 2.67 \times 10^{  -3} $ & $ 1.91 \times 10^{  -1} $ 
\\
Planck + SA(0.016) + BAO + $H_{0}$
 & $ 7.78 \times 10^{  -4} $ & $ 1.51 \times 10^{  -4} $ & $ 4.67 \times 10^{  -3} $ & $ 5.22 \times 10^{  -3} $ & $ 2.08 \times 10^{  -1} $ 
\\
\hline
\end{tabular} }
\caption{$f_{\textrm{sky}}=0.016$, normal hierarchy, with the residual foregrounds.}
\label{tab:fsk0016_fg_HieN}

\vspace{10pt}

\centering \scalebox{0.8}[0.77]{
\begin{tabular}{l|ccccc}
\hline \hline \\
& $\sigma( \Omega_{m}h^{2} )$ & $\sigma( \Omega_{b}h^{2} )$ &  $\sigma( \Omega_{\Lambda} )$ & $ \sigma( n_s ) $ & $\sigma( A_{s} \times 10^{10} )$ 
\\
\hline
Planck + PB-2(0.016) + BAO + $H_{0}$
 & $ 9.04 \times 10^{  -4} $ & $ 1.65 \times 10^{  -4} $ & $ 4.80 \times 10^{  -3} $ & $ 5.75 \times 10^{  -3} $ & $ 2.12 \times 10^{  -1} $ 
\\
+ SKA1 ($N_{\textrm{field}}=4$)
 & $ 7.35 \times 10^{  -4} $ & $ 1.54 \times 10^{  -4} $ & $ 4.45 \times 10^{  -3} $ & $ 5.34 \times 10^{  -3} $ & $ 2.09 \times 10^{  -1} $ 
\\
+ SKA1 ($N_{\textrm{field}}=8$)
 & $ 6.66 \times 10^{  -4} $ & $ 1.46 \times 10^{  -4} $ & $ 4.24 \times 10^{  -3} $ & $ 5.04 \times 10^{  -3} $ & $ 2.06 \times 10^{  -1} $ 
\\
+ SKA2 ($N_{\textrm{field}}=4$)
 & $ 5.06 \times 10^{  -4} $ & $ 1.35 \times 10^{  -4} $ & $ 3.02 \times 10^{  -3} $ & $ 3.44 \times 10^{  -3} $ & $ 1.97 \times 10^{  -1} $ 
\\
+ SKA2 ($N_{\textrm{field}}=8$)
 & $ 4.10 \times 10^{  -4} $ & $ 1.21 \times 10^{  -4} $ & $ 2.42 \times 10^{  -3} $ & $ 2.73 \times 10^{  -3} $ & $ 1.92 \times 10^{  -1} $ 
\\
Planck + SA(0.016) + BAO + $H_{0}$
 & $ 8.53 \times 10^{  -4} $ & $ 1.51 \times 10^{  -4} $ & $ 4.75 \times 10^{  -3} $ & $ 5.23 \times 10^{  -3} $ & $ 2.08 \times 10^{  -1} $ 
\\
\hline
\end{tabular} }
\caption{$f_{\textrm{sky}}=0.016$, inverted hierarchy, with the residual foregrounds.}
\label{tab:fsk0016_fg_HieI}

\end{table}

\begin{table}[htbp]
\centering \scalebox{0.8}[0.77]{
\begin{tabular}{l|ccccc}
\hline \hline \\
& $\sigma( \Omega_{m}h^{2} )$ & $\sigma( \Omega_{b}h^{2} )$ &  $\sigma( \Omega_{\Lambda} )$ & $ \sigma( n_s ) $ & $\sigma( A_{s} \times 10^{10} )$ 
\\
\hline
Planck + PB-2(0.2) + BAO + $H_{0}$
 & $ 7.17 \times 10^{  -4} $ & $ 1.18 \times 10^{  -4} $ & $ 4.56 \times 10^{  -3} $ & $ 4.13 \times 10^{  -3} $ & $ 1.99 \times 10^{  -1} $ 
\\
+ SKA1 ($N_{\textrm{field}}=4$)
 & $ 6.31 \times 10^{  -4} $ & $ 1.07 \times 10^{  -4} $ & $ 4.24 \times 10^{  -3} $ & $ 3.78 \times 10^{  -3} $ & $ 1.97 \times 10^{  -1} $ 
\\
+ SKA1 ($N_{\textrm{field}}=8$)
 & $ 5.75 \times 10^{  -4} $ & $ 1.00 \times 10^{  -4} $ & $ 3.99 \times 10^{  -3} $ & $ 3.53 \times 10^{  -3} $ & $ 1.95 \times 10^{  -1} $ 
\\
+ SKA2 ($N_{\textrm{field}}=4$)
 & $ 4.07 \times 10^{  -4} $ & $ 9.58 \times 10^{  -5} $ & $ 2.66 \times 10^{  -3} $ & $ 2.49 \times 10^{  -3} $ & $ 1.84 \times 10^{  -1} $ 
\\
+ SKA2 ($N_{\textrm{field}}=8$)
 & $ 3.20 \times 10^{  -4} $ & $ 8.49 \times 10^{  -5} $ & $ 2.08 \times 10^{  -3} $ & $ 1.98 \times 10^{  -3} $ & $ 1.76 \times 10^{  -1} $ 
\\
Planck + SA(0.2) + BAO + $H_{0}$
 & $ 6.96 \times 10^{  -4} $ & $ 1.14 \times 10^{  -4} $ & $ 4.52 \times 10^{  -3} $ & $ 3.94 \times 10^{  -3} $ & $ 1.97 \times 10^{  -1} $ 
\\
\hline
\end{tabular} }
\caption{$f_{\textrm{sky}}=0.2$, normal hierarchy, without the residual foregrounds.}
\label{tab:fsk02_HieN}

\vspace{10pt}

\centering \scalebox{0.8}[0.77]{
\begin{tabular}{l|ccccc}
\hline \hline \\
& $\sigma( \Omega_{m}h^{2} )$ & $\sigma( \Omega_{b}h^{2} )$ &  $\sigma( \Omega_{\Lambda} )$ & $ \sigma( n_s ) $ & $\sigma( A_{s} \times 10^{10} )$ 
\\
\hline
Planck + PB-2(0.2) + BAO + $H_{0}$
 & $ 7.68 \times 10^{  -4} $ & $ 1.18 \times 10^{  -4} $ & $ 4.62 \times 10^{  -3} $ & $ 4.14 \times 10^{  -3} $ & $ 1.99 \times 10^{  -1} $ 
\\
+ SKA1 ($N_{\textrm{field}}=4$)
 & $ 6.25 \times 10^{  -4} $ & $ 1.07 \times 10^{  -4} $ & $ 4.21 \times 10^{  -3} $ & $ 3.79 \times 10^{  -3} $ & $ 1.97 \times 10^{  -1} $ 
\\
+ SKA1 ($N_{\textrm{field}}=8$)
 & $ 5.61 \times 10^{  -4} $ & $ 1.00 \times 10^{  -4} $ & $ 3.95 \times 10^{  -3} $ & $ 3.54 \times 10^{  -3} $ & $ 1.95 \times 10^{  -1} $ 
\\
+ SKA2 ($N_{\textrm{field}}=4$)
 & $ 4.01 \times 10^{  -4} $ & $ 9.58 \times 10^{  -5} $ & $ 2.66 \times 10^{  -3} $ & $ 2.52 \times 10^{  -3} $ & $ 1.85 \times 10^{  -1} $ 
\\
+ SKA2 ($N_{\textrm{field}}=8$)
 & $ 3.15 \times 10^{  -4} $ & $ 8.48 \times 10^{  -5} $ & $ 2.09 \times 10^{  -3} $ & $ 2.00 \times 10^{  -3} $ & $ 1.77 \times 10^{  -1} $ 
\\
Planck + SA(0.2) + BAO + $H_{0}$
 & $ 7.45 \times 10^{  -4} $ & $ 1.14 \times 10^{  -4} $ & $ 4.58 \times 10^{  -3} $ & $ 3.95 \times 10^{  -3} $ & $ 1.97 \times 10^{  -1} $ 
\\
\hline
\end{tabular} }
\caption{$f_{\textrm{sky}}=0.2$, inverted hierarchy, without the residual foregrounds.}
\label{tab:fsk02_HieI}

\end{table}

\begin{table}[htbp]
\centering \scalebox{0.8}[0.77]{
\begin{tabular}{l|ccccc}
\hline \hline \\
& $\sigma( \Omega_{m}h^{2} )$ & $\sigma( \Omega_{b}h^{2} )$ &  $\sigma( \Omega_{\Lambda} )$ & $ \sigma( n_s ) $ & $\sigma( A_{s} \times 10^{10} )$ 
\\
\hline
Planck + PB-2(0.2) + BAO + $H_{0}$
 & $ 7.53 \times 10^{  -4} $ & $ 1.33 \times 10^{  -4} $ & $ 4.61 \times 10^{  -3} $ & $ 4.60 \times 10^{  -3} $ & $ 2.03 \times 10^{  -1} $ 
\\
+ SKA1 ($N_{\textrm{field}}=4$)
 & $ 6.92 \times 10^{  -4} $ & $ 1.27 \times 10^{  -4} $ & $ 4.40 \times 10^{  -3} $ & $ 4.36 \times 10^{  -3} $ & $ 2.01 \times 10^{  -1} $ 
\\
+ SKA1 ($N_{\textrm{field}}=8$)
 & $ 6.45 \times 10^{  -4} $ & $ 1.22 \times 10^{  -4} $ & $ 4.22 \times 10^{  -3} $ & $ 4.16 \times 10^{  -3} $ & $ 2.00 \times 10^{  -1} $ 
\\
+ SKA2 ($N_{\textrm{field}}=4$)
 & $ 4.84 \times 10^{  -4} $ & $ 1.16 \times 10^{  -4} $ & $ 2.99 \times 10^{  -3} $ & $ 3.05 \times 10^{  -3} $ & $ 1.91 \times 10^{  -1} $ 
\\
+ SKA2 ($N_{\textrm{field}}=8$)
 & $ 3.93 \times 10^{  -4} $ & $ 1.06 \times 10^{  -4} $ & $ 2.40 \times 10^{  -3} $ & $ 2.47 \times 10^{  -3} $ & $ 1.86 \times 10^{  -1} $ 
\\
Planck + SA(0.2) + BAO + $H_{0}$
 & $ 6.98 \times 10^{  -4} $ & $ 1.14 \times 10^{  -4} $ & $ 4.53 \times 10^{  -3} $ & $ 3.96 \times 10^{  -3} $ & $ 1.94 \times 10^{  -1} $ 
\\
\hline
\end{tabular} }
\caption{$f_{\textrm{sky}}=0.2$, normal hierarchy, with the residual foregrounds.}
\label{tab:fsk02_fg_HieN}

\vspace{10pt}

\centering \scalebox{0.8}[0.77]{
\begin{tabular}{l|ccccc}
\hline \hline \\
& $\sigma( \Omega_{m}h^{2} )$ & $\sigma( \Omega_{b}h^{2} )$ &  $\sigma( \Omega_{\Lambda} )$ & $ \sigma( n_s ) $ & $\sigma( A_{s} \times 10^{10} )$ 
\\
\hline
Planck + PB-2(0.2) + BAO + $H_{0}$
 & $ 8.15 \times 10^{  -4} $ & $ 1.33 \times 10^{  -4} $ & $ 4.69 \times 10^{  -3} $ & $ 4.61 \times 10^{  -3} $ & $ 2.03 \times 10^{  -1} $ 
\\
+ SKA1 ($N_{\textrm{field}}=4$)
 & $ 6.89 \times 10^{  -4} $ & $ 1.27 \times 10^{  -4} $ & $ 4.38 \times 10^{  -3} $ & $ 4.37 \times 10^{  -3} $ & $ 2.01 \times 10^{  -1} $ 
\\
+ SKA1 ($N_{\textrm{field}}=8$)
 & $ 6.33 \times 10^{  -4} $ & $ 1.22 \times 10^{  -4} $ & $ 4.19 \times 10^{  -3} $ & $ 4.18 \times 10^{  -3} $ & $ 2.00 \times 10^{  -1} $ 
\\
+ SKA2 ($N_{\textrm{field}}=4$)
 & $ 4.77 \times 10^{  -4} $ & $ 1.16 \times 10^{  -4} $ & $ 3.00 \times 10^{  -3} $ & $ 3.09 \times 10^{  -3} $ & $ 1.92 \times 10^{  -1} $ 
\\
+ SKA2 ($N_{\textrm{field}}=8$)
 & $ 3.88 \times 10^{  -4} $ & $ 1.06 \times 10^{  -4} $ & $ 2.41 \times 10^{  -3} $ & $ 2.52 \times 10^{  -3} $ & $ 1.87 \times 10^{  -1} $ 
\\
Planck + SA(0.2) + BAO + $H_{0}$
 & $ 7.47 \times 10^{  -4} $ & $ 1.14 \times 10^{  -4} $ & $ 4.59 \times 10^{  -3} $ & $ 3.97 \times 10^{  -3} $ & $ 1.95 \times 10^{  -1} $ 
\\
\hline
\end{tabular} }
\caption{$f_{\textrm{sky}}=0.2$, inverted hierarchy, with the residual foregrounds.}
\label{tab:fsk02_fg_HieI}

\end{table}

\begin{table}[htbp]
\centering \scalebox{0.8}[0.77]{
\begin{tabular}{l|ccccc}
\hline \hline \\
& $\sigma( \Omega_{m}h^{2} )$ & $\sigma( \Omega_{b}h^{2} )$ &  $\sigma( \Omega_{\Lambda} )$ & $ \sigma( n_s ) $ & $\sigma( A_{s} \times 10^{10} )$ 
\\
\hline
Planck + PB-2(0.65) + BAO + $H_{0}$
 & $ 6.53 \times 10^{  -4} $ & $ 9.58 \times 10^{  -5} $ & $ 4.40 \times 10^{  -3} $ & $ 3.44 \times 10^{  -3} $ & $ 1.91 \times 10^{  -1} $ 
\\
+ SKA1 ($N_{\textrm{field}}=4$)
 & $ 5.91 \times 10^{  -4} $ & $ 9.04 \times 10^{  -5} $ & $ 4.13 \times 10^{  -3} $ & $ 3.23 \times 10^{  -3} $ & $ 1.89 \times 10^{  -1} $ 
\\
+ SKA1 ($N_{\textrm{field}}=8$)
 & $ 5.47 \times 10^{  -4} $ & $ 8.61 \times 10^{  -5} $ & $ 3.91 \times 10^{  -3} $ & $ 3.07 \times 10^{  -3} $ & $ 1.88 \times 10^{  -1} $ 
\\
+ SKA2 ($N_{\textrm{field}}=4$)
 & $ 3.88 \times 10^{  -4} $ & $ 8.29 \times 10^{  -5} $ & $ 2.64 \times 10^{  -3} $ & $ 2.30 \times 10^{  -3} $ & $ 1.78 \times 10^{  -1} $ 
\\
+ SKA2 ($N_{\textrm{field}}=8$)
 & $ 3.08 \times 10^{  -4} $ & $ 7.54 \times 10^{  -5} $ & $ 2.07 \times 10^{  -3} $ & $ 1.87 \times 10^{  -3} $ & $ 1.71 \times 10^{  -1} $ 
\\
Planck + SA(0.65) + BAO + $H_{0}$
 & $ 6.30 \times 10^{  -4} $ & $ 8.83 \times 10^{  -5} $ & $ 4.33 \times 10^{  -3} $ & $ 3.20 \times 10^{  -3} $ & $ 1.87 \times 10^{  -1} $ 
\\
\hline
\end{tabular} }
\caption{$f_{\textrm{sky}}=0.65$, normal hierarchy, without the residual foregrounds.}
\label{tab:fsk065_HieN}

\vspace{10pt}

\centering \scalebox{0.8}[0.77]{
\begin{tabular}{l|ccccc}
\hline \hline \\
& $\sigma( \Omega_{m}h^{2} )$ & $\sigma( \Omega_{b}h^{2} )$ &  $\sigma( \Omega_{\Lambda} )$ & $ \sigma( n_s ) $ & $\sigma( A_{s} \times 10^{10} )$ 
\\
\hline
Planck + PB-2(0.65) + BAO + $H_{0}$
 & $ 6.94 \times 10^{  -4} $ & $ 9.60 \times 10^{  -5} $ & $ 4.44 \times 10^{  -3} $ & $ 3.44 \times 10^{  -3} $ & $ 1.91 \times 10^{  -1} $ 
\\
+ SKA1 ($N_{\textrm{field}}=4$)
 & $ 5.88 \times 10^{  -4} $ & $ 9.04 \times 10^{  -5} $ & $ 4.10 \times 10^{  -3} $ & $ 3.24 \times 10^{  -3} $ & $ 1.89 \times 10^{  -1} $ 
\\
+ SKA1 ($N_{\textrm{field}}=8$)
 & $ 5.35 \times 10^{  -4} $ & $ 8.61 \times 10^{  -5} $ & $ 3.88 \times 10^{  -3} $ & $ 3.08 \times 10^{  -3} $ & $ 1.88 \times 10^{  -1} $ 
\\
+ SKA2 ($N_{\textrm{field}}=4$)
 & $ 3.82 \times 10^{  -4} $ & $ 8.29 \times 10^{  -5} $ & $ 2.64 \times 10^{  -3} $ & $ 2.32 \times 10^{  -3} $ & $ 1.79 \times 10^{  -1} $ 
\\
+ SKA2 ($N_{\textrm{field}}=8$)
 & $ 3.03 \times 10^{  -4} $ & $ 7.54 \times 10^{  -5} $ & $ 2.08 \times 10^{  -3} $ & $ 1.90 \times 10^{  -3} $ & $ 1.72 \times 10^{  -1} $ 
\\
Planck + SA(0.65) + BAO + $H_{0}$
 & $ 6.64 \times 10^{  -4} $ & $ 8.84 \times 10^{  -5} $ & $ 4.37 \times 10^{  -3} $ & $ 3.20 \times 10^{  -3} $ & $ 1.88 \times 10^{  -1} $ 
\\
\hline
\end{tabular} }
\caption{$f_{\textrm{sky}}=0.65$, inverted hierarchy, without the residual foregrounds.}
\label{tab:fsk065_HieI}

\end{table}

\begin{table}[htbp]
\centering \scalebox{0.8}[0.77]{
\begin{tabular}{l|ccccc}
\hline \hline \\
& $\sigma( \Omega_{m}h^{2} )$ & $\sigma( \Omega_{b}h^{2} )$ &  $\sigma( \Omega_{\Lambda} )$ & $ \sigma( n_s ) $ & $\sigma( A_{s} \times 10^{10} )$ 
\\
\hline
Planck + PB-2(0.65) + BAO + $H_{0}$
 & $ 6.81 \times 10^{  -4} $ & $ 1.08 \times 10^{  -4} $ & $ 4.47 \times 10^{  -3} $ & $ 3.81 \times 10^{  -3} $ & $ 1.97 \times 10^{  -1} $ 
\\
+ SKA1 ($N_{\textrm{field}}=4$)
 & $ 6.39 \times 10^{  -4} $ & $ 1.05 \times 10^{  -4} $ & $ 4.29 \times 10^{  -3} $ & $ 3.66 \times 10^{  -3} $ & $ 1.96 \times 10^{  -1} $ 
\\
+ SKA1 ($N_{\textrm{field}}=8$)
 & $ 6.04 \times 10^{  -4} $ & $ 1.02 \times 10^{  -4} $ & $ 4.13 \times 10^{  -3} $ & $ 3.54 \times 10^{  -3} $ & $ 1.95 \times 10^{  -1} $ 
\\
+ SKA2 ($N_{\textrm{field}}=4$)
 & $ 4.54 \times 10^{  -4} $ & $ 9.81 \times 10^{  -5} $ & $ 2.96 \times 10^{  -3} $ & $ 2.76 \times 10^{  -3} $ & $ 1.88 \times 10^{  -1} $ 
\\
+ SKA2 ($N_{\textrm{field}}=8$)
 & $ 3.72 \times 10^{  -4} $ & $ 9.18 \times 10^{  -5} $ & $ 2.38 \times 10^{  -3} $ & $ 2.29 \times 10^{  -3} $ & $ 1.82 \times 10^{  -1} $ 
\\
Planck + SA(0.65) + BAO + $H_{0}$
 & $ 6.32 \times 10^{  -4} $ & $ 8.89 \times 10^{  -5} $ & $ 4.33 \times 10^{  -3} $ & $ 3.22 \times 10^{  -3} $ & $ 1.86 \times 10^{  -1} $ 
\\
\hline
\end{tabular} }
\caption{$f_{\textrm{sky}}=0.65$, normal hierarchy, with the residual foregrounds.}
\label{tab:fsk065_fg_HieN}

\vspace{10pt}

\centering \scalebox{0.8}[0.77]{
\begin{tabular}{l|ccccc}
\hline \hline \\
& $\sigma( \Omega_{m}h^{2} )$ & $\sigma( \Omega_{b}h^{2} )$ &  $\sigma( \Omega_{\Lambda} )$ & $ \sigma( n_s ) $ & $\sigma( A_{s} \times 10^{10} )$ 
\\
\hline
Planck + PB-2(0.65) + BAO + $H_{0}$
 & $ 7.32 \times 10^{  -4} $ & $ 1.09 \times 10^{  -4} $ & $ 4.53 \times 10^{  -3} $ & $ 3.81 \times 10^{  -3} $ & $ 1.97 \times 10^{  -1} $ 
\\
+ SKA1 ($N_{\textrm{field}}=4$)
 & $ 6.39 \times 10^{  -4} $ & $ 1.05 \times 10^{  -4} $ & $ 4.28 \times 10^{  -3} $ & $ 3.67 \times 10^{  -3} $ & $ 1.96 \times 10^{  -1} $ 
\\
+ SKA1 ($N_{\textrm{field}}=8$)
 & $ 5.95 \times 10^{  -4} $ & $ 1.02 \times 10^{  -4} $ & $ 4.10 \times 10^{  -3} $ & $ 3.55 \times 10^{  -3} $ & $ 1.96 \times 10^{  -1} $ 
\\
+ SKA2 ($N_{\textrm{field}}=4$)
 & $ 4.47 \times 10^{  -4} $ & $ 9.83 \times 10^{  -5} $ & $ 2.96 \times 10^{  -3} $ & $ 2.79 \times 10^{  -3} $ & $ 1.88 \times 10^{  -1} $ 
\\
+ SKA2 ($N_{\textrm{field}}=8$)
 & $ 3.67 \times 10^{  -4} $ & $ 9.20 \times 10^{  -5} $ & $ 2.39 \times 10^{  -3} $ & $ 2.33 \times 10^{  -3} $ & $ 1.83 \times 10^{  -1} $ 
\\
Planck + SA(0.65) + BAO + $H_{0}$
 & $ 6.67 \times 10^{  -4} $ & $ 8.90 \times 10^{  -5} $ & $ 4.37 \times 10^{  -3} $ & $ 3.22 \times 10^{  -3} $ & $ 1.86 \times 10^{  -1} $ 
\\
\hline
\end{tabular} }
\caption{$f_{\textrm{sky}}=0.65$, inverted hierarchy, with the residual foregrounds.}
\label{tab:fsk065_fg_HieI}

\end{table}

\begin{table}[htbp]
\centering \scalebox{0.8}[0.77]{
\begin{tabular}{l|ccccc}
\hline \hline \\
& $\sigma( \Omega_{m}h^{2} )$ & $\sigma( \Omega_{b}h^{2} )$ &  $\sigma( \Omega_{\Lambda} )$ & $ \sigma( n_s ) $ & $\sigma( A_{s} \times 10^{10} )$ 
\\
\hline
Planck + PB-2(0.016) + BAO + $H_{0}$
 & $ 8.28 \times 10^{  -4} $ & $ 1.65 \times 10^{  -4} $ & $ 4.72 \times 10^{  -3} $ & $ 5.74 \times 10^{  -3} $ & $ 2.12 \times 10^{  -1} $ 
\\
+ SKA1 ($N_{\textrm{field}}=4$)
 & $ 6.84 \times 10^{  -4} $ & $ 1.44 \times 10^{  -4} $ & $ 4.34 \times 10^{  -3} $ & $ 5.04 \times 10^{  -3} $ & $ 2.09 \times 10^{  -1} $ 
\\
+ SKA1 ($N_{\textrm{field}}=8$)
 & $ 6.09 \times 10^{  -4} $ & $ 1.31 \times 10^{  -4} $ & $ 4.10 \times 10^{  -3} $ & $ 4.59 \times 10^{  -3} $ & $ 2.06 \times 10^{  -1} $ 
\\
+ SKA2 ($N_{\textrm{field}}=4$)
 & $ 4.35 \times 10^{  -4} $ & $ 1.20 \times 10^{  -4} $ & $ 2.69 \times 10^{  -3} $ & $ 2.89 \times 10^{  -3} $ & $ 1.95 \times 10^{  -1} $ 
\\
+ SKA2 ($N_{\textrm{field}}=8$)
 & $ 3.41 \times 10^{  -4} $ & $ 1.02 \times 10^{  -4} $ & $ 2.10 \times 10^{  -3} $ & $ 2.22 \times 10^{  -3} $ & $ 1.90 \times 10^{  -1} $ 
\\
Planck + SA(0.016) + BAO + $H_{0}$
 & $ 7.78 \times 10^{  -4} $ & $ 1.51 \times 10^{  -4} $ & $ 4.67 \times 10^{  -3} $ & $ 5.22 \times 10^{  -3} $ & $ 2.08 \times 10^{  -1} $ 
\\
\hline
\end{tabular} }
\caption{$f_{\textrm{sky}}=0.016$, normal hierarchy,  with the residual foregrounds of CMB, without those of 21 cm line.}
\label{tab:fsk0016_fg_HieN_SKAnoFG}

\vspace{10pt}

\centering \scalebox{0.8}[0.77]{
\begin{tabular}{l|ccccc}
\hline \hline \\
& $\sigma( \Omega_{m}h^{2} )$ & $\sigma( \Omega_{b}h^{2} )$ &  $\sigma( \Omega_{\Lambda} )$ & $ \sigma( n_s ) $ & $\sigma( A_{s} \times 10^{10} )$ 
\\
\hline
Planck + PB-2(0.016) + BAO + $H_{0}$
 & $ 9.04 \times 10^{  -4} $ & $ 1.65 \times 10^{  -4} $ & $ 4.80 \times 10^{  -3} $ & $ 5.75 \times 10^{  -3} $ & $ 2.12 \times 10^{  -1} $ 
\\
+ SKA1 ($N_{\textrm{field}}=4$)
 & $ 6.71 \times 10^{  -4} $ & $ 1.44 \times 10^{  -4} $ & $ 4.31 \times 10^{  -3} $ & $ 5.06 \times 10^{  -3} $ & $ 2.09 \times 10^{  -1} $ 
\\
+ SKA1 ($N_{\textrm{field}}=8$)
 & $ 5.89 \times 10^{  -4} $ & $ 1.31 \times 10^{  -4} $ & $ 4.06 \times 10^{  -3} $ & $ 4.62 \times 10^{  -3} $ & $ 2.06 \times 10^{  -1} $ 
\\
+ SKA2 ($N_{\textrm{field}}=4$)
 & $ 4.29 \times 10^{  -4} $ & $ 1.20 \times 10^{  -4} $ & $ 2.70 \times 10^{  -3} $ & $ 2.94 \times 10^{  -3} $ & $ 1.96 \times 10^{  -1} $ 
\\
+ SKA2 ($N_{\textrm{field}}=8$)
 & $ 3.36 \times 10^{  -4} $ & $ 1.02 \times 10^{  -4} $ & $ 2.12 \times 10^{  -3} $ & $ 2.26 \times 10^{  -3} $ & $ 1.91 \times 10^{  -1} $ 
\\
Planck + SA(0.016) + BAO + $H_{0}$
 & $ 8.53 \times 10^{  -4} $ & $ 1.51 \times 10^{  -4} $ & $ 4.75 \times 10^{  -3} $ & $ 5.23 \times 10^{  -3} $ & $ 2.08 \times 10^{  -1} $ 
\\
\hline
\end{tabular} }
\caption{$f_{\textrm{sky}}=0.016$, inverted hierarchy,  with the residual foregrounds of CMB, without those of 21 cm line.}
\label{tab:fsk0016_fg_HieI_SKAnoFG}

\end{table}

\begin{table}[htbp]
\centering \scalebox{0.8}[0.77]{
\begin{tabular}{l|ccccc}
\hline \hline \\
& $\sigma( \Omega_{m}h^{2} )$ & $\sigma( \Omega_{b}h^{2} )$ &  $\sigma( \Omega_{\Lambda} )$ & $ \sigma( n_s ) $ & $\sigma( A_{s} \times 10^{10} )$ 
\\
\hline
Planck + PB-2(0.2) + BAO + $H_{0}$
 & $ 7.53 \times 10^{  -4} $ & $ 1.33 \times 10^{  -4} $ & $ 4.61 \times 10^{  -3} $ & $ 4.60 \times 10^{  -3} $ & $ 2.03 \times 10^{  -1} $ 
\\
+ SKA1 ($N_{\textrm{field}}=4$)
 & $ 6.49 \times 10^{  -4} $ & $ 1.21 \times 10^{  -4} $ & $ 4.28 \times 10^{  -3} $ & $ 4.18 \times 10^{  -3} $ & $ 2.02 \times 10^{  -1} $ 
\\
+ SKA1 ($N_{\textrm{field}}=8$)
 & $ 5.86 \times 10^{  -4} $ & $ 1.12 \times 10^{  -4} $ & $ 4.05 \times 10^{  -3} $ & $ 3.89 \times 10^{  -3} $ & $ 2.00 \times 10^{  -1} $ 
\\
+ SKA2 ($N_{\textrm{field}}=4$)
 & $ 4.16 \times 10^{  -4} $ & $ 1.05 \times 10^{  -4} $ & $ 2.68 \times 10^{  -3} $ & $ 2.65 \times 10^{  -3} $ & $ 1.90 \times 10^{  -1} $ 
\\
+ SKA2 ($N_{\textrm{field}}=8$)
 & $ 3.27 \times 10^{  -4} $ & $ 9.21 \times 10^{  -5} $ & $ 2.09 \times 10^{  -3} $ & $ 2.09 \times 10^{  -3} $ & $ 1.85 \times 10^{  -1} $ 
\\
Planck + SA(0.2) + BAO + $H_{0}$
 & $ 6.98 \times 10^{  -4} $ & $ 1.14 \times 10^{  -4} $ & $ 4.53 \times 10^{  -3} $ & $ 3.96 \times 10^{  -3} $ & $ 1.94 \times 10^{  -1} $ 
\\
\hline
\end{tabular} }
\caption{$f_{\textrm{sky}}=0.2$, normal hierarchy, with the residual foregrounds of CMB, without those of 21 cm line.}
\label{tab:fsk02_fg_HieN_SKAnoFG}

\vspace{10pt}

\centering \scalebox{0.8}[0.77]{
\begin{tabular}{l|ccccc}
\hline \hline \\
& $\sigma( \Omega_{m}h^{2} )$ & $\sigma( \Omega_{b}h^{2} )$ &  $\sigma( \Omega_{\Lambda} )$ & $ \sigma( n_s ) $ & $\sigma( A_{s} \times 10^{10} )$ 
\\
\hline
Planck + PB-2(0.2) + BAO + $H_{0}$
 & $ 8.15 \times 10^{  -4} $ & $ 1.33 \times 10^{  -4} $ & $ 4.69 \times 10^{  -3} $ & $ 4.61 \times 10^{  -3} $ & $ 2.03 \times 10^{  -1} $ 
\\
+ SKA1 ($N_{\textrm{field}}=4$)
 & $ 6.39 \times 10^{  -4} $ & $ 1.21 \times 10^{  -4} $ & $ 4.25 \times 10^{  -3} $ & $ 4.19 \times 10^{  -3} $ & $ 2.03 \times 10^{  -1} $ 
\\
+ SKA1 ($N_{\textrm{field}}=8$)
 & $ 5.69 \times 10^{  -4} $ & $ 1.12 \times 10^{  -4} $ & $ 4.01 \times 10^{  -3} $ & $ 3.91 \times 10^{  -3} $ & $ 2.00 \times 10^{  -1} $ 
\\
+ SKA2 ($N_{\textrm{field}}=4$)
 & $ 4.10 \times 10^{  -4} $ & $ 1.06 \times 10^{  -4} $ & $ 2.68 \times 10^{  -3} $ & $ 2.69 \times 10^{  -3} $ & $ 1.91 \times 10^{  -1} $ 
\\
+ SKA2 ($N_{\textrm{field}}=8$)
 & $ 3.22 \times 10^{  -4} $ & $ 9.23 \times 10^{  -5} $ & $ 2.11 \times 10^{  -3} $ & $ 2.13 \times 10^{  -3} $ & $ 1.85 \times 10^{  -1} $ 
\\
Planck + SA(0.2) + BAO + $H_{0}$
 & $ 7.47 \times 10^{  -4} $ & $ 1.14 \times 10^{  -4} $ & $ 4.59 \times 10^{  -3} $ & $ 3.97 \times 10^{  -3} $ & $ 1.95 \times 10^{  -1} $ 
\\
\hline
\end{tabular} }
\caption{$f_{\textrm{sky}}=0.2$, inverted hierarchy, with the residual foregrounds of CMB, without those of 21 cm line.}
\label{tab:fsk02_fg_HieI_SKAnoFG}

\end{table}

\begin{table}[htbp]
\centering \scalebox{0.8}[0.77]{
\begin{tabular}{l|ccccc}
\hline \hline \\
& $\sigma( \Omega_{m}h^{2} )$ & $\sigma( \Omega_{b}h^{2} )$ &  $\sigma( \Omega_{\Lambda} )$ & $ \sigma( n_s ) $ & $\sigma( A_{s} \times 10^{10} )$ 
\\
\hline
Planck + PB-2(0.65) + BAO + $H_{0}$
 & $ 6.81 \times 10^{  -4} $ & $ 1.08 \times 10^{  -4} $ & $ 4.47 \times 10^{  -3} $ & $ 3.81 \times 10^{  -3} $ & $ 1.97 \times 10^{  -1} $ 
\\
+ SKA1 ($N_{\textrm{field}}=4$)
 & $ 6.09 \times 10^{  -4} $ & $ 1.01 \times 10^{  -4} $ & $ 4.19 \times 10^{  -3} $ & $ 3.55 \times 10^{  -3} $ & $ 1.98 \times 10^{  -1} $ 
\\
+ SKA1 ($N_{\textrm{field}}=8$)
 & $ 5.60 \times 10^{  -4} $ & $ 9.58 \times 10^{  -5} $ & $ 3.98 \times 10^{  -3} $ & $ 3.36 \times 10^{  -3} $ & $ 1.96 \times 10^{  -1} $ 
\\
+ SKA2 ($N_{\textrm{field}}=4$)
 & $ 3.97 \times 10^{  -4} $ & $ 9.15 \times 10^{  -5} $ & $ 2.66 \times 10^{  -3} $ & $ 2.44 \times 10^{  -3} $ & $ 1.87 \times 10^{  -1} $ 
\\
+ SKA2 ($N_{\textrm{field}}=8$)
 & $ 3.14 \times 10^{  -4} $ & $ 8.21 \times 10^{  -5} $ & $ 2.08 \times 10^{  -3} $ & $ 1.97 \times 10^{  -3} $ & $ 1.80 \times 10^{  -1} $ 
\\
Planck + SA(0.65) + BAO + $H_{0}$
 & $ 6.32 \times 10^{  -4} $ & $ 8.89 \times 10^{  -5} $ & $ 4.33 \times 10^{  -3} $ & $ 3.22 \times 10^{  -3} $ & $ 1.86 \times 10^{  -1} $ 
\\
\hline
\end{tabular} }
\caption{$f_{\textrm{sky}}=0.65$, normal hierarchy, with the residual foregrounds of CMB, without those of 21 cm line.}
\label{tab:fsk065_fg_HieN_SKAnoFG}

\vspace{10pt}

\centering \scalebox{0.8}[0.77]{
\begin{tabular}{l|ccccc}
\hline \hline \\
& $\sigma( \Omega_{m}h^{2} )$ & $\sigma( \Omega_{b}h^{2} )$ &  $\sigma( \Omega_{\Lambda} )$ & $ \sigma( n_s ) $ & $\sigma( A_{s} \times 10^{10} )$ 
\\
\hline
Planck + PB-2(0.65) + BAO + $H_{0}$
 & $ 7.32 \times 10^{  -4} $ & $ 1.09 \times 10^{  -4} $ & $ 4.53 \times 10^{  -3} $ & $ 3.81 \times 10^{  -3} $ & $ 1.97 \times 10^{  -1} $ 
\\
+ SKA1 ($N_{\textrm{field}}=4$)
 & $ 6.03 \times 10^{  -4} $ & $ 1.01 \times 10^{  -4} $ & $ 4.16 \times 10^{  -3} $ & $ 3.56 \times 10^{  -3} $ & $ 1.98 \times 10^{  -1} $ 
\\
+ SKA1 ($N_{\textrm{field}}=8$)
 & $ 5.46 \times 10^{  -4} $ & $ 9.58 \times 10^{  -5} $ & $ 3.94 \times 10^{  -3} $ & $ 3.37 \times 10^{  -3} $ & $ 1.96 \times 10^{  -1} $ 
\\
+ SKA2 ($N_{\textrm{field}}=4$)
 & $ 3.91 \times 10^{  -4} $ & $ 9.16 \times 10^{  -5} $ & $ 2.66 \times 10^{  -3} $ & $ 2.47 \times 10^{  -3} $ & $ 1.87 \times 10^{  -1} $ 
\\
+ SKA2 ($N_{\textrm{field}}=8$)
 & $ 3.09 \times 10^{  -4} $ & $ 8.22 \times 10^{  -5} $ & $ 2.09 \times 10^{  -3} $ & $ 1.99 \times 10^{  -3} $ & $ 1.81 \times 10^{  -1} $ 
\\
Planck + SA(0.65) + BAO + $H_{0}$
 & $ 6.67 \times 10^{  -4} $ & $ 8.90 \times 10^{  -5} $ & $ 4.37 \times 10^{  -3} $ & $ 3.22 \times 10^{  -3} $ & $ 1.86 \times 10^{  -1} $ 
\\
\hline
\end{tabular} }
\caption{$f_{\textrm{sky}}=0.65$, inverted hierarchy, with the residual foregrounds of CMB, without those of 21 cm line.}
\label{tab:fsk065_fg_HieI_SKAnoFG}

\end{table}

%%%%%%%%%%%%%%%%%%%%%%%%%%%%%%%%%%%%%%%%%%%%%%%%%%%%%%%%%%%%%%%%%%%%%
\section{Conclusions}
\label{sec:conclusion}
%%%%%%%%%%%%%%%%%%%%%%%%%%%%%%%%%%%%%%%%%%%%%%%%%%%%%%%%%%%%%%%%%%%%%

In this paper,
we have studied how well we can constrain the total neutrino mass $\Sigma
m_{\nu}$, the effective number of neutrino species $N_{\nu}$
and the neutrino mass hierarchy by using 21 cm line (SKA) and CMB 
(Planck + \textsc{Polarbear}-2 or Simons Array) observations.
It is essential to combine the 21 cm line observation 
with the precise CMB polarization observation 
to break various degeneracies in cosmological parameters
when we perform multiple-parameter fittings.

About the constraints on the $\Sigma m_{\nu}$--$N_{\nu}$ plane, 
we have found that there is a significant improvement in 
the sensitivity to $\Sigma m_{\nu}$ and $N_{\nu}$ 
by adding the information of BAO observation to that of CMB. 
However, for a fiducial value $\Sigma m_{\nu}=0.1$ eV, 
it is impossible to detect the nonzero neutrino mass
at 2$\sigma$ level even by using the combination of Simons Array and DESI.
On the other hand, by adding the information of 21 cm observation (SKA) 
to that of CMB, we have found that there is a substantial improvement.
By using Planck + Simons Array + BAO~(DESI) + SKA phase~1, we can detect
the nonzero neutrino mass if the value satisfies $\Sigma m_{\nu} \geq $ 0.1 eV
(but it is necessary to remove foregrounds with high degree of accuracy).
For a fiducial value $\Sigma m_{\nu}=0.06$ eV, which corresponds to
the lowest value in the normal hierarchy of the neutrino total mass, 
we need the sensitivity of SKA phase~2 with
relatively large sky coverage and 
very strong foreground removal
in order to detect the nonzero neutrino mass at 2$\sigma$ level.

As for the determination of the neutrino mass hierarchy,
we have used the parameter $r_{\nu}= (m_{3} - m_{1})/\Sigma m_{\nu}$,
and studied how to discriminate a true mass hierarchy from the other
by constraining $r_{\nu}$.  
As was clearly shown in Fig.~\ref{fig:hie_ellipse_fsky0016}-\ref{fig:hie_ellipse_m8}, 
by adopting the combination of Planck + Simons Array + BAO~(DESI) + SKA phase~2, 
%we will be able to determine the hierarchy to be inverted or normal at 2$\sigma$
%for $ \Sigma m_{\nu}\lesssim 0.1$~eV or $\lesssim 0.06$~eV, respectively.
we will be able to determine the hierarchy to be inverted or normal at 2$\sigma$
unless the mass structure is  degenerated.
By using Omniscope, we can discriminate any mass hierarchies
up to $\Sigma m_{\nu}\sim 0.1 $ eV \cite{Oyama:2012tq}.

%for $ \Sigma m_{\nu}\lesssim 0.1$~eV or $\lesssim 0.06$~eV, respectively.
%

%Our results indicate that the 21~cm line and CMB polarization observation 
%will become powerful probe of the neutrino properties 
%and the origin of matter in the Universe.
%
%Our results indicate that the 21~cm line and CMB polarization observation 
Our results indicate that 
combining the 21 cm line observations with the CMB polarization observations
%have the powerful capability of proving the neutrino property 
has strong impacts on the determinations of the neutrino property 
and the origin of matter in the Universe.

\newpage

%%%%%%%%%%%%%%%%%%%%%%%%%%%%%%%%%%%%%%%%%%%%%%%%%%%%%%%%%%%%%%%%%%%%%
\section*{Acknowledgments}
%%%%%%%%%%%%%%%%%%%%%%%%%%%%%%%%%%%%%%%%%%%%%%%%%%%%%%%%%%%%%%%%%%%%%

We thank Kiyotomo Ichiki for a useful correspondence about specications of SKA, and
Maresuke Shiraishi for useful comments about treatments of the residual foregrounds of CMB.
This work is supported in part by
MEXT KAKENHI Grant Numbers 15H05889 (K.K.), 15H05891 (M.H.),
JSPS KAKENHI Grant Numbers  26105520 (K.K.), 26247042 (K.K.) and 26220709 (M.H.).
The work of K.K. is also supported by the Center for the Promotion of Integrated Science
(CPIS) of Sokendai (1HB5804100).

\section*{Note added}

While finalizing this manuscript, Ref.\cite{Liu:2015txa} appeared which has some
overlaps with this work.

%%%%%%%%%%%%%%%%%%%%%%%%%%%%%%%%%%%%%%%%%%%%%%%%%%%%%%%%%%%%%%%%%%%%%
%\appendix
%\bigskip
%\bigskip
%\noindent 
%{\Large \bf Appendix}

%\section{1$\sigma$ error tables}

%%%%%%%%%%%%%%%%%%%%%%%%%%%%%%%%%%%%%%%%%%%%%%%%%%%%%%%%%%%%%%%%%%%%%

\newpage

%%%%%%%%%%%%%%%%%%%%%%%%%%%%%%%%%%%%%%%%%%%%%%%%%%%%%%%%%%%%%%%
% References
%%%%%%%%%%%%%%%%%%%%%%%%%%%%%%%%%%%%%%%%%%%%%%%%%%%%%%%%%%%%%%%

%%%%%%%%%%%%%%%%%%%%%%%%%%%%%%%%%%%%%%%%%%%%%%%%%%%%%%%%%%%%%%%
%%%%%%%%%%%%%%%%%%%%%%%%%%%%%%%%%%%%%%%%%%%%%%%%%%%%%%%%%%%%%%%

\end{document}